\documentclass[10pt]{article}
\usepackage{url}
\usepackage{hyperref}
\usepackage{subfigure}
\usepackage{booktabs}
\usepackage{graphicx}
\usepackage{float}
\usepackage{amssymb}
\usepackage{amsmath}
\usepackage{array}
\usepackage{caption}
\usepackage{ctable}
\usepackage{chngcntr}
\usepackage{xcolor}

\begin{document}

\title{\large \bf Parametric Resonance in $\phi^4$ Preheating: An Exact Numerical Study}

        \author{\normalsize \sc Hrisikesh Thakur\thanks{\tt hthakur@iitg.ac.in} \ {\rm and} Malay K. Nandy\thanks{\tt mknandy@iitg.ac.in \rm (Corresponding Author)}\\
        \normalsize \it Department of Physics, Indian Institute of Technology Guwahati,
        \normalsize Guwahati 781 039, India}

        \date{\normalsize (01 July 2026)}
\maketitle

\begin{abstract}

Preheating after inflation proceeds through parametric resonance, leading to efficient particle production in scalar field models. In this work, we investigate the structure of parametric resonance in the $\phi^4$ chaotic inflationary model during the preheating phase by performing a fully numerical analysis of the coupled dynamical equations governing the inflaton field and the mode function of the produced particles, thereby avoiding the approximations commonly employed in earlier studies. Our results reveal resonance patterns that differ significantly from those obtained with approximate analytical treatments. In the weak coupling regime, short-wavelength modes rapidly settle into oscillations with nearly constant amplitude, while the corresponding occupation numbers approach saturation. However, the long-wavelength modes exhibit gradual amplitude growth, with occupation numbers transitioning into a non-linear oscillatory regime. As the coupling strength increases, the dynamics becomes increasingly nonlinear, leading to the emergence of stochastic behavior. In the strong coupling regime, short-wavelength modes display a step-like (staircase) evolution in the occupation number, indicative of intermittent bursts of particle production. However, the long-wavelength modes exhibit a more gradual, monotonic growth with small superimposed fluctuations. These findings highlight the rich, coupling-dependent, structure of parametric resonance in the quartic inflationary model and underscore the importance of exact numerical treatment in accurately capturing preheating dynamics.
\\


\end{abstract}

\tableofcontents

\section{Introduction}\label{sec_intro}

The inflationary paradigms of Guth \cite{Guth1981}, Starobinsky \cite{Starobinsky1980}, Kazanas \cite{Kazanas1980}, and Sato \cite{Sato1981} were proposed in the early 80's in order to resolve the flatness, horizon and monopole problems of the standard big-bang cosmology. Subsequently, Linde \cite{Linde1982} and Albrecht and Steinhardt \cite{Albrecht1982} proposed the new inflation theory which was further revised by Linde \cite{Linde1983} leading to the chaotic inflation theory. Chaotic inflation has been the most successful one where the inflaton field undergoes slow-roll phase so that a sufficient amount of inflation is achieved.

Building on the foundation of chaotic inflation, subsequent research focused on the detailed dynamics of the inflationary phase. Inflation has been analyzed as a supercooled phase transition in the early Universe \cite{Hawking1982}, and the slow-roll nature of the inflaton field has been studied extensively \cite{Steinhardt1984, Liddle1992, Liddle1994}. These investigations aimed to understand the conditions under which inflation could not only resolve cosmological puzzles but also yield observable predictions, such as the amplitude of density perturbations and the thermal history of the Universe. Steinhardt and Turner \cite{Steinhardt1984} developed a general framework for characterizing the inflaton potential within de Sitter space, identifying criteria necessary for prolonged inflation, generation of suitable density fluctuations, and sufficiently high reheating temperatures to explain the observed baryon asymmetry.

Liddle and collaborators later formalized the slow-roll approximation \cite{Liddle1994}, providing a systematic treatment of the inflaton's dynamics under conditions where its potential dominates over kinetic energy. Furthermore, considerable attention has been given to understanding the onset of slow-roll inflation, with numerous studies examining the sensitivity of inflationary dynamics to initial conditions \cite{Coughlan1985, Linde1985, Albrecht1985, Albrecht1985b, Albrecht1987}. Several early comprehensive reviews provide overviews of these developments and continue to serve as valuable references for the field \cite{Linde2005, Linde2007, Olive1990, Baumann2011, Martin2014}.

Complementing the earlier analytical and numerical efforts, subsequent studies explored the onset of inflation in increasingly realistic and complicated settings. Kurki-Suonio et al. \cite{Kurki-Suonio1987} extended the work of Albrecht et al. \cite{Albrecht1985b, Albrecht1987} by simulating a dynamical spacetime coupled to both scalar field and fluid components. Their results were largely consistent, though they observed that for a $\phi^4$-type potential, inflation failed to occur. Goldwirth and Piran \cite{Goldwirth1990} further demonstrated that small initial inhomogeneities reduce the amount of inflation, while sufficiently large ones can inhibit it altogether. They also noted that inflation fails to occur when only sub-horizon modes—specifically those with wavelengths less than three times the horizon size—are excited. However, in a subsequent analysis, Goldwirth and Piran \cite{Goldwirth1992} showed that in the case of chaotic inflation, inflationary dynamics are generically robust and largely insensitive to initial conditions. Supporting this, Kurki-Suonio et al. \cite{Kurki-Suonio-1993} confirmed via numerical simulations that sufficient inflation can still be achieved in inhomogeneous settings.

In a related study, Iguchi and Ishihara \cite{Iguchi1997} examined the onset of inflation in a locally closed Friedmann universe with initial inhomogeneities. Their results suggested that, for inflation to commence, the spatial curvature must be below a critical threshold over regions several times larger than the local curvature radius. Vachaspati and Trodden \cite{Vachaspati1999} argued that chaotic inflation could still proceed without requiring large-scale homogeneity at the outset, lending further credibility to the naturalness of inflationary initial conditions. More recently, Easther et al. \cite{Easther2014} explored multifield inflation and found that small subhorizon inhomogeneities can significantly influence the onset of inflation. In an even more striking result, East et al. \cite{East2016} demonstrated that inflation can emerge even from highly inhomogeneous initial configurations, provided that the fluctuations are confined to a sufficiently flat region of the inflaton potential. Importantly, their analysis showed that the presence of an initially homogeneous Hubble-sized patch is not a necessary condition for inflation to begin.

While inflation provides elegant solutions to the horizon, flatness, and monopole problems of standard cosmology, it practically results in a cold and empty Universe, devoid of the matter and radiation necessary for the subsequent hot big-bang evolution. Therefore, to explain the existence of observed matter and radiation in the Universe, a post-inflationary mechanism must exist to populate the Universe with matter and radiation and to establish thermal equilibrium.

The entire mechanism of particle creation and their thermalization following the inflationary period is usually known as reheating of the Universe. Since thermalisation takes place by energy exchange among the particles during their collisions, the created particles do not acquire thermal equilibrium in the initial phase of reheating. This initial phase is often called the {\em preheating} phase of the Universe.

A major breakthrough in the understanding of the preheating mechanism came from the work of Kofman, Linde, and Starobinsky \cite{Kofman1997}. They showed that the initial preheating process happens due to explosive particle creation via parametric resonance. They took the chaotic model in its quadratic form [$V(\phi) = \frac{1}{2} m^2\phi^2$] and allowed the inflaton field $\phi$ to decay into a bosonic field $\chi$ and showed that the mode function of the $\chi$-field evolves according to the Mathieu equation with its frequency term governed by the sinusoidal oscillation of the inflaton field. With different conditions on the parameters involved, their numerical solutions of the Mathieu equation yielded narrow, broad and stochastic resonances.

In a later study, Greene et al. \cite{Greene1997} took the model of chaotic inflation in its quartic form [$V(\phi) = \frac{1}{4}\lambda \phi^4$]. Being quartic, the oscillations of the inflaton field in the reheating era is no longer sinusoidal, prohibiting the construction of a closed analytic form of the frequency term in the corresponding Mathieu-type equation for the mode function in this case. To circumvent this difficulty, they simplified the dynamical equations, expressed with respect to the conformal time, by considering average over many oscillations so as to neglect the intractable terms in those equations. The resulting equation for the mode function turned out to be \text{Lam\'e}-type with the frequency term given by an elliptic cosine function varying periodically in conformal time. This analytic approximation for the construction of the \text{Lam\'e}-type equation facilitated a direct numerical study which showed resonance patterns quite different from those obtained in the quadratic case, indicating that the nature of preheating is highly sensitive to the form of the equation governing the parametric resonance.

Kaiser \cite{kaiser1998} reconsidered the problem of preheating in the context of the quartic inflaton potential, focusing on massless fields interacting via a term of the form $\frac{1}{2} g^2 \phi^2 \chi^2$. While earlier work by Greene et al. investigated preheating in a spatially flat background, Kaiser was motivated to explore this process in an open universe to contrast the dynamics between the two scenarios. To derive analytical solutions for the evolution of the fields $\phi$ and $\chi$, the Hartree approximation was employed—valid in the linear regime before nonlinear effects become significant. This framework enabled analytical calculation of the Floquet exponent, demonstrating that resonant modes can grow significantly faster in an open universe compared to the flat case. Furthermore, the averaged scale factor was shown to evolve like that of a radiation-dominated universe \cite{kaiser1997}, resulting in a vanishing Ricci scalar and a conformal equivalence between preheating with minimally coupled massless fields in an expanding universe and the corresponding dynamics in Minkowski spacetime.

In this context, Calzetta and Hu \cite{Calzetta1995} studied how classical stochastic behavior emerges from quantum fluctuations in the early universe. Within the framework of stochastic nonlinear dynamics and for a variety of scalar potentials including $\lambda \phi^4$, they demonstrated that decoherence of the quantum mean field by its own fluctuations results in classical stochastic perturbations. This led to the prediction that the resulting density contrast is of the order of the quartic coupling, suggesting $\lambda \sim 10^{-6}$. Building on these ideas, Bassett \cite{Bassett1999} argued that for moderate coupling values $g < 10^{-3}$, the system exhibits only weak parametric resonance, thereby constraining the efficiency of preheating in such regimes.

Easther and Parry \cite{Easther2000} investigated the impact of nonlinear gravitational effects on the preheating phase in the quartic inflation model. By considering inhomogeneities in the gravitational metric along a single spatial direction, they found that the overall dynamics remains qualitatively similar to those obtained in a rigid (non-dynamical) spacetime background. Their analysis is divided into two cases based on initial conditions: in the first, only a single mode of the field perturbation has a non-zero initial amplitude; in the second, all modes are initially excited. For the first case, they observed that the decay of perturbations seen in rigid spacetime, after resonant growth ceases, does not occur when nonlinear gravitational effects are included. In the second case, long-wavelength modes grow earlier and more strongly in the presence of nonlinear gravitational effects than in the rigid case, though their growth halts once the resonance ends.

Jin and Tsujikawa \cite{Jin_2006} investigated the emergence of chaotic dynamics during the post-inflationary preheating phase in power-law inflationary models of the form $ V(\phi) \sim \phi^n $, considering the cases $ n = 2 $ and $ n = 4 $ with an interaction term $ \frac{1}{2} g^2 \phi^2 \chi^2 $. Their work was motivated by a suggestion from Podolsky and Starobinsky \cite{Podolsky2002}, who proposed that parametric resonance could lead to chaotic behavior when the coupling ratio $ g^2 / \lambda $ is not too large, potentially modifying the standard picture of preheating. Jin and Tsujikawa derived conditions under which chaos arises, showing that the resonant amplification of the $\chi$-field can drive significant mixing with the inflaton, thereby inducing a chaotic instability. In such regimes, the conventional Floquet analysis based on the Lamé equation becomes invalid due to the breakdown of periodicity. Through the use of fractal maps, they confirmed the presence of chaos in parameter ranges where $ g^2 / \lambda $ remains moderate.

Nambu and Araki \cite{Nambu_2006} studied the nonlinear evolution of field fluctuations during the preheating phase driven by parametric resonance in a quartic potential with an interaction term $ g^2 \phi^2 \chi^2 $. Focusing specifically on the case $ g^2/\lambda = 2 $, they examined the dynamics of the longest wavelength modes ($ k \sim 0 $) within the strong resonance band. As the massless $\chi$-field grows and its amplitude becomes comparable to that of the inflaton, the interaction term dominates, driving the system into a chaotic regime. Particular attention was given to the evolution of the power spectrum of curvature perturbations during this phase, with an emphasis on long-wavelength nonlinear fluctuations. Their analysis showed that in
\begin{figure}
 \centering
 \includegraphics[width=.6\textwidth]{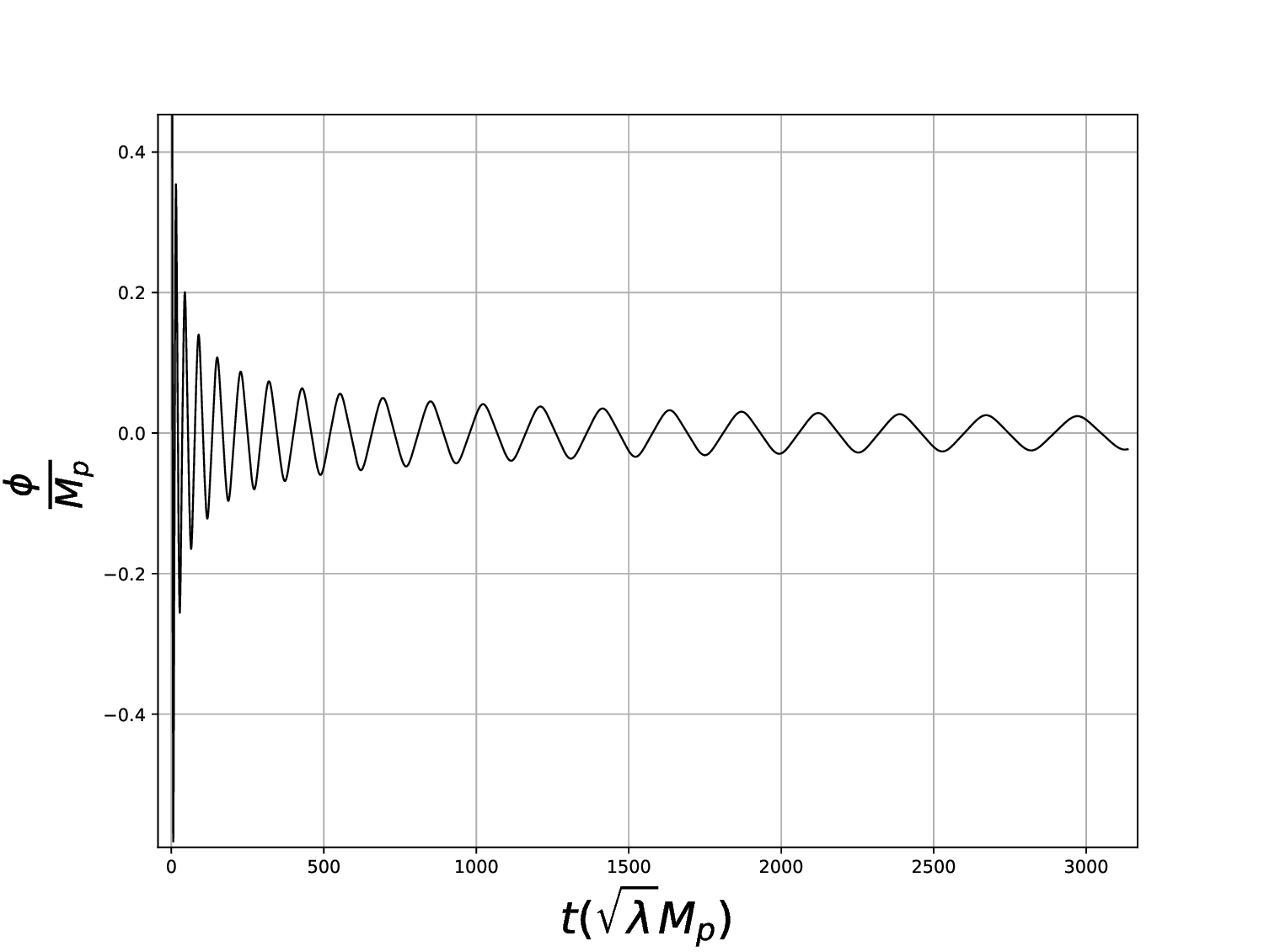}
 \caption{Damped oscillation of the inflaton field $\phi(t)$ in the preheating regime.}
 \label{fig:p4vt}
\end{figure}
the nonlinear regime, the amplification of fluctuations eventually saturates, and the power spectrum evolves significantly as the system transitions into chaotic behavior.

Suyama and Yokoyama \cite{Suyama_2007} investigated a two-field preheating scenario with the potential $\lambda \phi^4 + \frac{1}{2} g^2 \phi^2 \chi^2$, considering coupling ratios in the range $g^2/\lambda = 1 \sim 3$. They numerically solved the coupled field equations for varying initial values of the $\chi$-field to study the post-preheating evolution of curvature perturbations. Their results revealed significant deviations in the final large-scale curvature perturbations from those predicted by the Hartree approximation, highlighting the importance of nonlinear dynamics in shaping the outcome of preheating.

Jedamzik et al.~\cite{Jedamzik_2010} examined the evolution of scalar perturbations in the post-inflationary oscillatory phase, focusing on the behavior of the Mukhanov variable for both quadratic and quartic inflaton potentials. In the quadratic case, the dynamics of the Mukhanov variable is governed by the Mathieu equation, whereas in the quartic case, the corresponding equation takes the form of a Lam\'e equation driven by a Jacobi elliptic cosine function. These analytical forms allow the identification of instability bands from Floquet charts. However, numerical integration of the full equations of motion revealed deviations from the expected band structure, indicating a shift in the location and nature of the resonance bands due to nonlinear effects.

For a general inflaton potential $ V(\phi) $ of complicated form, it is often not possible to recast the equations governing parametric resonance into an exact Mathieu-type form due to the complicated and nonlinear nature of the resulting dynamics. One notable example is the quartic potential, where the governing equation for mode functions can be reduced to the form of a Lam\'e equation {\em only} under specific approximations, as demonstrated in Ref.~\cite{Greene1997}.

In the present work, we focus on the post-inflationary dynamics governed by the quartic inflaton potential ($\phi^4$) coupled with another scalar field through an interaction term $g^2\phi^2\chi^2$ with the aim of obtaining the {\em exact} structure of parametric resonance during the preheating phase that follows the end of inflation. We mainly investigate how the resonant amplification of quantum fluctuations associated with particle production is influenced by the inherent nonlinear oscillatory behavior of the inflaton field. Consequently, we concentrate on the coupled dynamical equations governing the evolution of the mode functions of the produced particles, which are sourced by the exact nature of background {\em non-sinusoidal} oscillations of the inflaton in the preheating regime without resorting to any approximations.

The dynamical equations are then solved {\em exactly} via numerical integration, allowing us to capture the full nonlinear evolution of the inflaton and the produced field mode without analytical approximations such as the Hartree or Lam\'e-type approaches. Our results reveal a resonance structure that differs significantly from the one found in the $\phi^4$ inflationary model, as studied earlier with approximations to construct the differential equation governing the evolution of the mode function exhibiting sustained growth over successive oscillations.

\begin{figure}[H]

    \includegraphics[width=.47\textwidth, height=5cm]{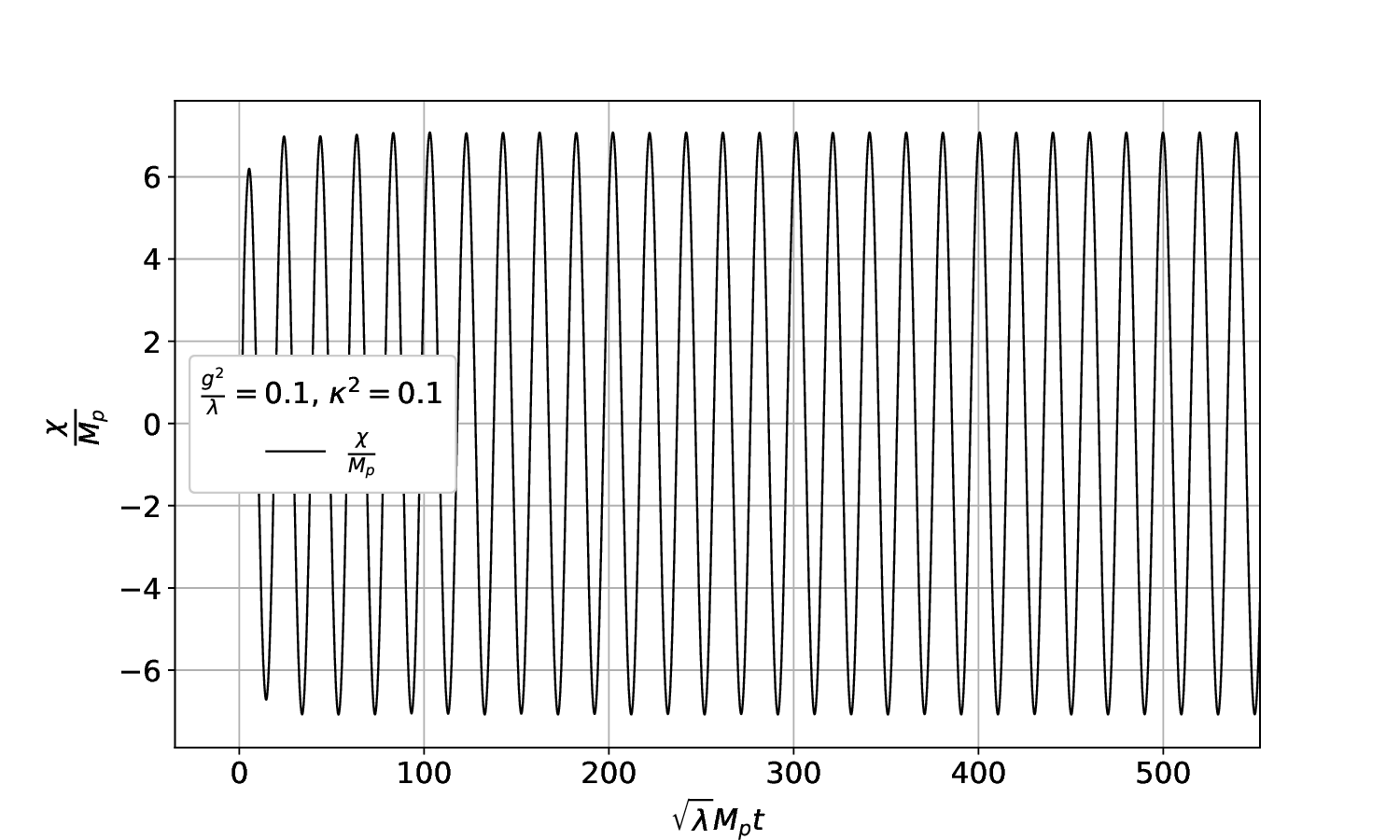}
    \includegraphics[width=.47\textwidth, height=5cm]{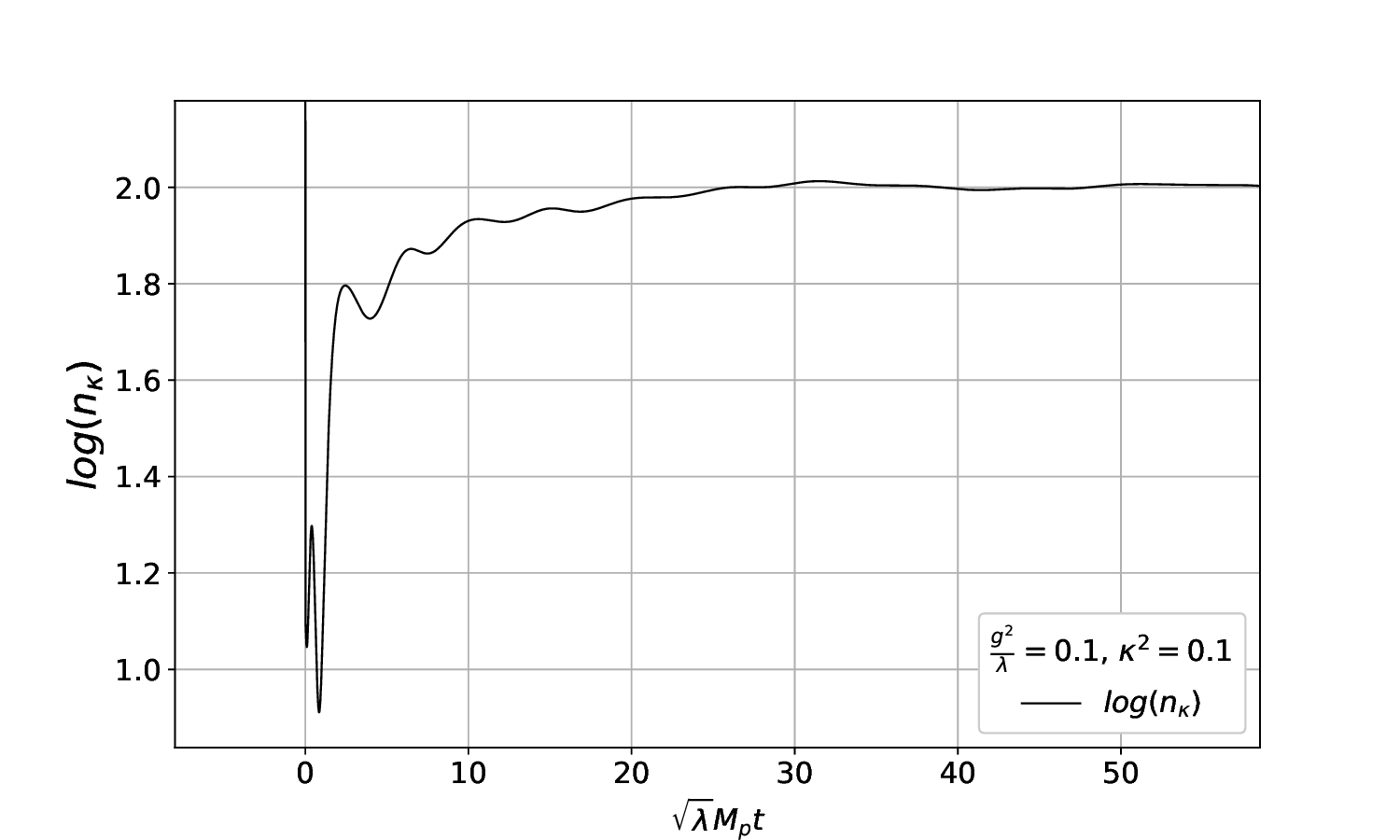}
    \includegraphics[width=.47\textwidth, height=5cm]{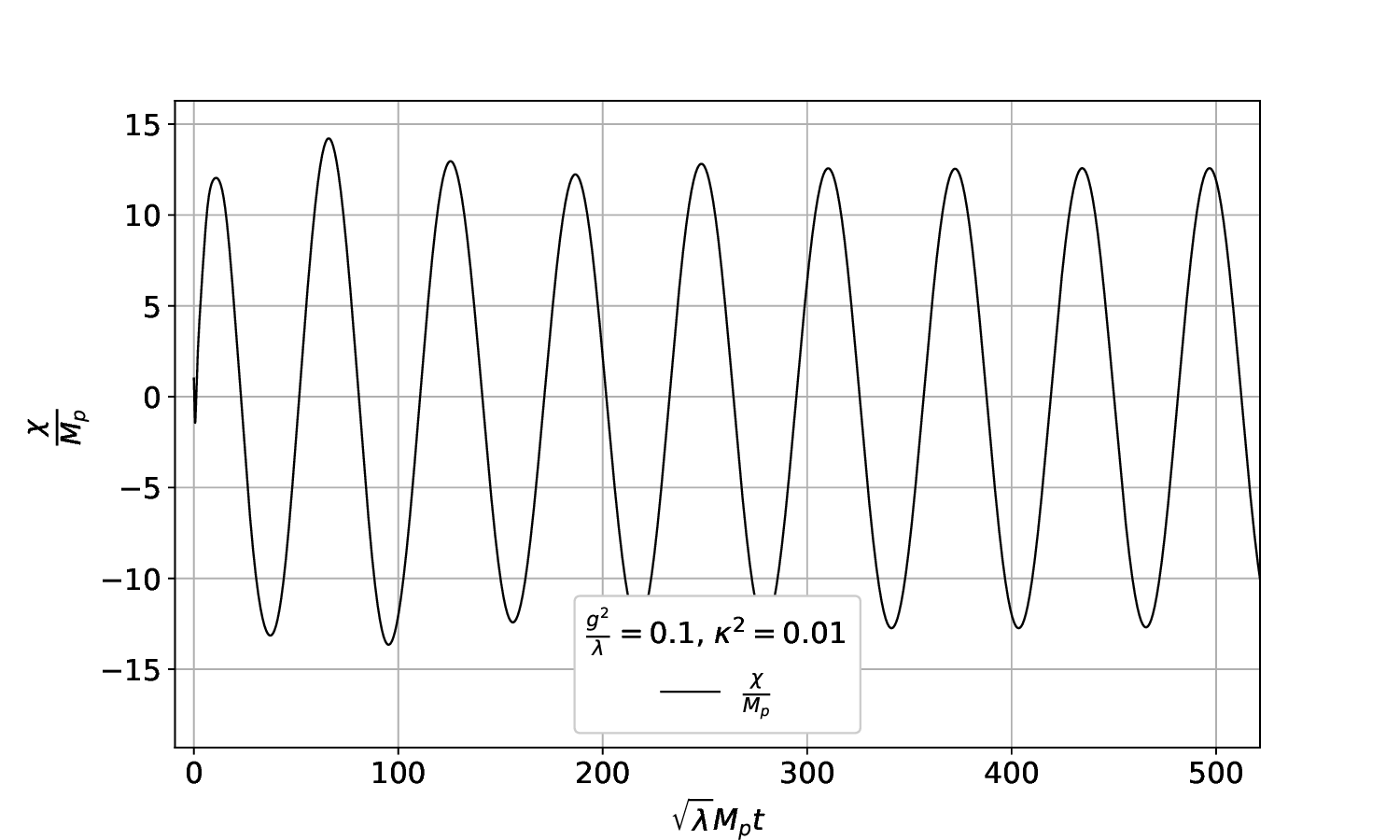}
    \includegraphics[width=.47\textwidth, height=5cm]{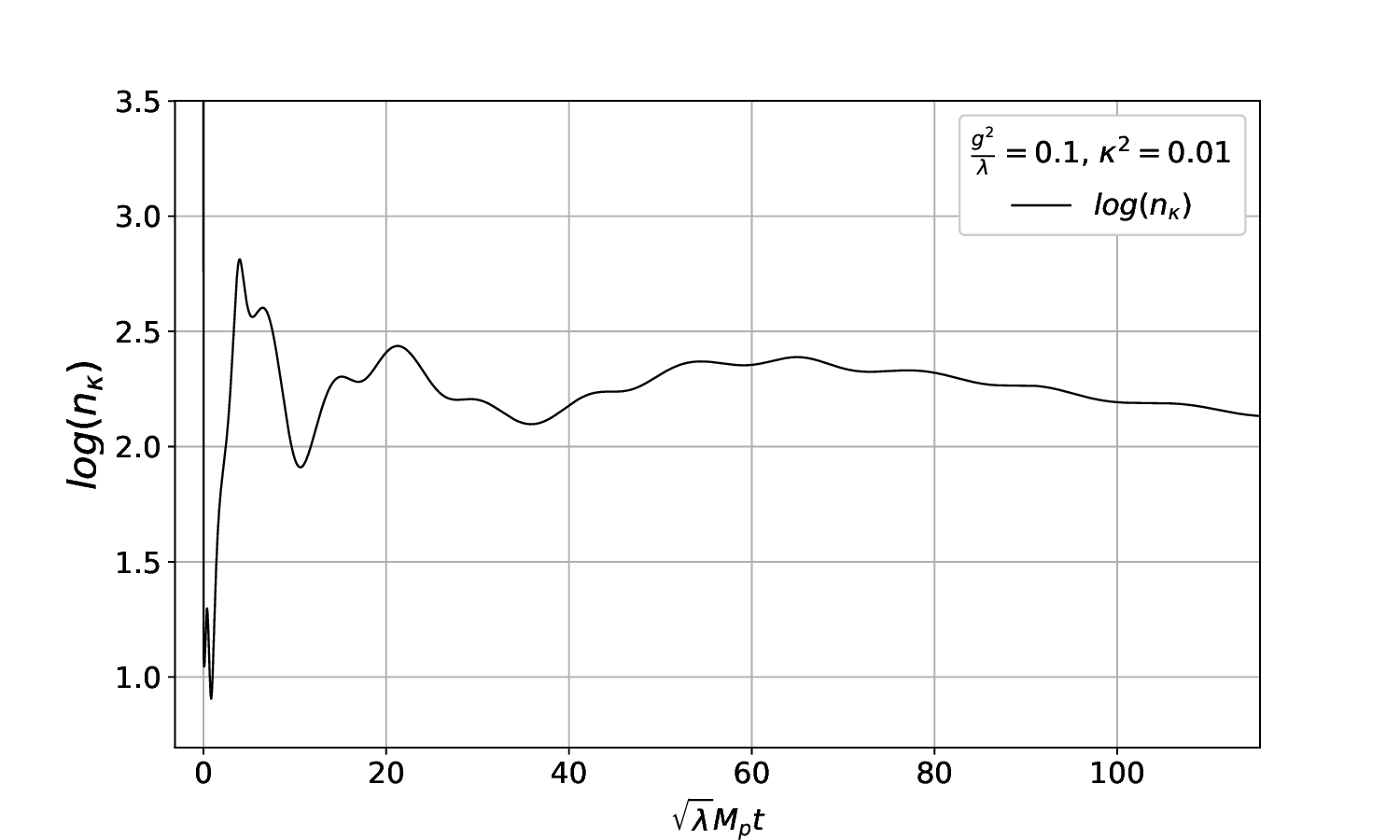}
    \includegraphics[width=.47\textwidth, height=5cm]{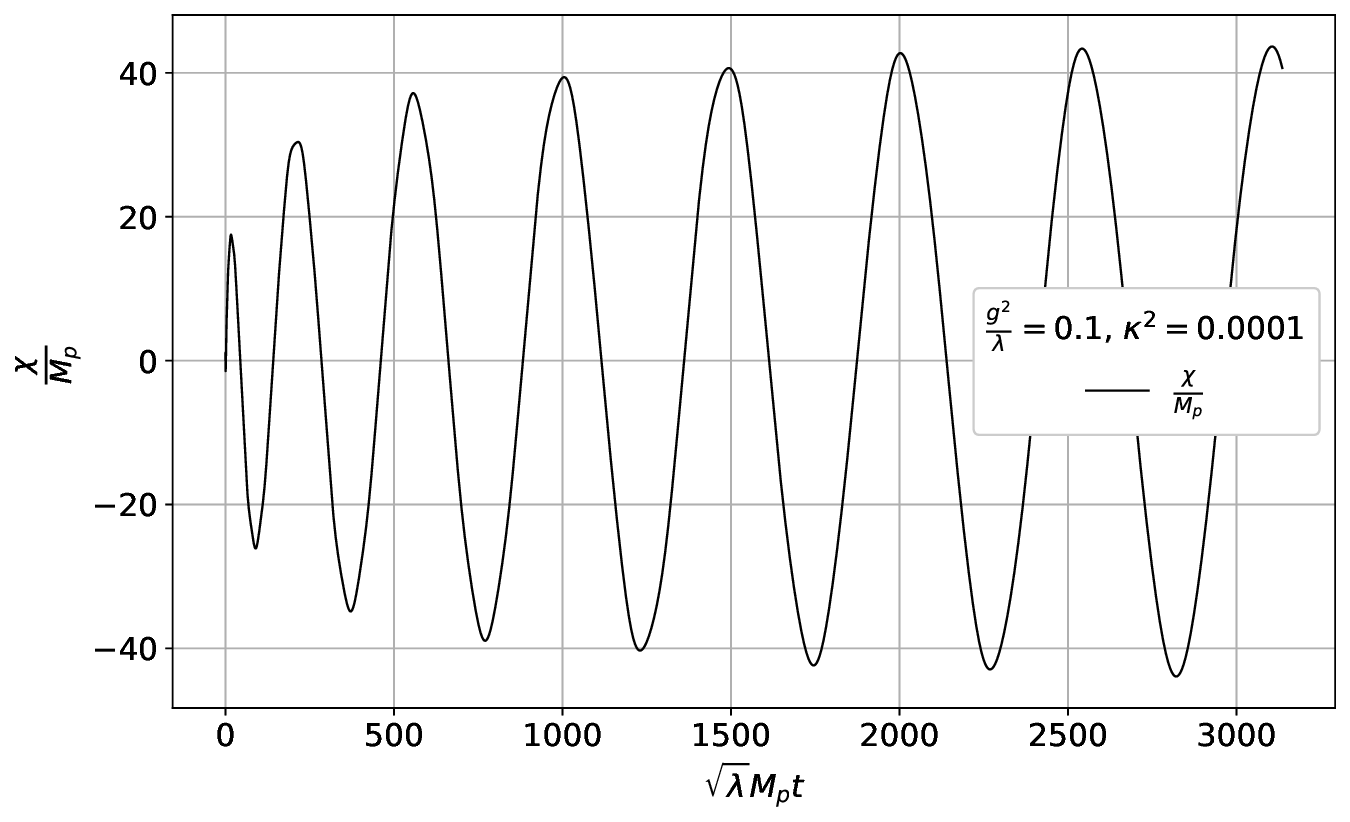}
    \includegraphics[width=.47\textwidth, height=5cm]{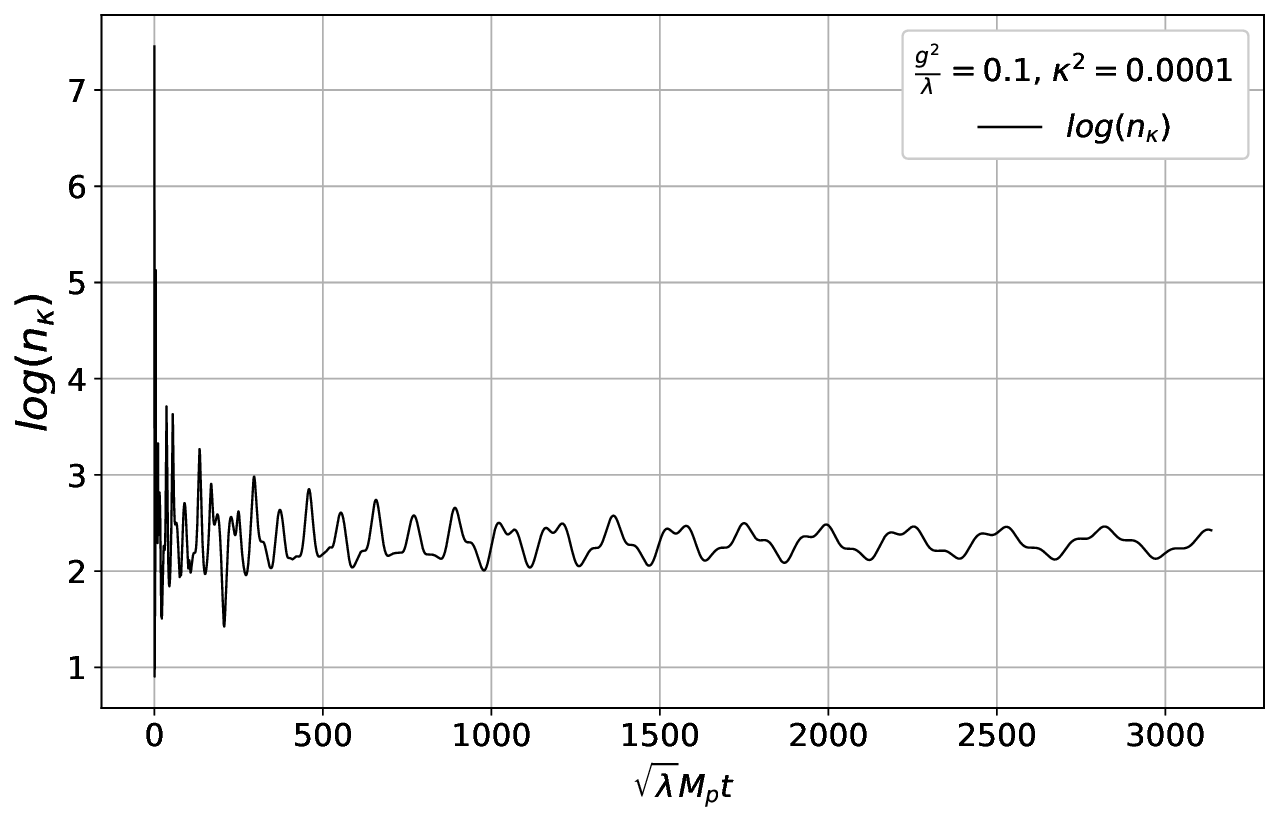}
    \includegraphics[width=.47\textwidth, height=5cm]{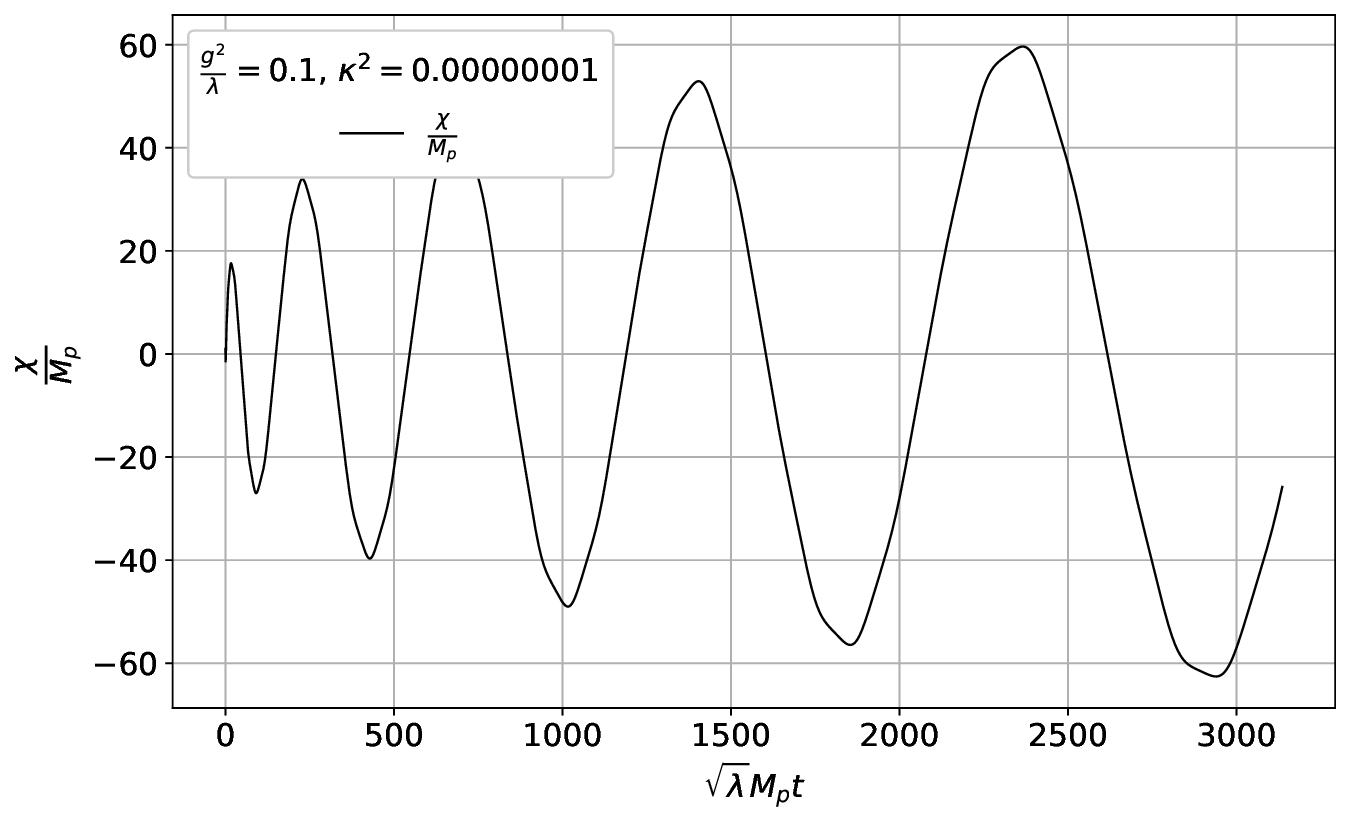}
    \includegraphics[width=.47\textwidth, height=5cm]{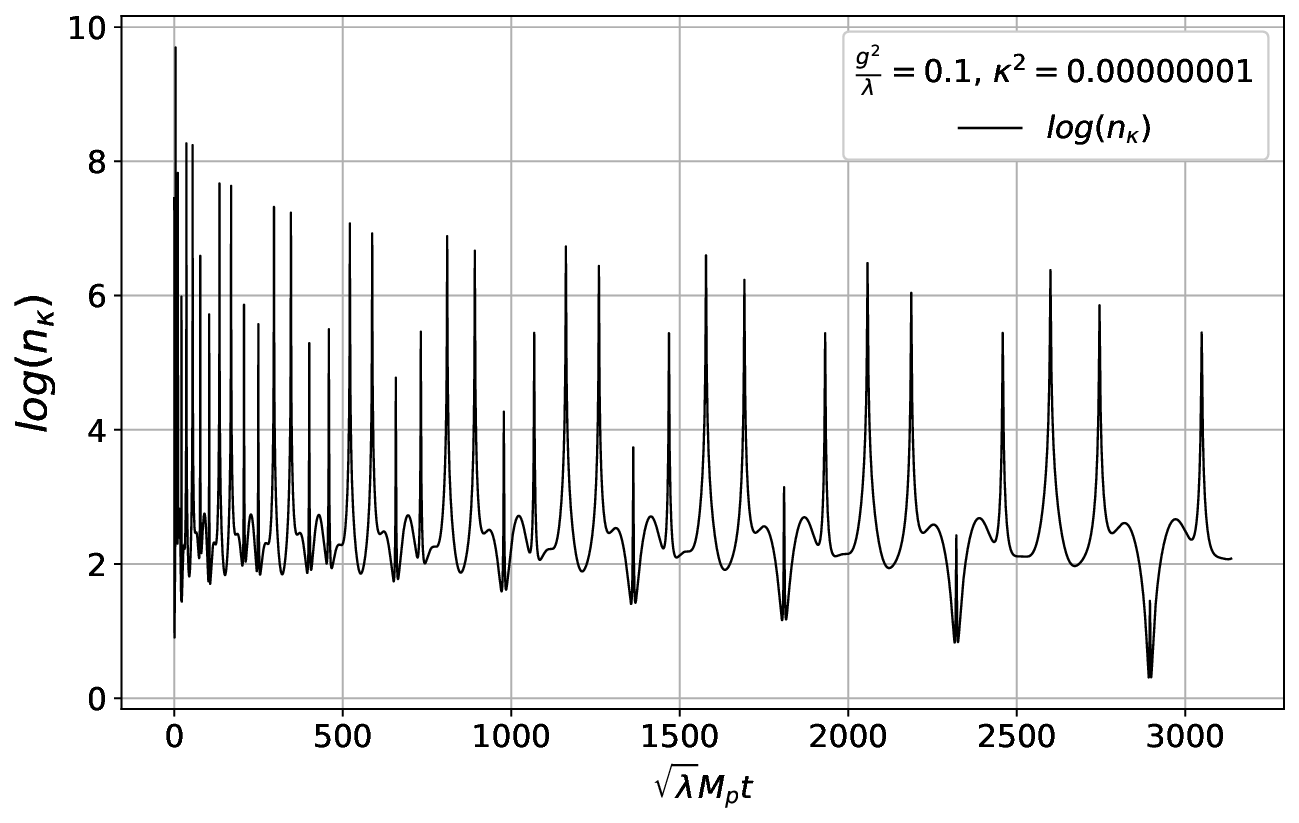}
    \caption{Left panels: Time evolution of the mode function $\chi_k(t)$ for different values of $\kappa^2$. Right panels: Time evolution of the log of occupation number $n_k(t)$ for different values of $\kappa^2$. In all panels, $\frac{g^2}{\lambda}=0.1$.}
    \label{fig:g_0.1_k_0.1ZOOM}

\end{figure}

\begin{figure}[H]
\centering
    \includegraphics[width=0.7\textwidth]{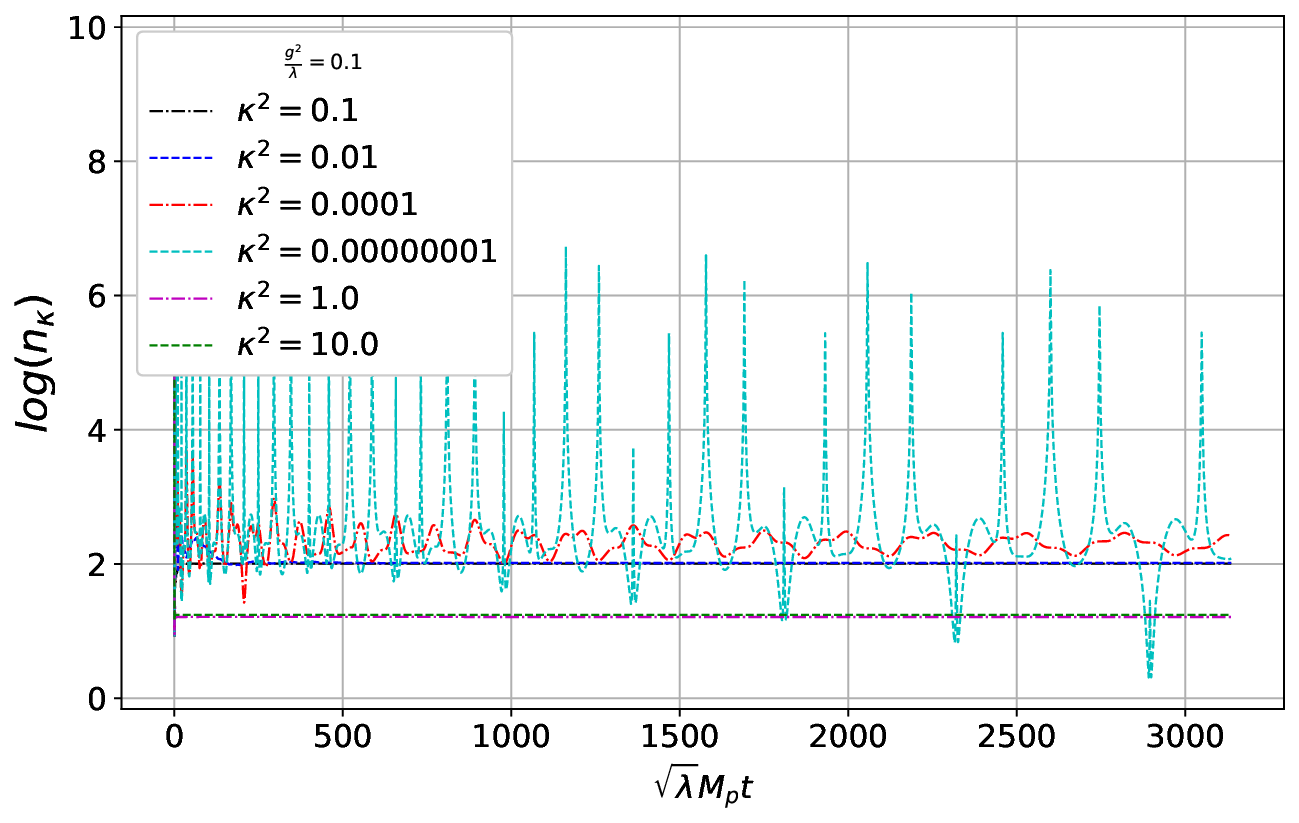}
    \caption{Comparison of the evolution of occupation number $n_k(t)$ for different values of $\kappa^2$, with $\frac{g^2}{\lambda} = 0.1.$}
    \label{fig:g_0.1_PARTICLEno}
\end{figure}

Although the quartic inflaton potential considered in this study is not fully consistent with the latest observational constraints from Planck, BICEP and Keck experiments \cite{Planck2018, Bicep2015}, it serves as an important theoretical laboratory for understanding {\em exact} nonlinear preheating dynamics. The simplicity of the $\phi^4$ model allows us to develop a robust {\em numerical methodology} that captures the full nonlinear interplay between the inflaton and the coupled field-mode without relying on approximations. Importantly, this approach is readily extendible to more realistic and observationally favoured potentials, which are usually of complicated structure, such as the Starobinsky model or the $\alpha$-attractor frameworks. Thus, while our focus is on the quartic case, the techniques and insights presented here lay the groundwork for future studies of preheating dynamics in models with more complicated potentials compatible with cosmological observations.

The rest of the paper is organized as follows. In Section \ref{sec_Dynamics}, we present the dynamics of the inflaton field in the post-inflationary regime. Section \ref{sec_phi4} introduces the interaction between the inflaton $\phi$ and a bosonic field $\chi$, allowing for the transfer of energy from the $\phi$-field into the modes of the $\chi$-field during parametric oscillations following the inflationary phase. In this section, we also lay out the coupled set of nonlinear dynamical equations governing the parametric resonance phase. In Section \ref{sec_NumericalIntegration}, we numerically integrate the dynamical equations and discuss the results for different coupling ratios in the weak and strong coupling regimes, illustrating key features with graphical representations. Finally, in Section \ref{sec_conclusion}, we conclude the paper with a discussion.

\section{Dynamics of the Inflaton Field}\label{sec_Dynamics}

The Einstein Hilbert action, written as $S = \frac{1}{16 \pi G} \int R \sqrt{-g} \, d^4x + S_\phi$, yields the Einstein field equations
\begin{equation}
\label{eqn:field}
    R_{\mu\nu} = 8\pi G(T_{\mu\nu} - \frac{1}{2}g_{\mu\nu} T),
\end{equation}

\begin{figure}[H]

    \includegraphics[width=.47\textwidth, height=5cm]{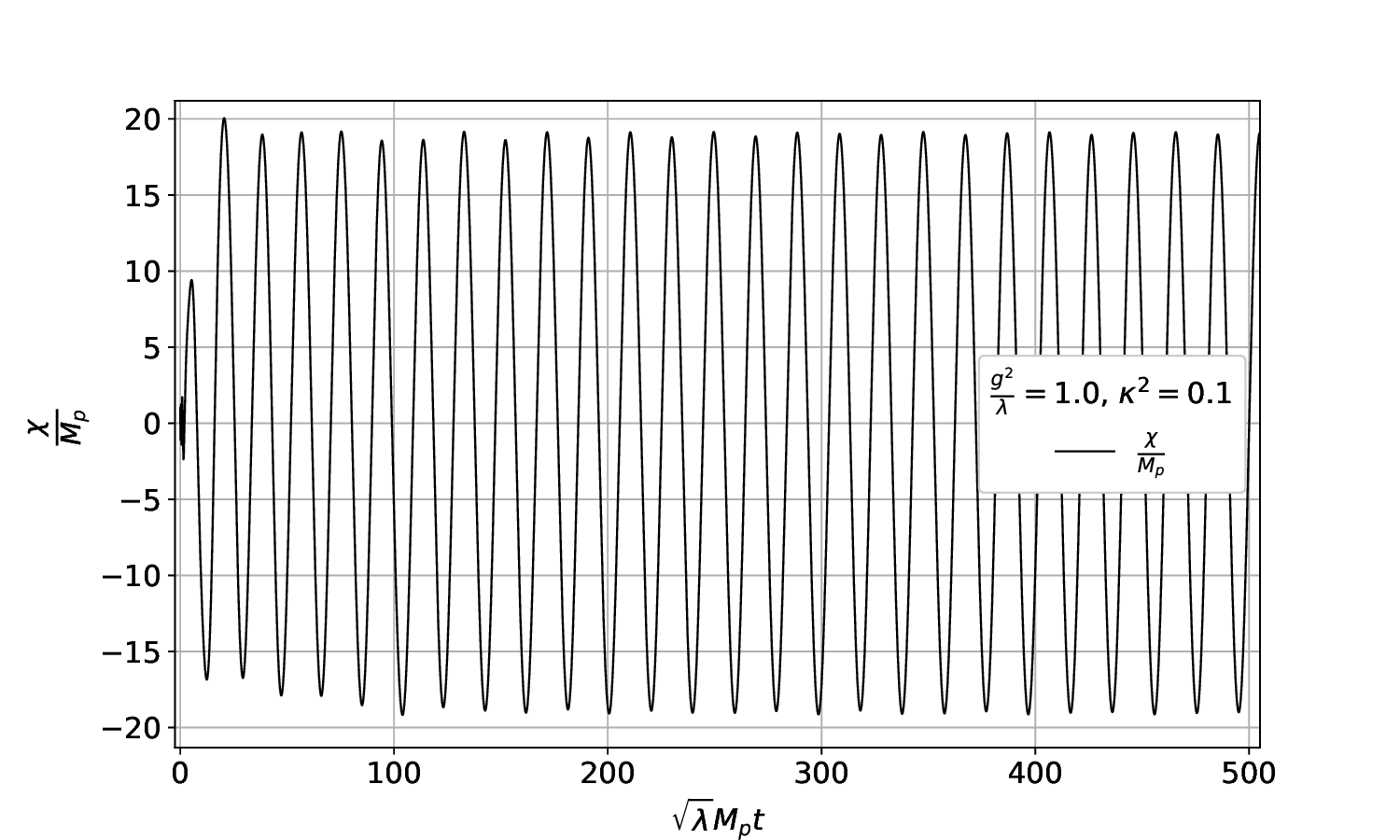}
    \includegraphics[width=.47\textwidth, height=5cm]{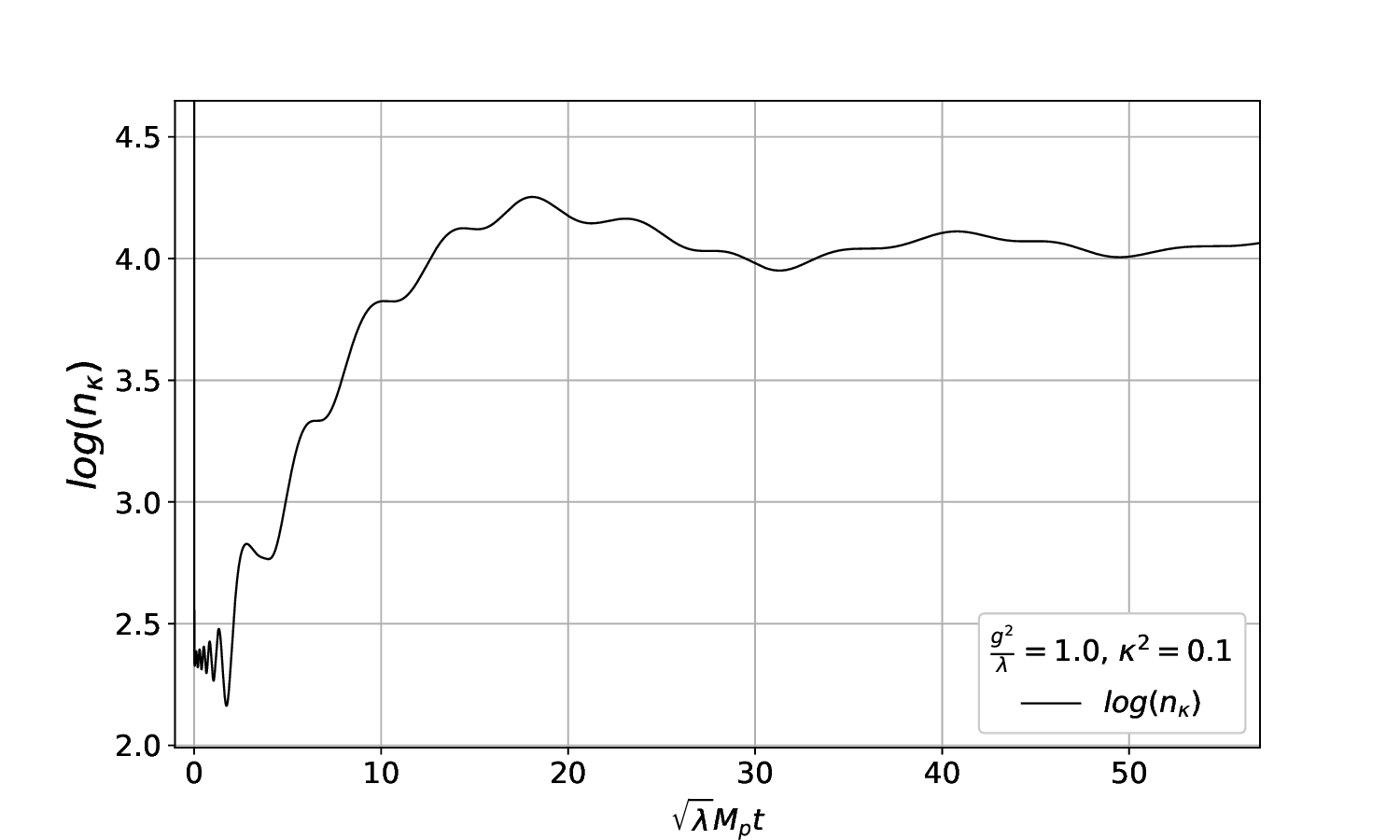}
    \includegraphics[width=.47\textwidth, height=5cm]{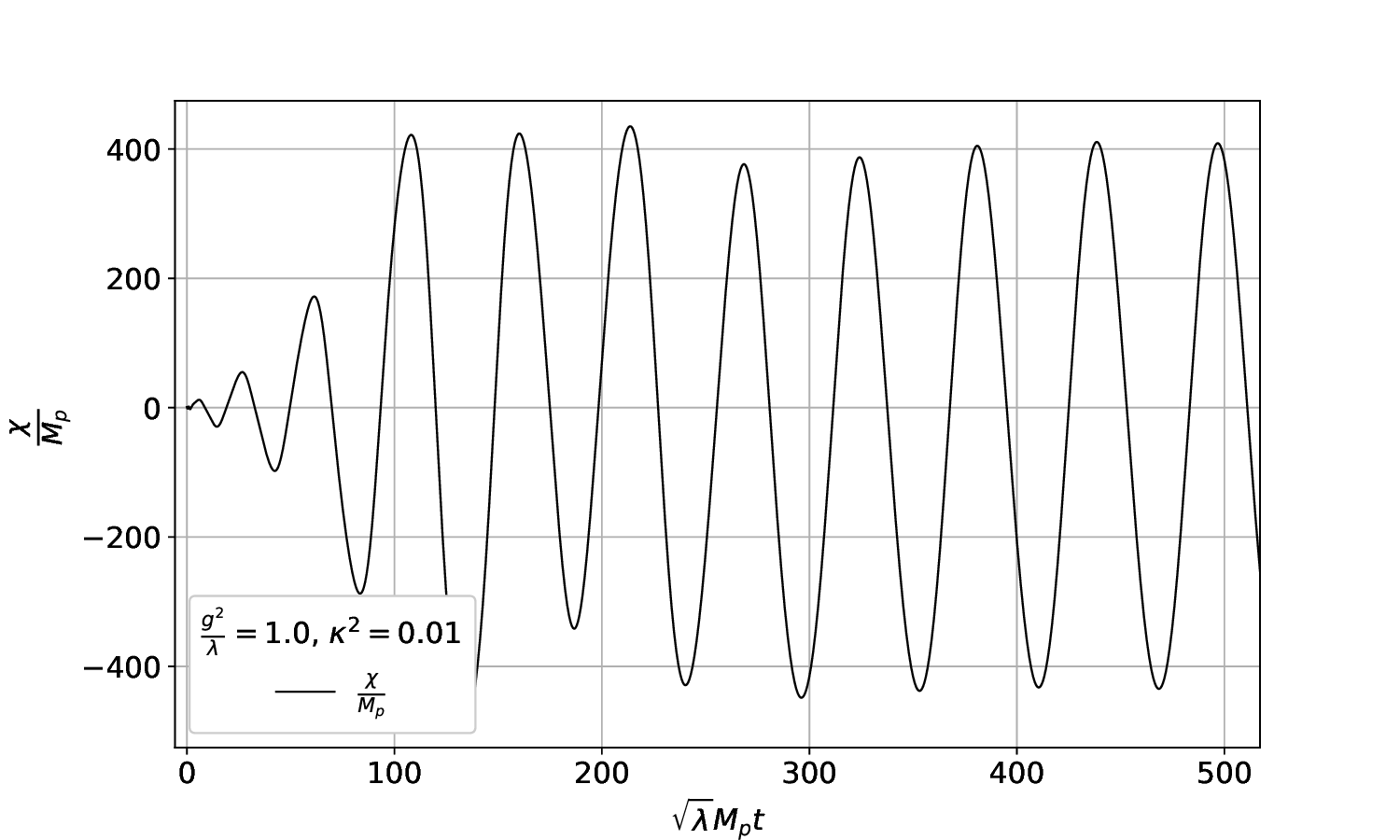}
    \includegraphics[width=.47\textwidth, height=5cm]{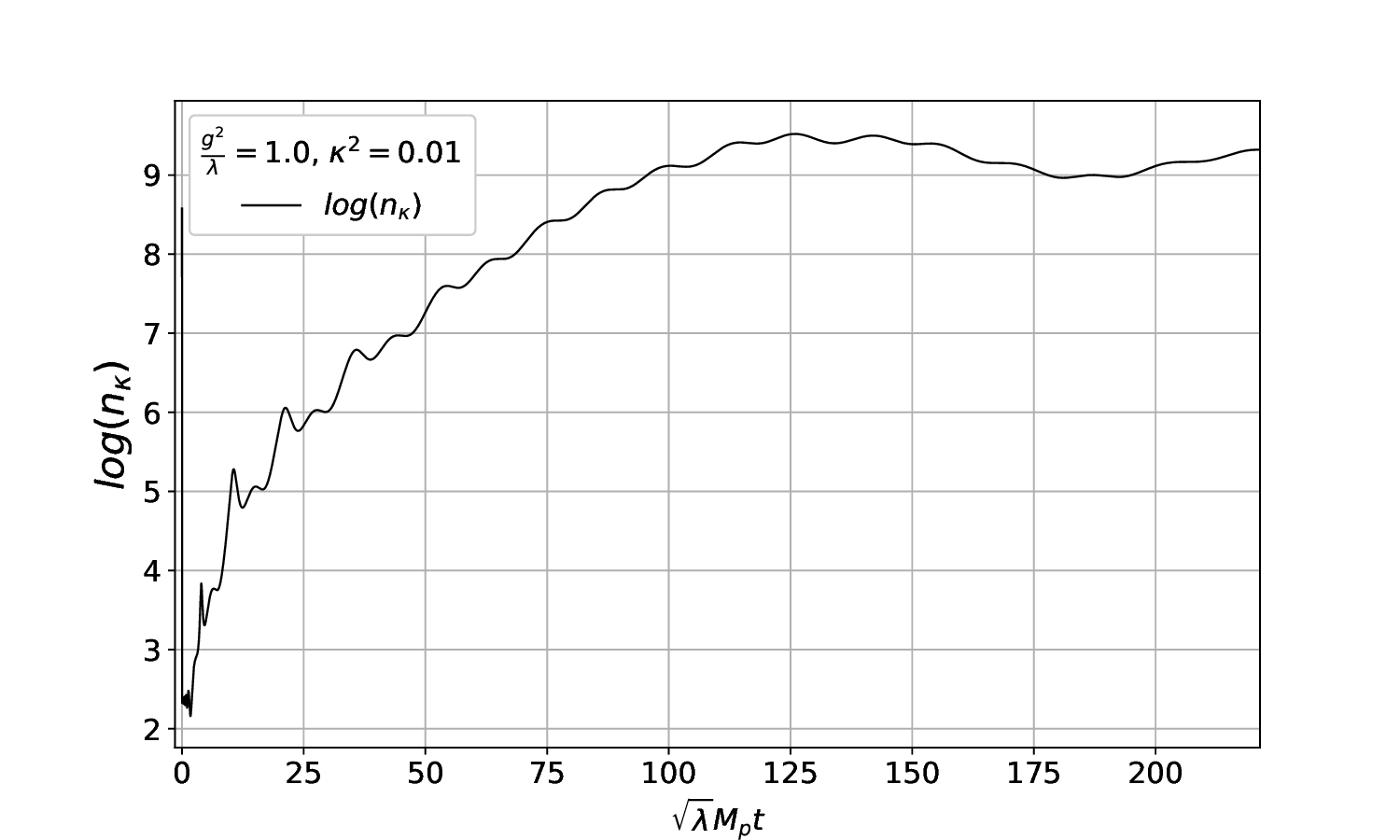}
    \includegraphics[width=.47\textwidth, height=5cm]{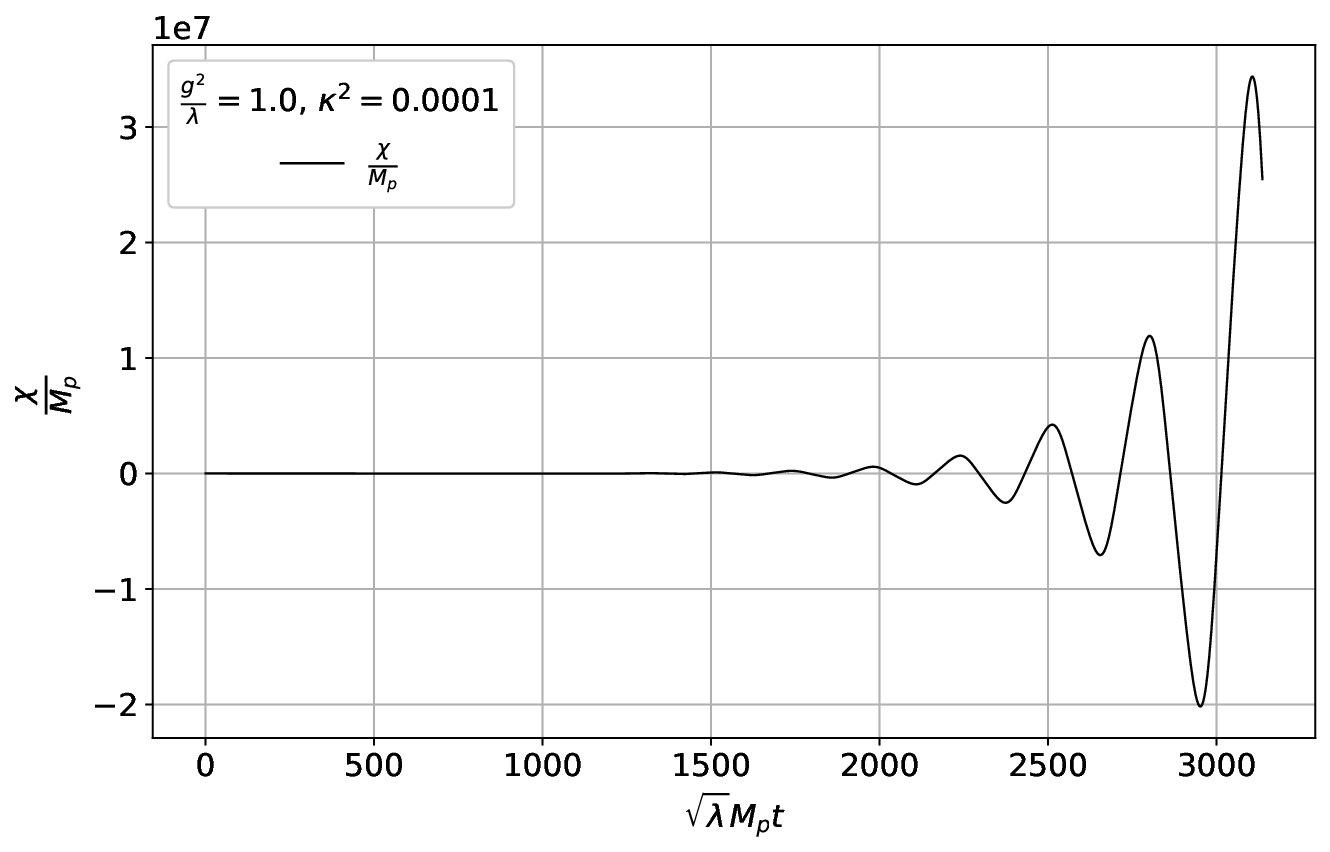}
    \includegraphics[width=.47\textwidth, height=5cm]{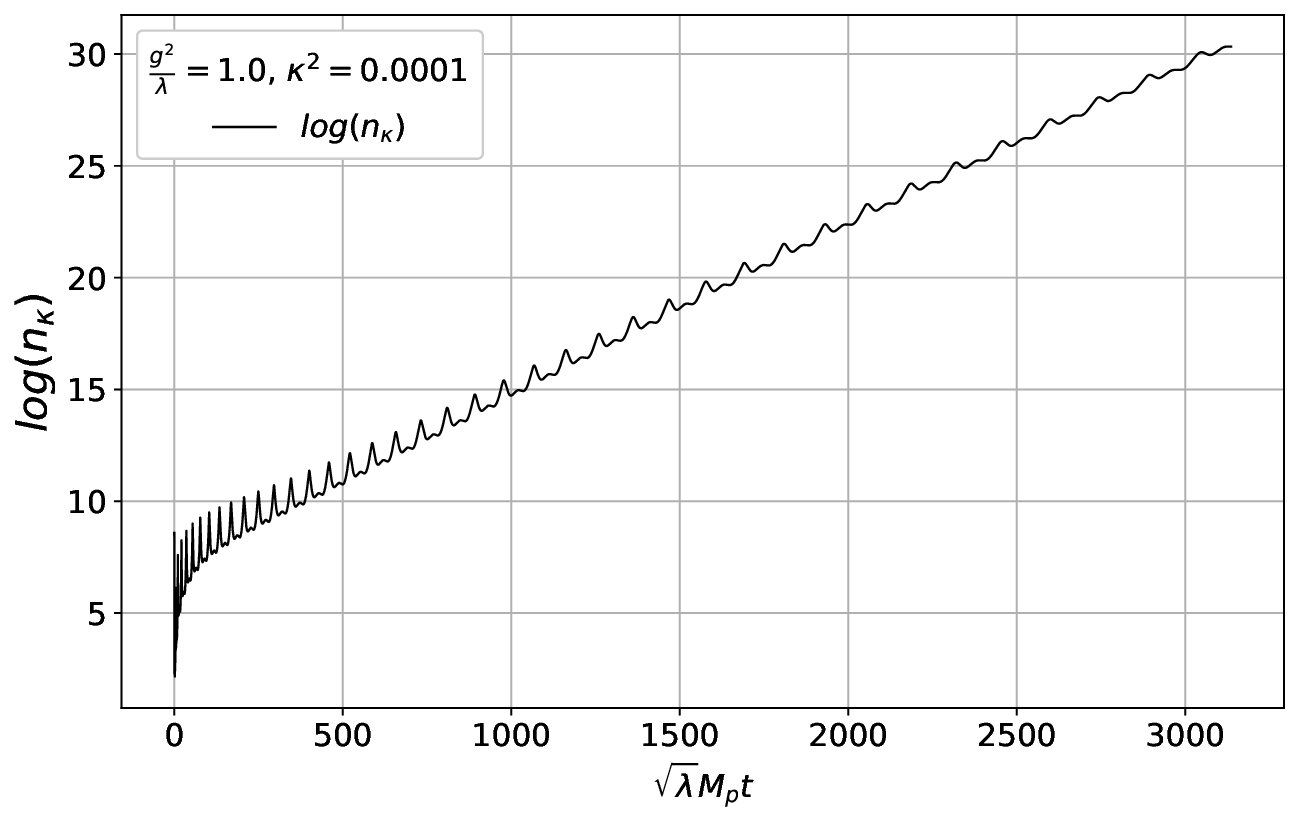}
    \includegraphics[width=.47\textwidth, height=5cm]{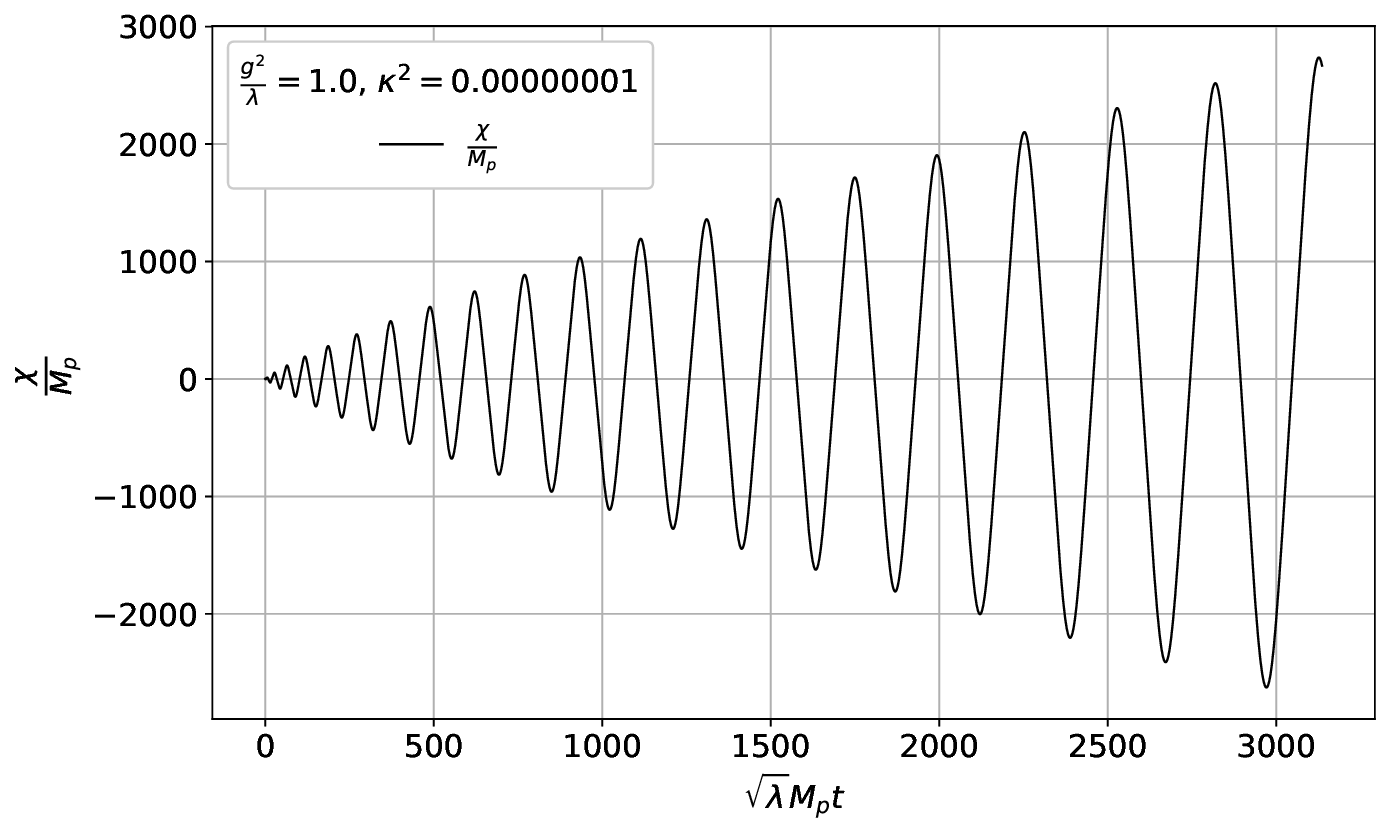}
    \includegraphics[width=.47\textwidth, height=5cm]{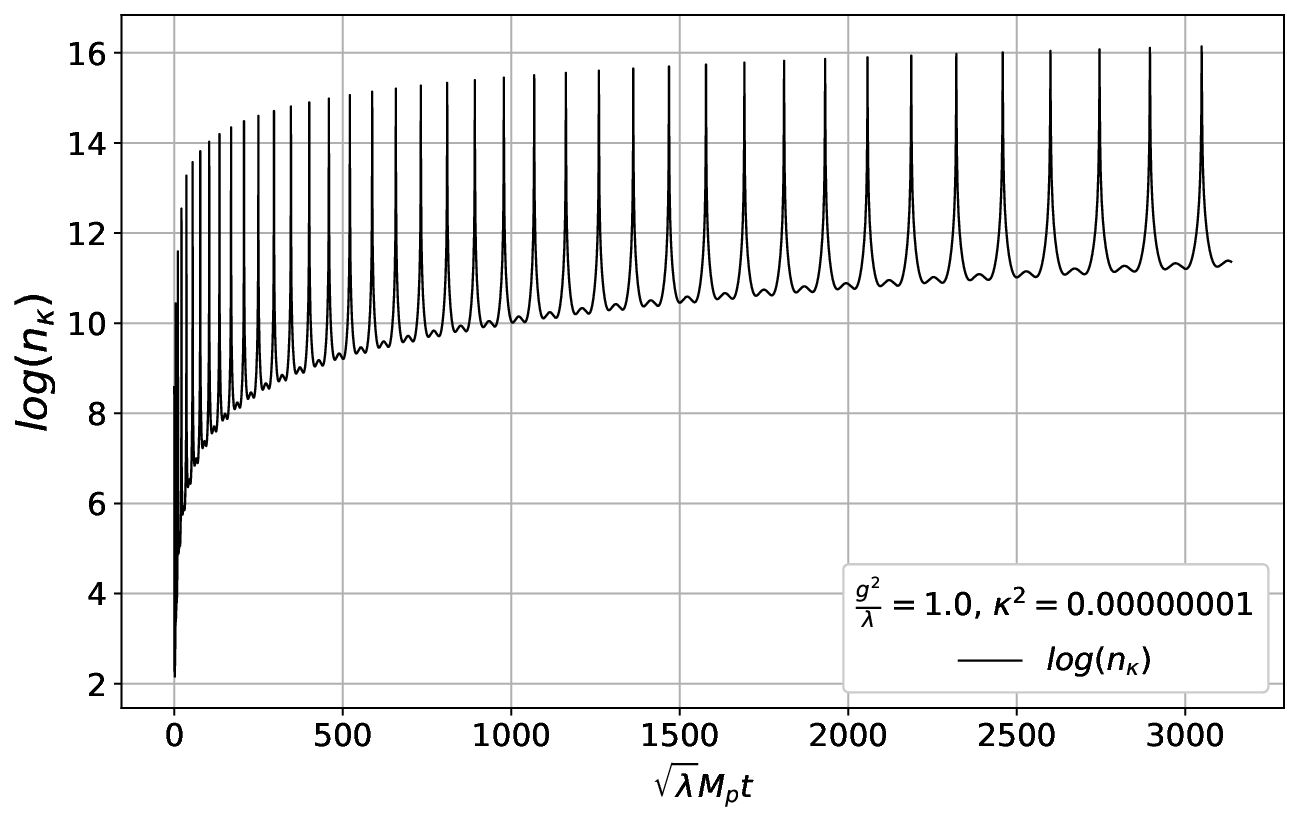}
    \caption{Left panels: Time evolution of the mode function $\chi_k(t)$ for different values of $\kappa^2$. Right panels: Time evolution of the log of occupation number $n_k(t)$ for different values of $\kappa^2$. In all panels, $\frac{g^2}{\lambda}= 1.0$.}
    \label{fig:g_1_k_0.1ZOOM}

\end{figure}
\begin{figure}[H]
\centering
    \includegraphics[width=0.7\textwidth]{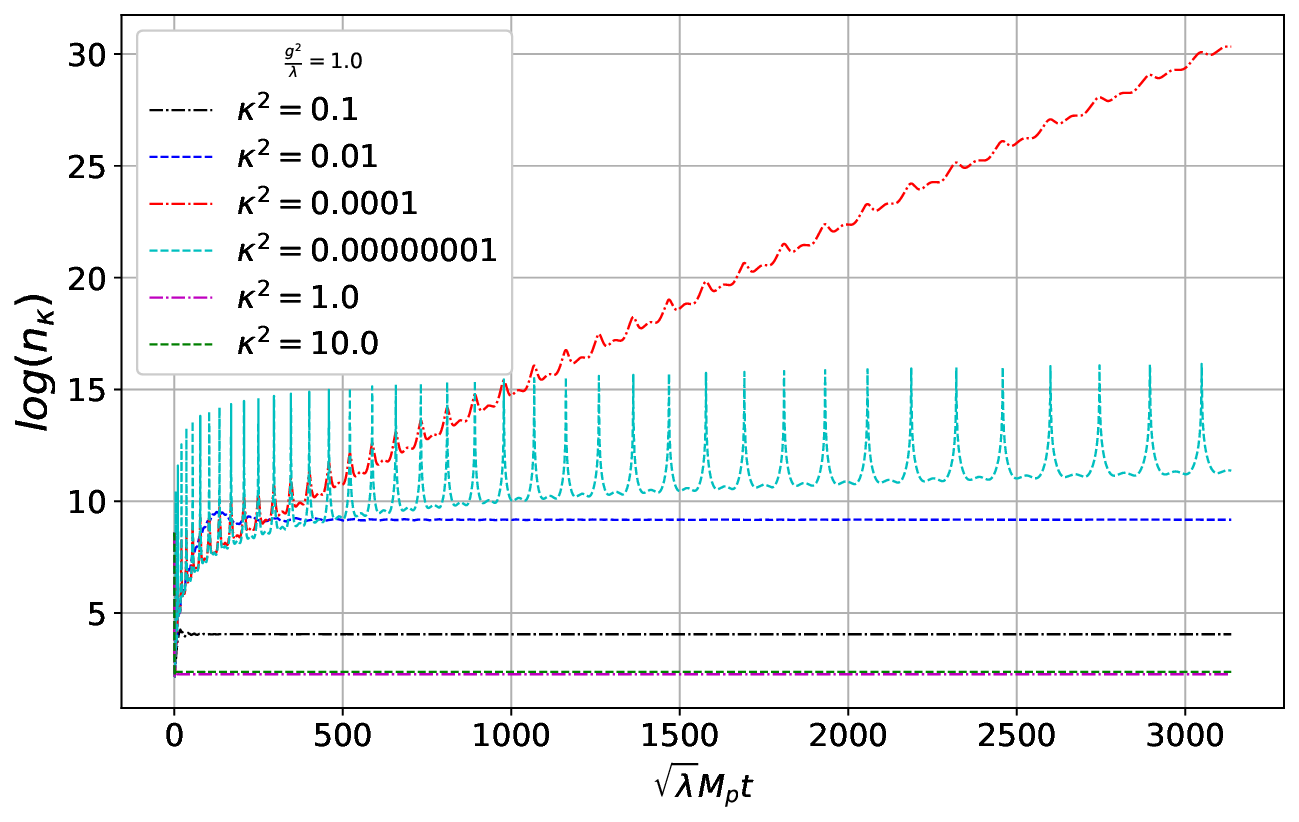}
    \caption{Comparison of the evolution of occupation number $n_k(t)$ for different values of $\kappa^2$, with $\frac{g^2}{\lambda} = 1.0.$}
    \label{fig:g_1_PARTICLEno}
\end{figure}
\noindent where $R_{\mu \nu}$ is the Ricci tensor and $T_{\mu\nu} \equiv$ diag($\rho, P, P, P$) is the energy momentum tensor in the proper frame, so that
\begin{equation}
\label{eqn:EMtensor}
T_{\mu\nu} = (\rho + P) u_\mu u_\nu - Pg_{\mu\nu}
\end{equation}
in an arbitrary frame, $u^\mu$ being the four-velocity.

We consider the spacetime of Friedmann-L\^emaitre-Robertson-Walker (FLRW) universe, with the metric
\begin{equation}
    ds^2 = -dt^2 + a(t) \left[\frac{dr^2}{1-K r^2} + r^2 \left(d\theta^2 + \sin^2\theta\, d\varphi^2\right) \right],
\end{equation}
and we shall take the case of a flat universe, $K=0$. With this choice of the metric, equation \ref{eqn:field} and the energy momentum tensor \ref{eqn:EMtensor} yield the Friedmann equations
\begin{equation}
\label{eqn:Fried1}
 H^2 = \frac{8 \pi G}{3} \rho
\end{equation}
and
\begin{equation}
\label{eqn:Fried2}
 \frac{\ddot{a}}{a} = -\frac{4 \pi G}{3}(\rho + 3P),
\end{equation}
where $H = \frac{\dot{a}}{a}$ is the Hubble rate, $\rho$ and $P$ are the energy density and pressure of the field $\phi$.

The scalar field $\phi$ is identified as the inflaton with the Lagrangian density
\begin{equation}
\label{eqn:L-phi}
\mathcal{L}_{\phi}= -\frac{1}{2} g^{\mu\nu}\phi_{,\mu} \phi_{,\nu} - V(\phi)
\end{equation}
where $V(\phi)$ is the potential energy density of the inflaton field. With the assumption of homogeneity and isotropy, the equation of motion for $\phi(t)$ is given by the Klein Gordon equation,
\begin{equation}
\label{eqn:KG}
\ddot{\phi}+ 3H\dot{\phi} + V_{,\phi}=0.
\end{equation}

Equations \ref{eqn:EMtensor} and \ref{eqn:L-phi} lead to the energy density and pressure of the inflaton field as
\begin{equation}
\label{eqn:EnergyDen}
\rho = \frac{1}{2}\dot{\phi}^2 + V(\phi)
\end{equation}
\nopagebreak

\begin{figure}[H]

    \includegraphics[width=.47\textwidth, height=5cm]{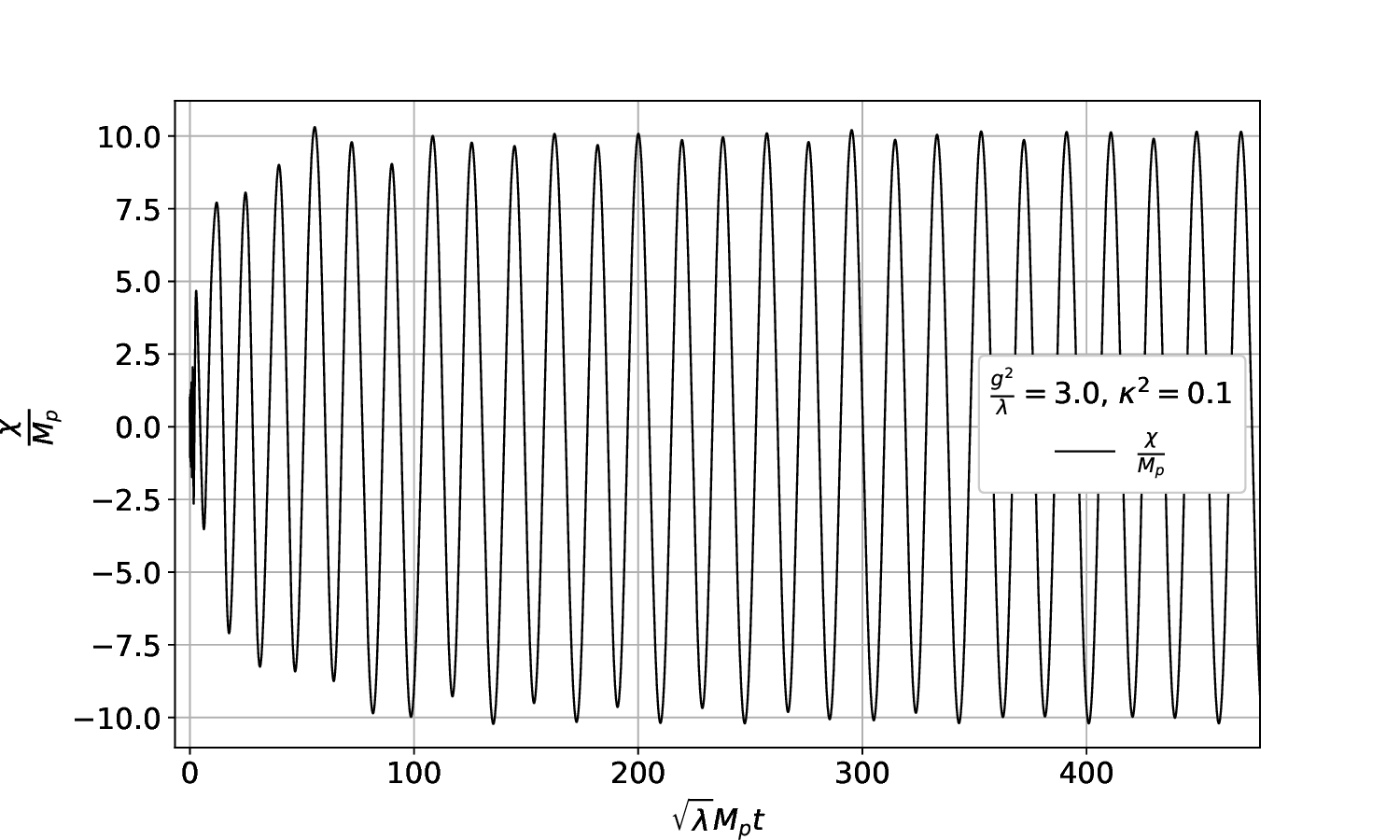}
    \includegraphics[width=.47\textwidth, height=5cm]{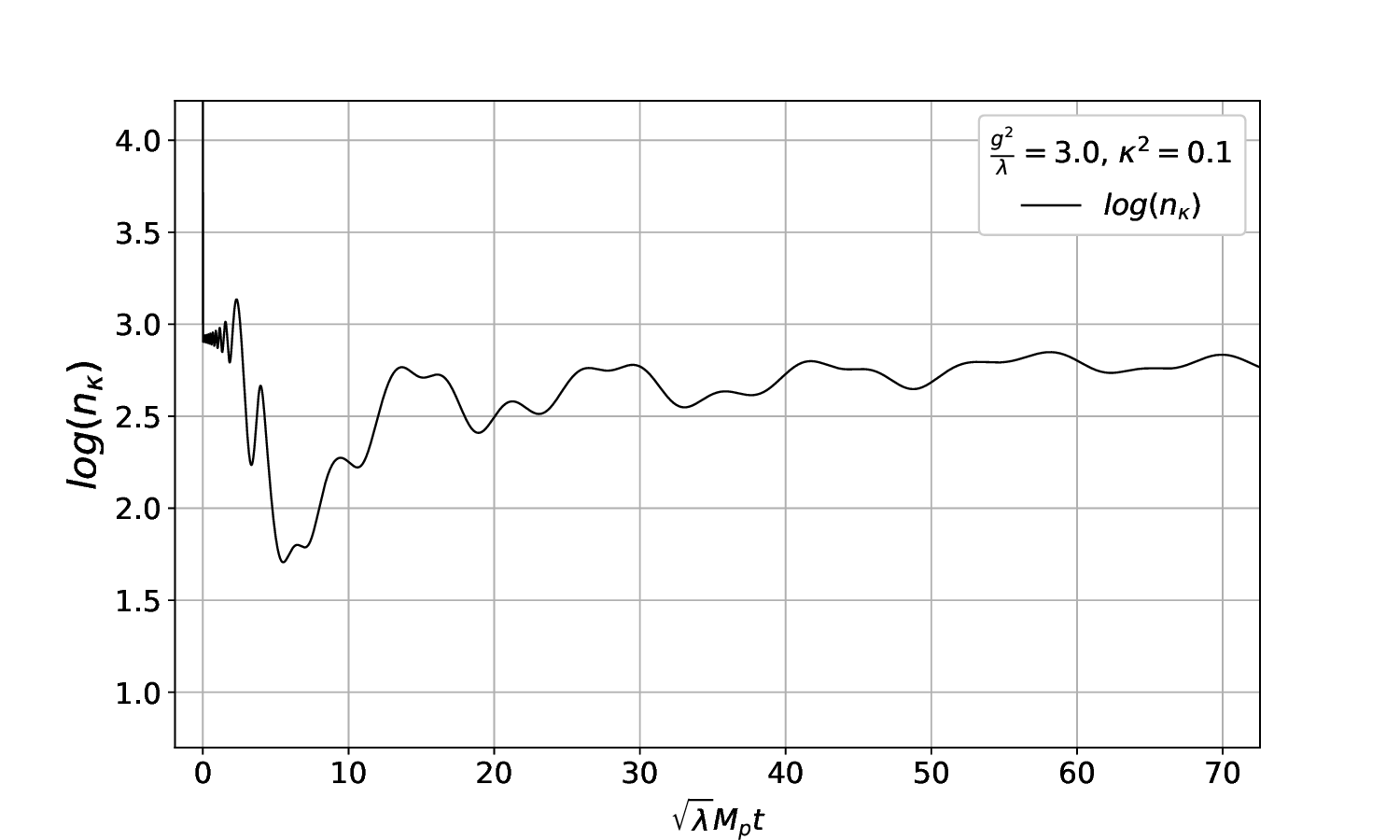}
    \includegraphics[width=.47\textwidth, height=5cm]{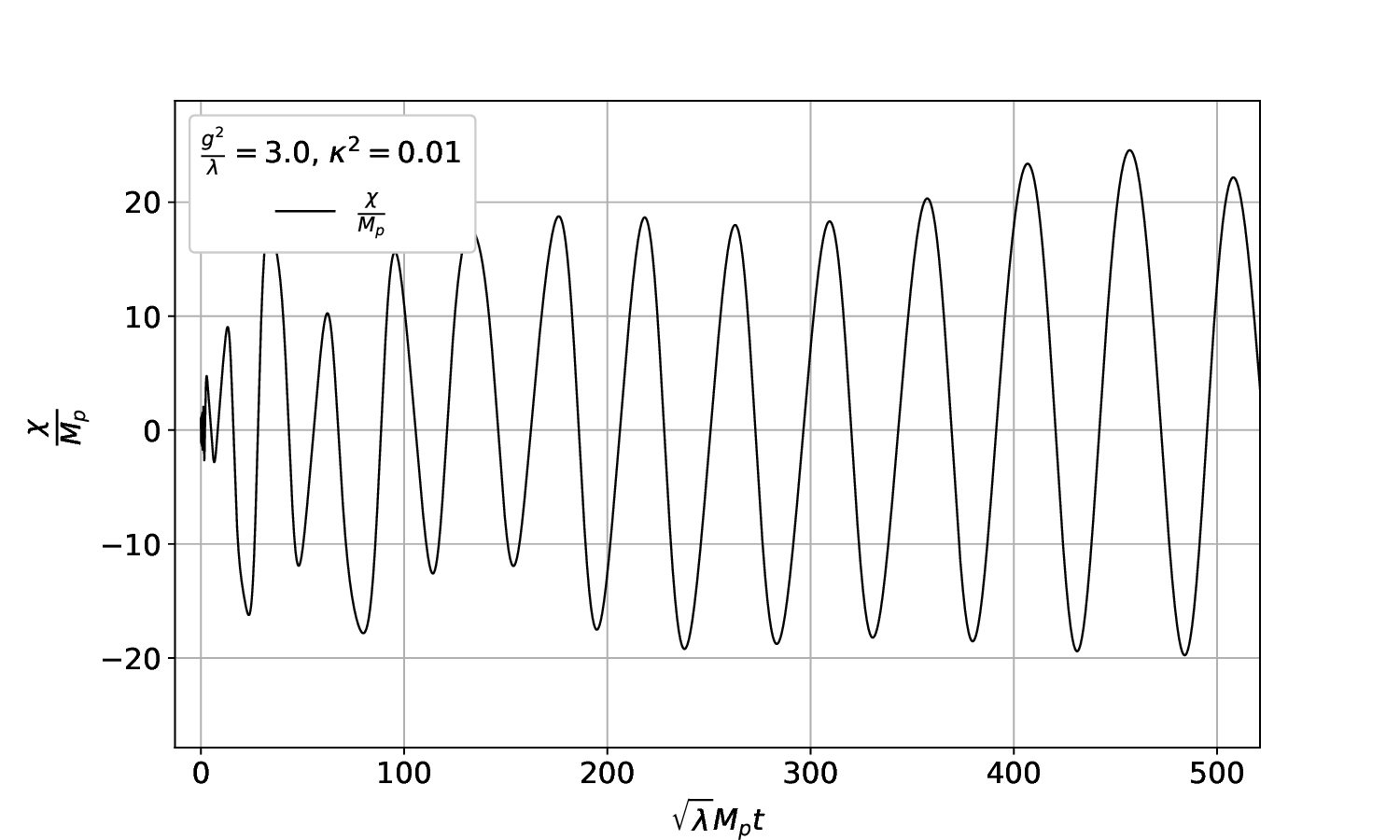}
    \includegraphics[width=.47\textwidth, height=5cm]{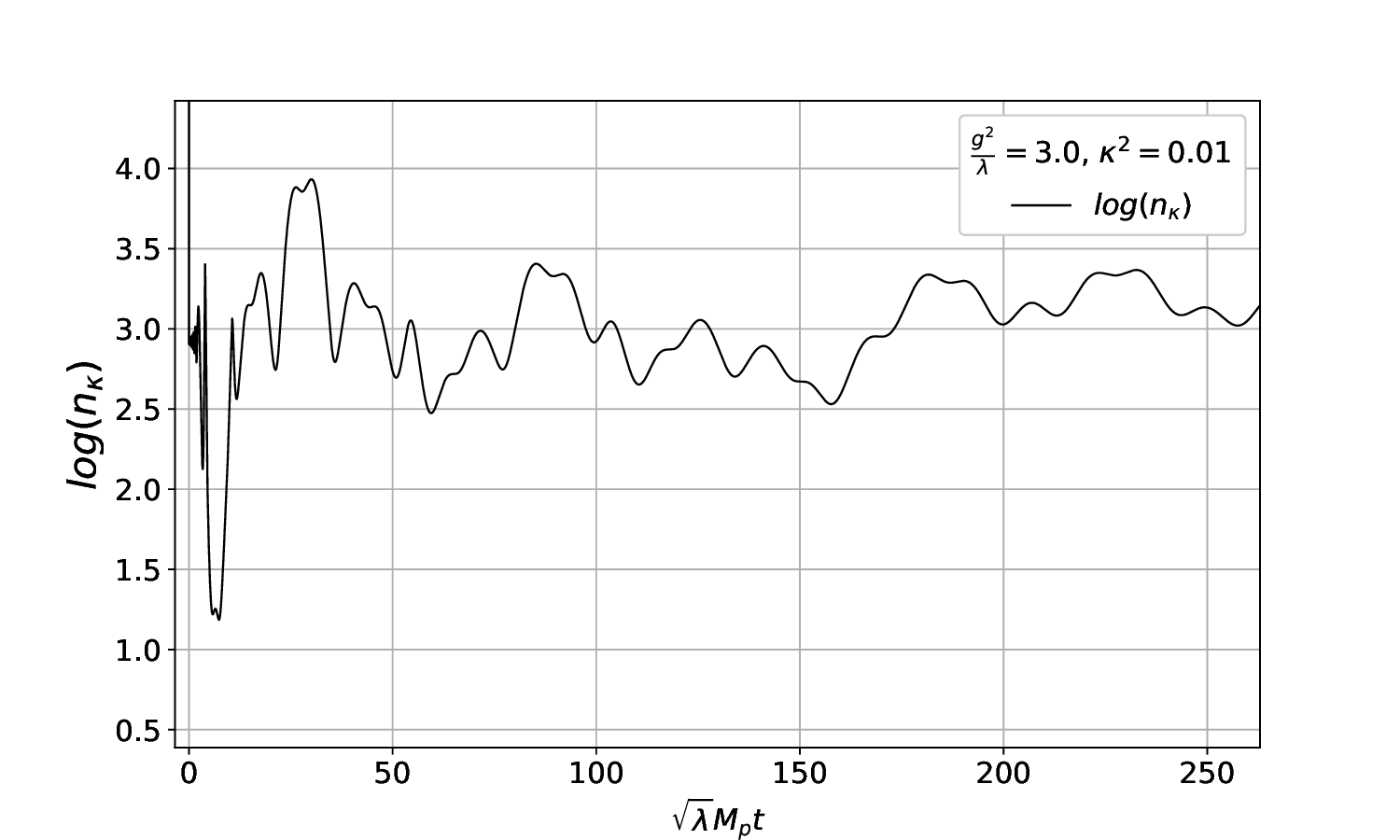}
    \includegraphics[width=.47\textwidth, height=5cm]{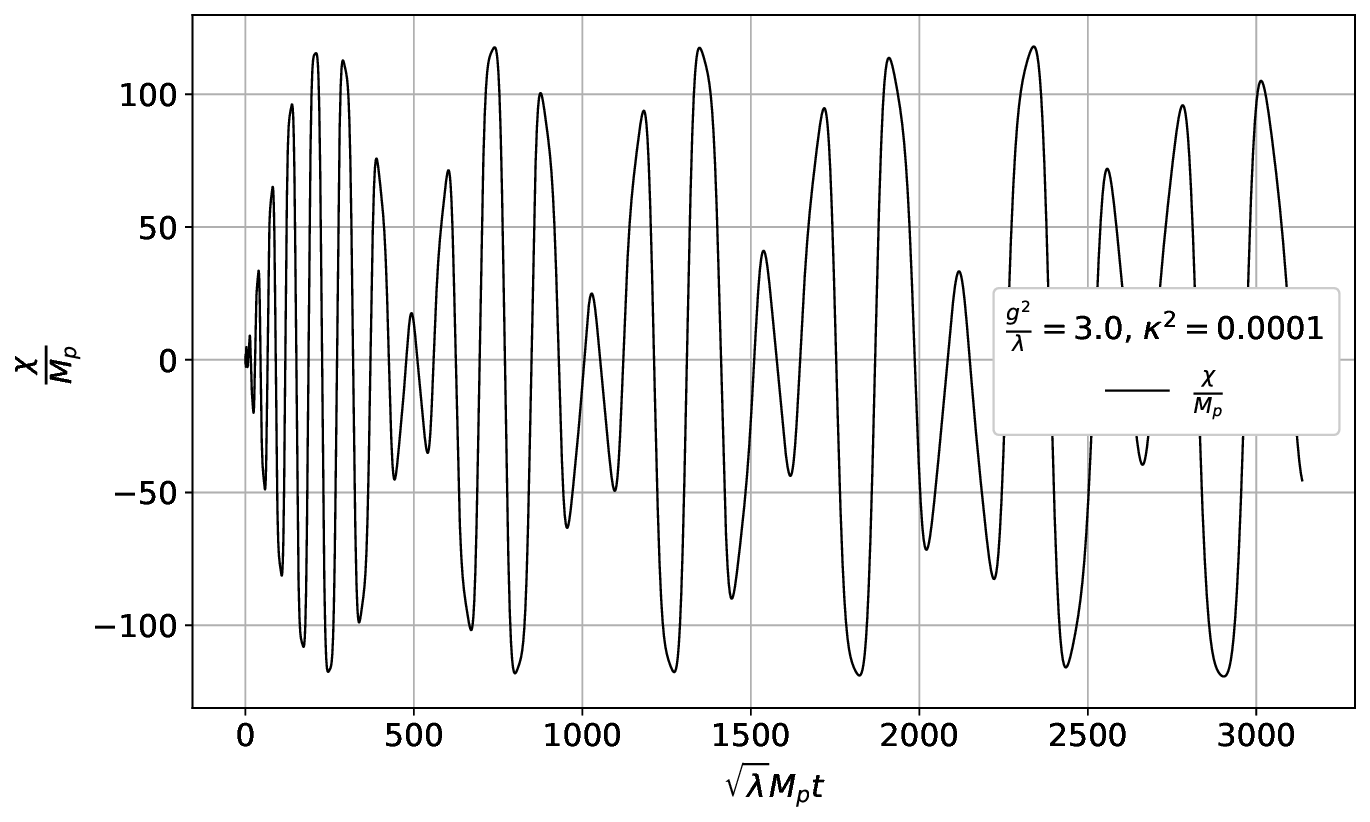}
    \includegraphics[width=.47\textwidth, height=5cm]{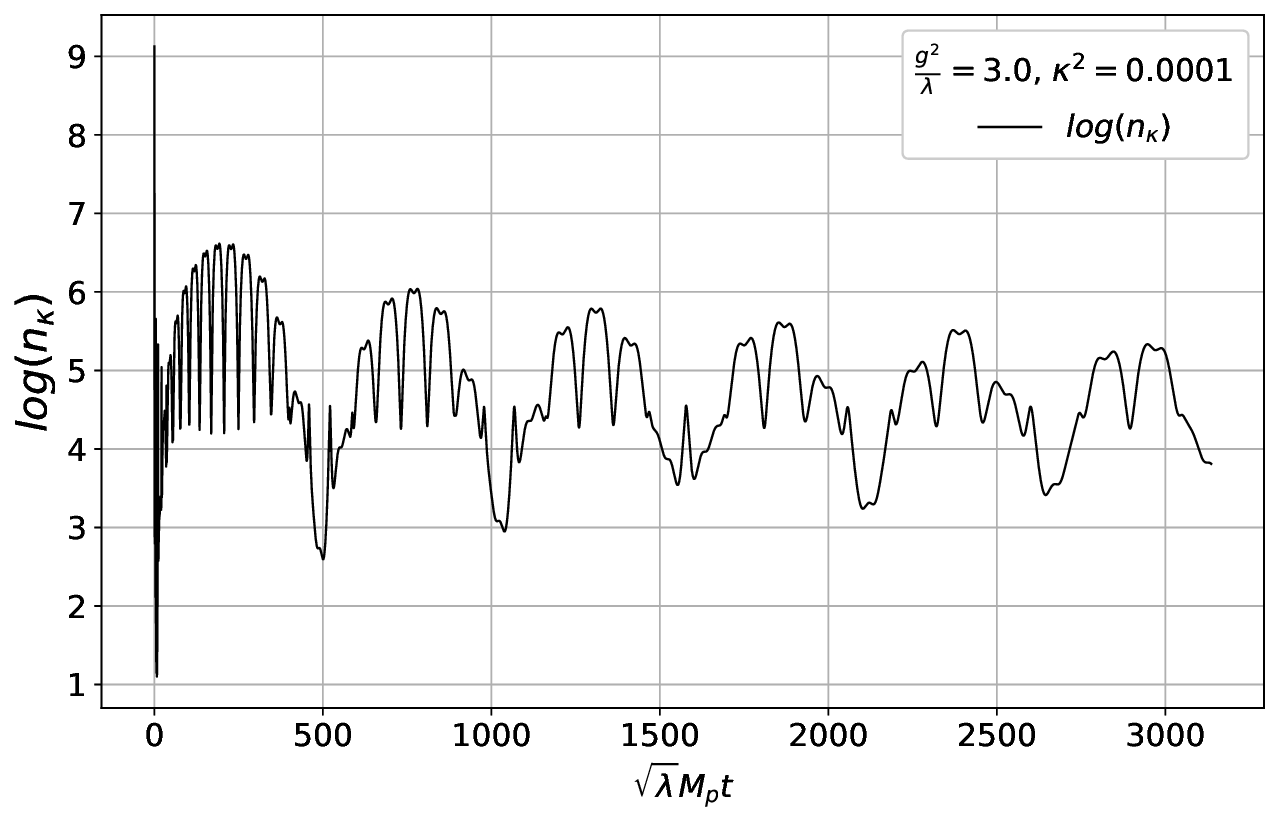}
    \includegraphics[width=.47\textwidth, height=5cm]{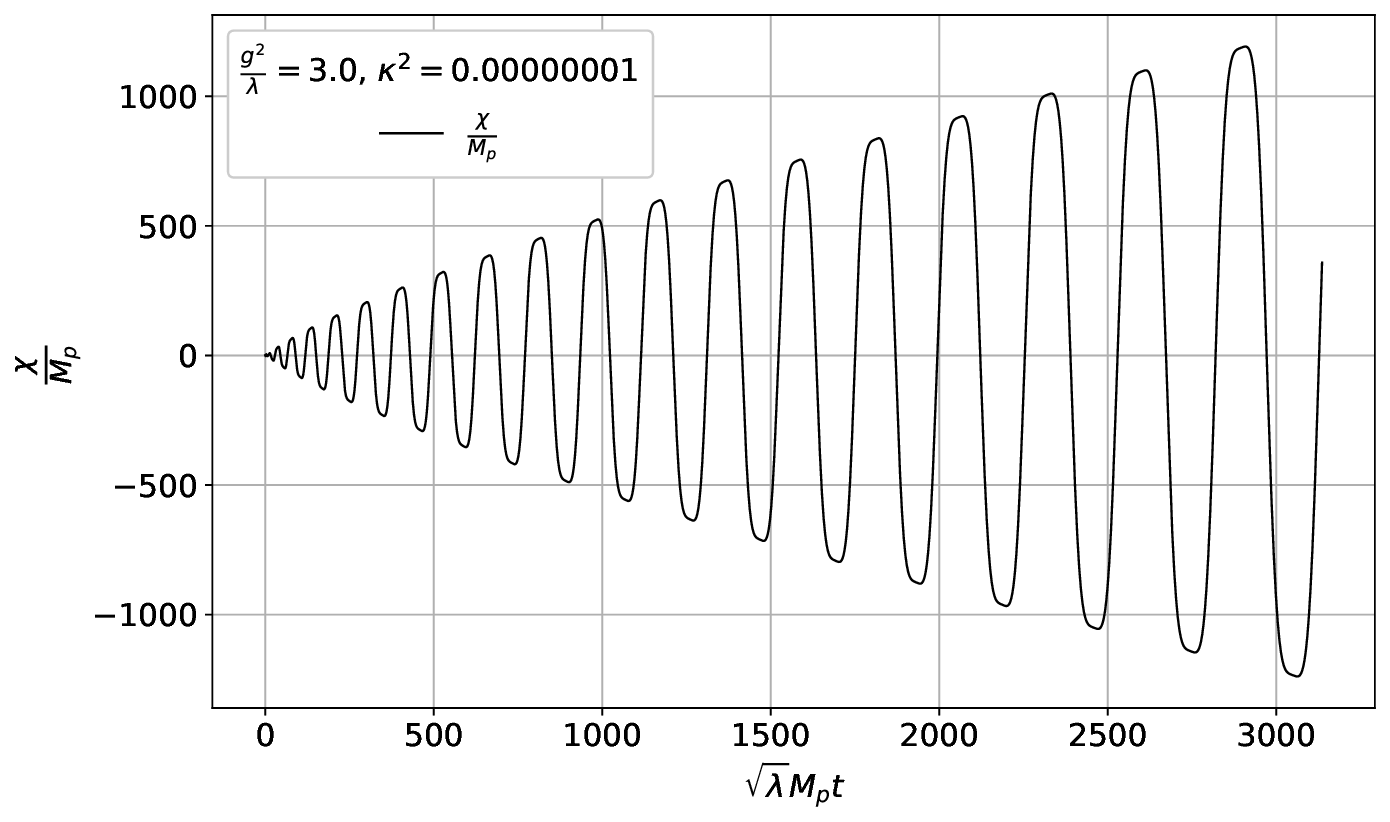}
    \includegraphics[width=.47\textwidth, height=5cm]{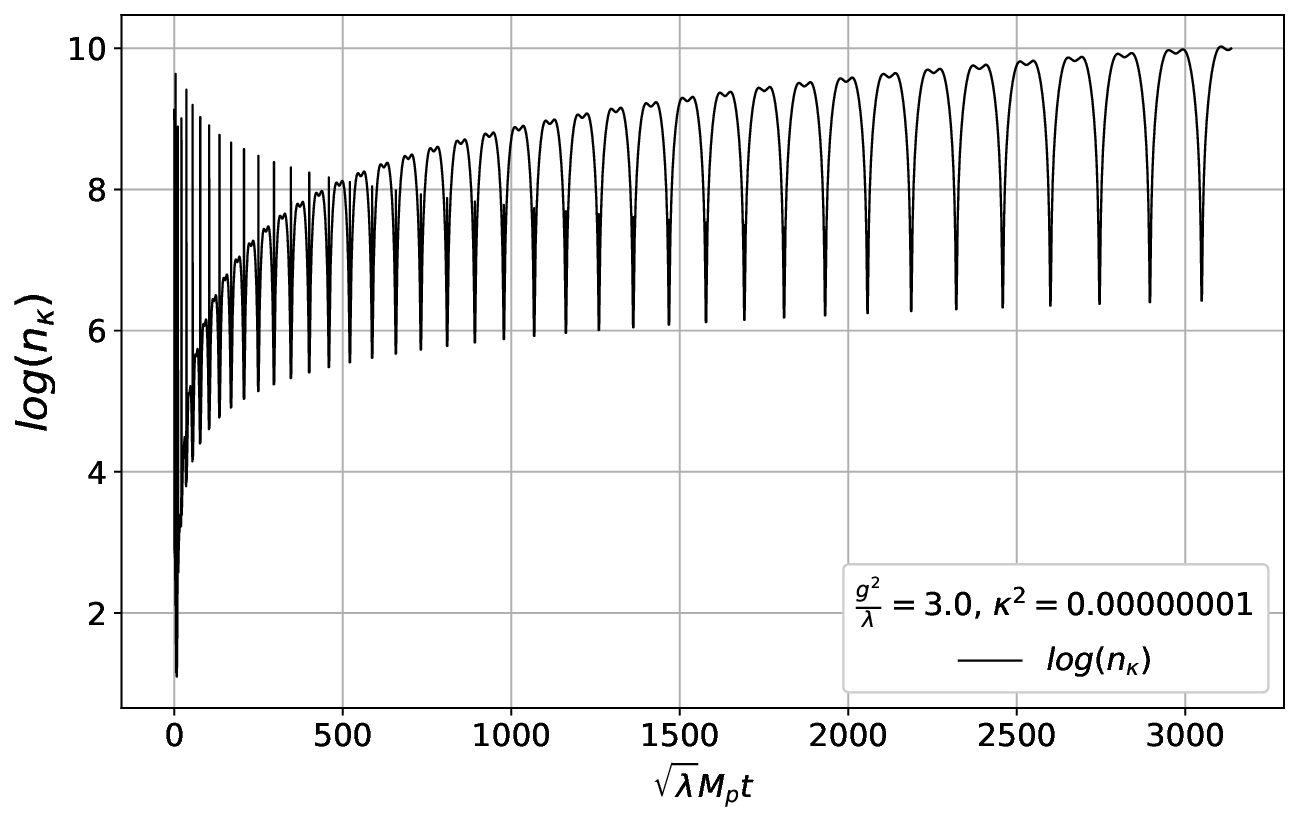}
    \caption{Left panels: Time evolution of the mode function $\chi_k(t)$ for different values of $\kappa^2$. Right panels: Time evolution of the log of occupation number $n_k(t)$ for different values of $\kappa^2$. In all panels, $\frac{g^2}{\lambda}= 3.0$.}
    \label{fig:g_3_k_0.1ZOOM}

\end{figure}

\begin{figure}[H]
\centering
    \includegraphics[width=0.7\textwidth]{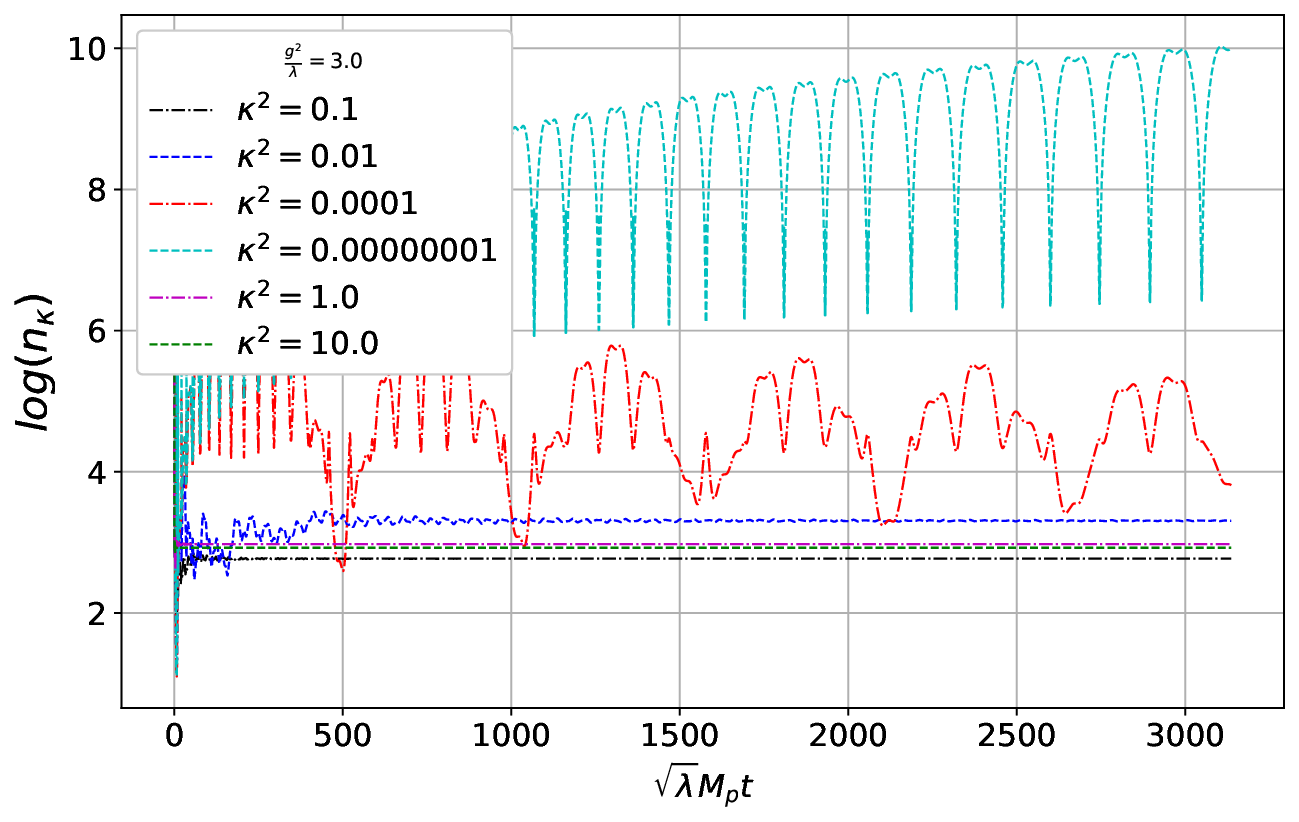}
    \caption{Comparison of the evolution of occupation number $n_k(t)$ for different values of $\kappa^2$, with $\frac{g^2}{\lambda}= 3.0$.}
    \label{fig:g_3_PARTICLEno}
\end{figure}

\noindent and
\begin{equation}
\label{eqn:Pressure}
P = \frac{1}{2}\dot{\phi}^2 - V(\phi).
\end{equation}

Using the above expressions, the Friedmann equations \ref{eqn:Fried1} and \ref{eqn:Fried2} take the forms
\begin{equation}
\label{eqn:Fried1a}
H^2 = \frac{1}{3M_{P}^2}\left[\frac{1}{2}\dot{\phi}^2 + V(\phi)\right]
\end{equation}
and
\begin{equation}
\label{eqn:Fried2a}
    \frac{\ddot{a}}{a} = -\frac{1}{3M_{P}^2}\left[\dot{\phi}^2 - V(\phi)\right]
\end{equation}
where $M_P = 1/\sqrt{8 \pi G} = 2.4 \times 10^{18}$GeV is the Planck mass.

For a general inflaton potential $V(\phi)$, it is often not possible to recast the equations governing parametric resonance into an exact Mathieu-type form due to the nonlinear nature of the resulting dynamics. One notable example is the quartic potential, where the governing equation for mode functions takes the form of a Lam\'e equation only under specific approximations, as shown in Ref.~\cite{Greene1997}.

For simplicity, while still capturing key nonlinear features, we focus on the quartic inflaton model, characterized by the potential
\begin{equation}
\label{eqn:phi4-potential}
V(\phi) = \frac{1}{4} \lambda \phi^4.
\end{equation}
This choice offers a well-motivated and physically relevant framework for studying parametric resonance, while still necessitating numerical techniques for an exact and accurate treatment without making any approximations in the resulting dynamical equations. Developing a robust numerical methodology in this context not only enables us to study the quartic case with precision, but also provides a foundation for extending the analysis to more complicated inflationary potentials where analytical approaches may be entirely inapplicable.\nopagebreak

In the context of the quartic potential \ref{eqn:phi4-potential}, the Friedmann equations \ref{eqn:Fried1a} and \ref{eqn:Fried2a} and the Klein Gordon equation \ref{eqn:KG} take the forms

\begin{figure}[H]
    \includegraphics[width=.47\textwidth, height=5cm]{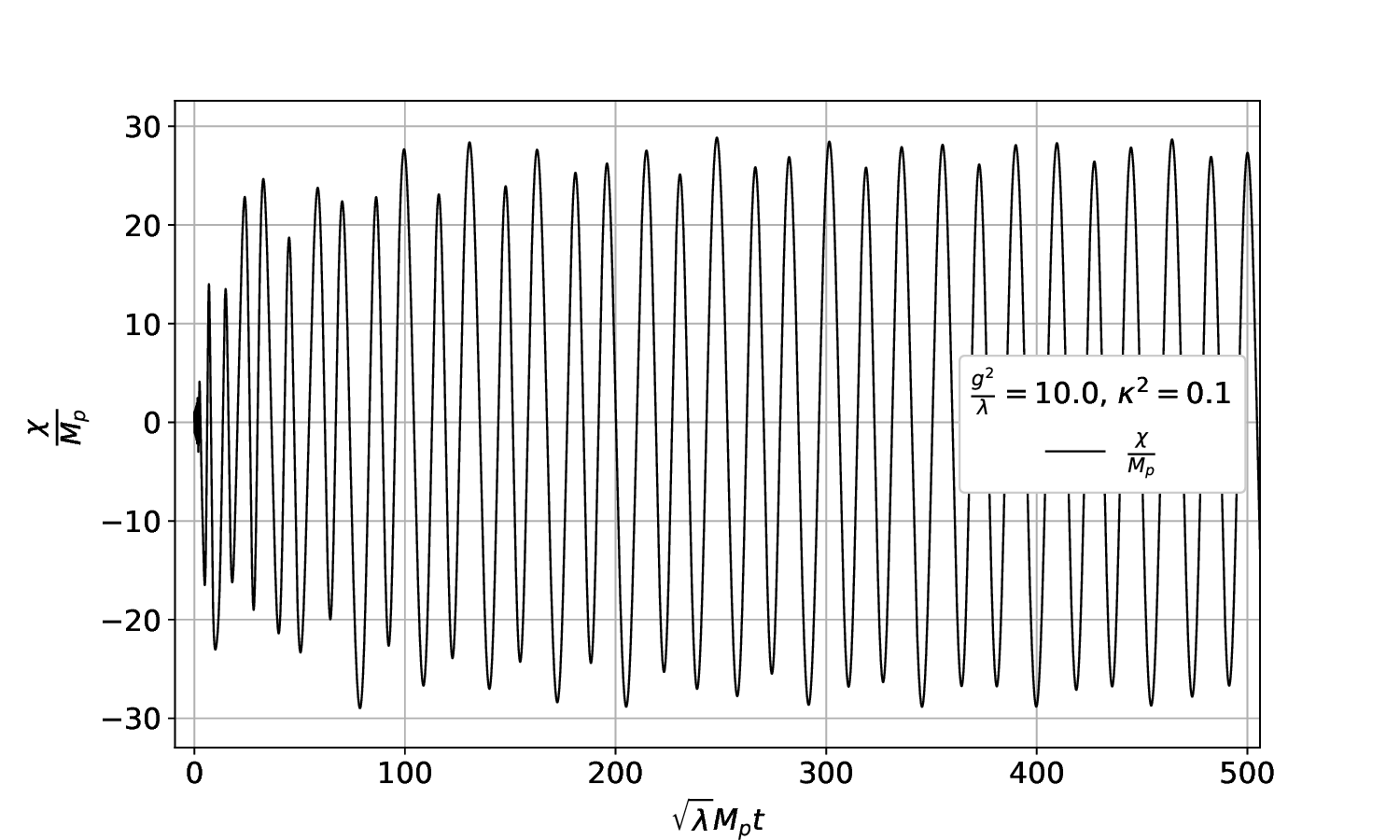}
    \includegraphics[width=.47\textwidth, height=5cm]{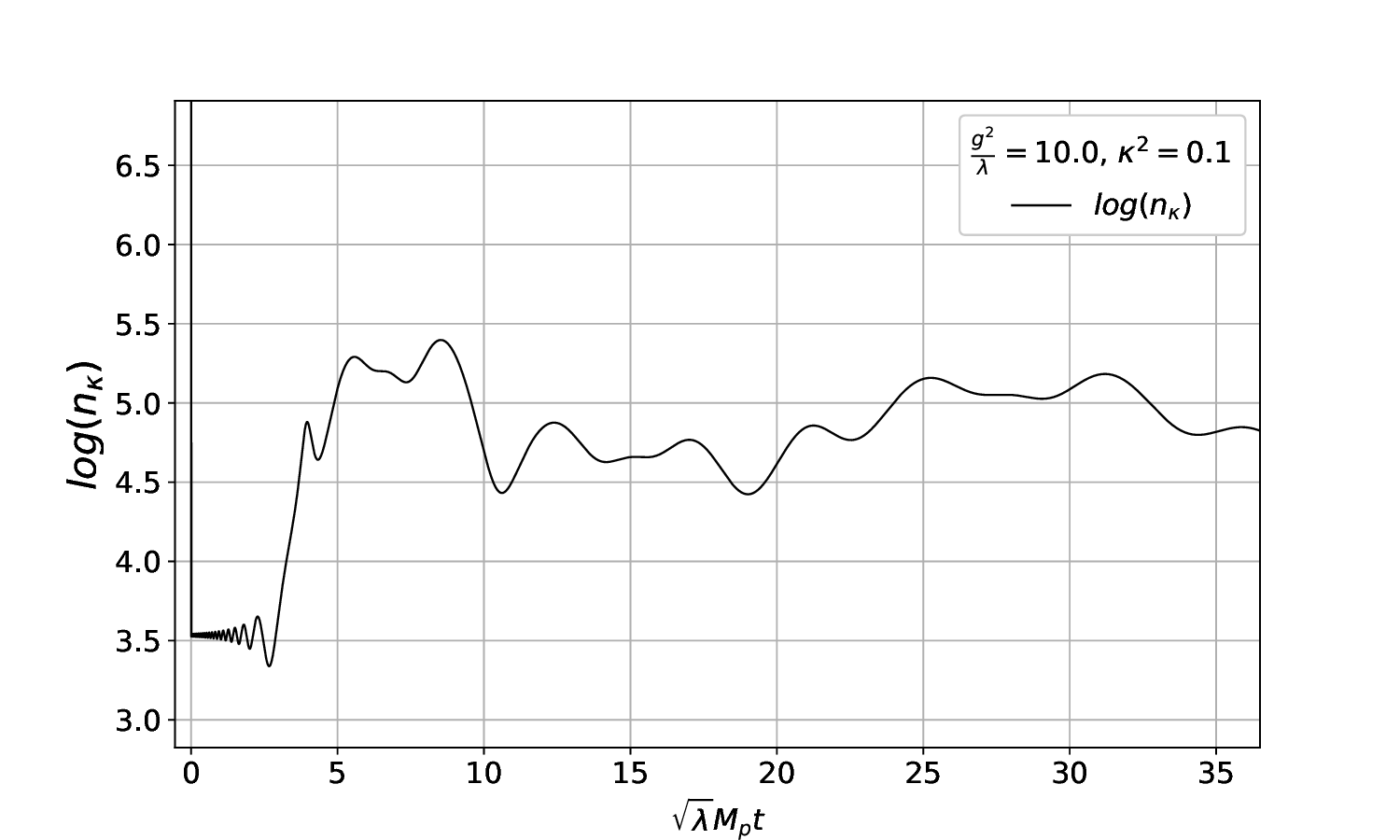}
    \includegraphics[width=.47\textwidth, height=5cm]{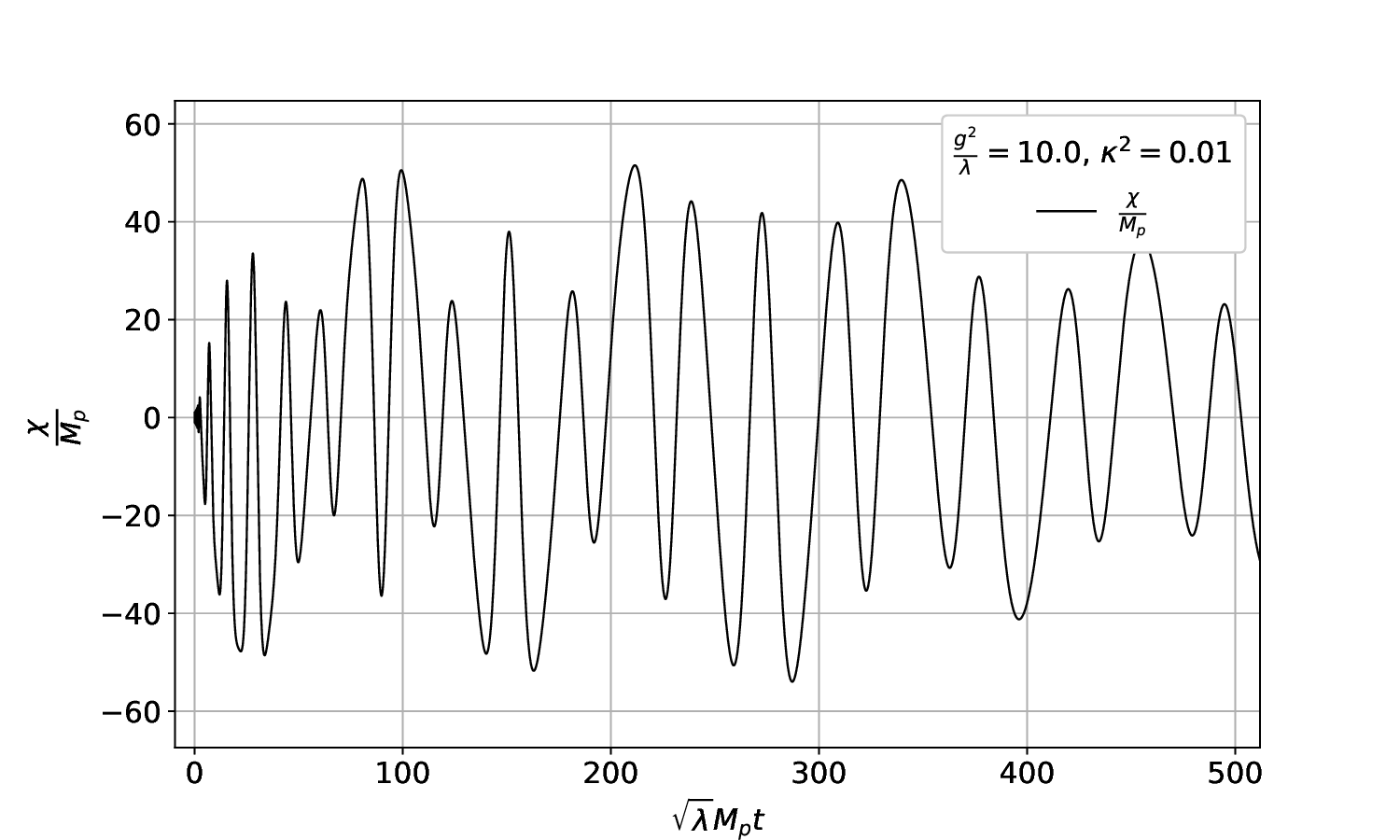}
    \includegraphics[width=.47\textwidth, height=5cm]{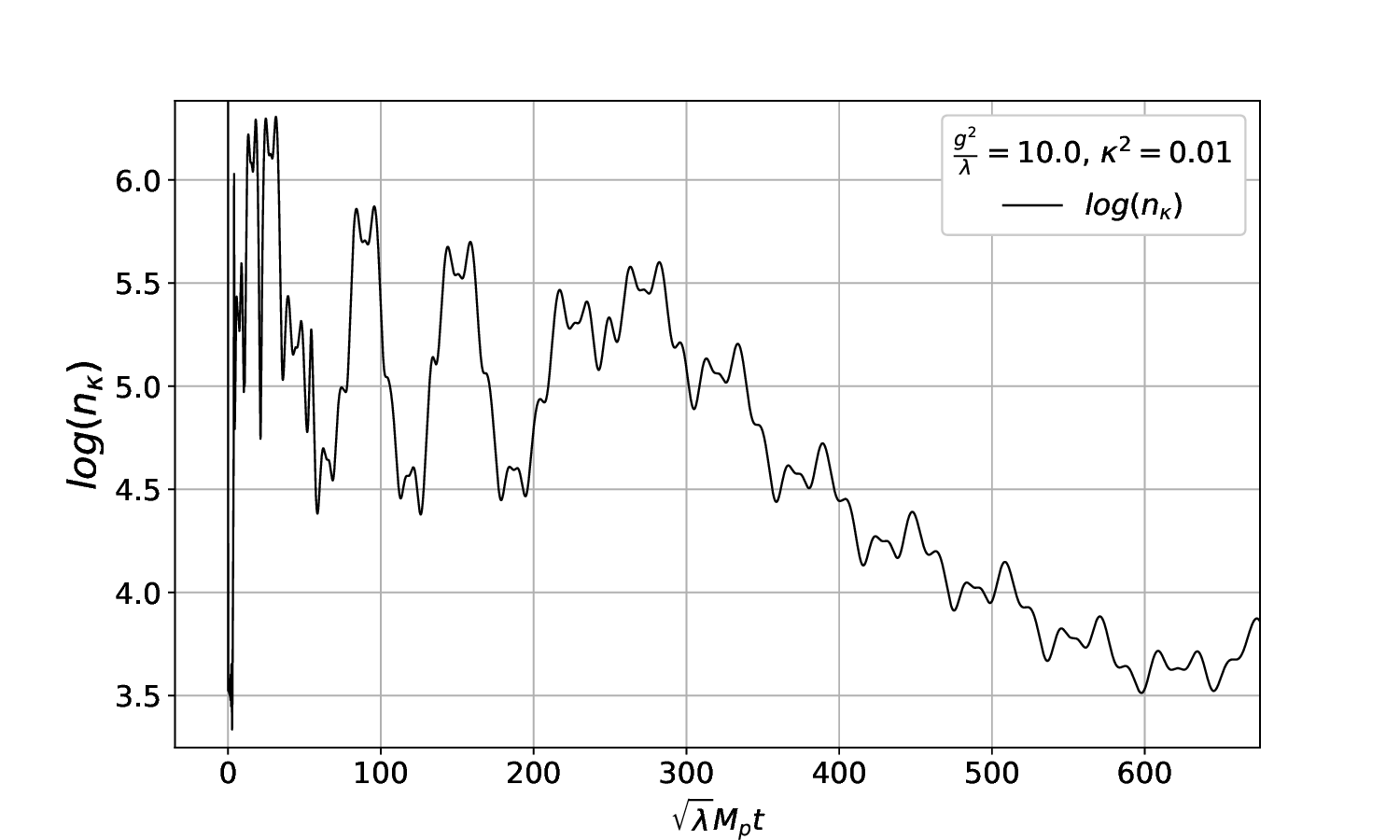}
    \includegraphics[width=.47\textwidth, height=5cm]{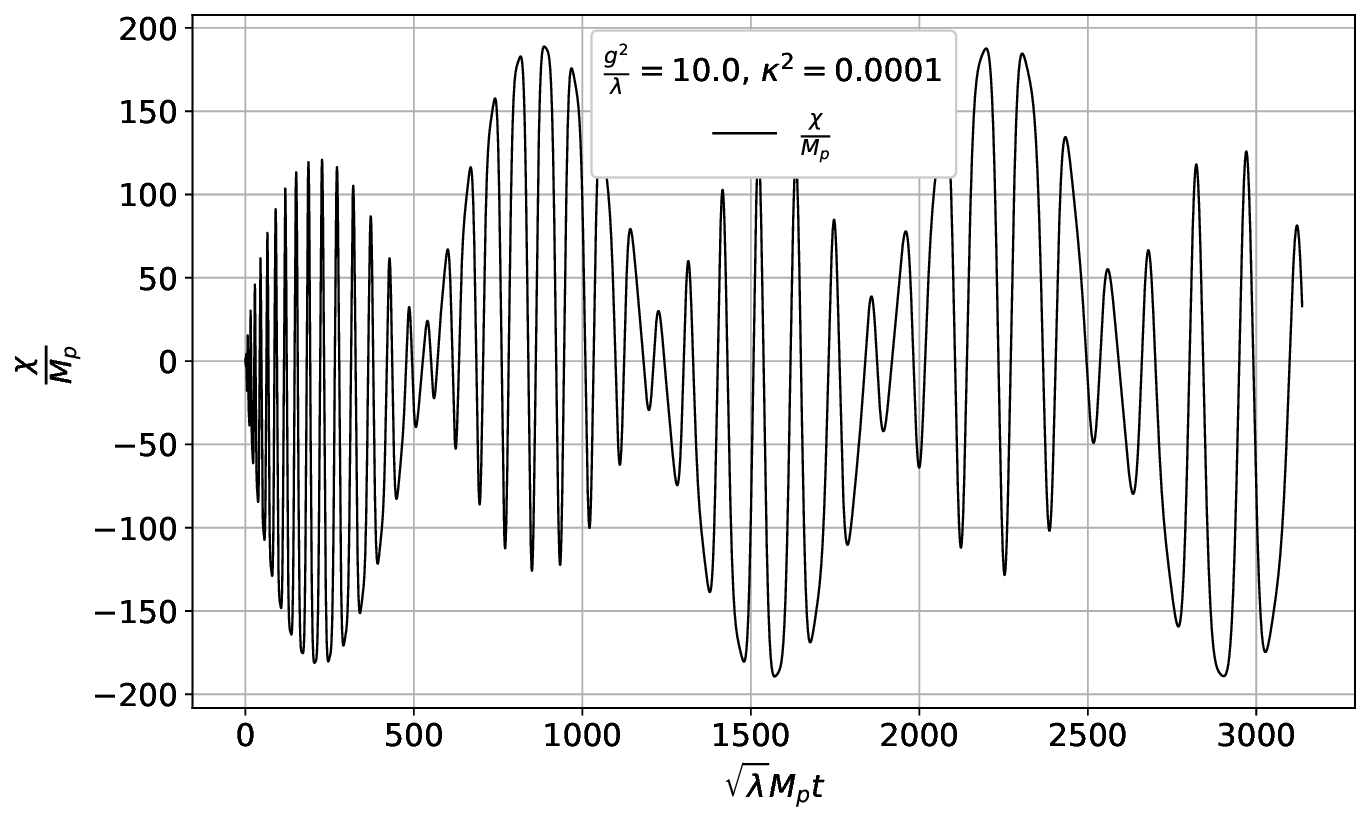}
    \includegraphics[width=.47\textwidth, height=5cm]{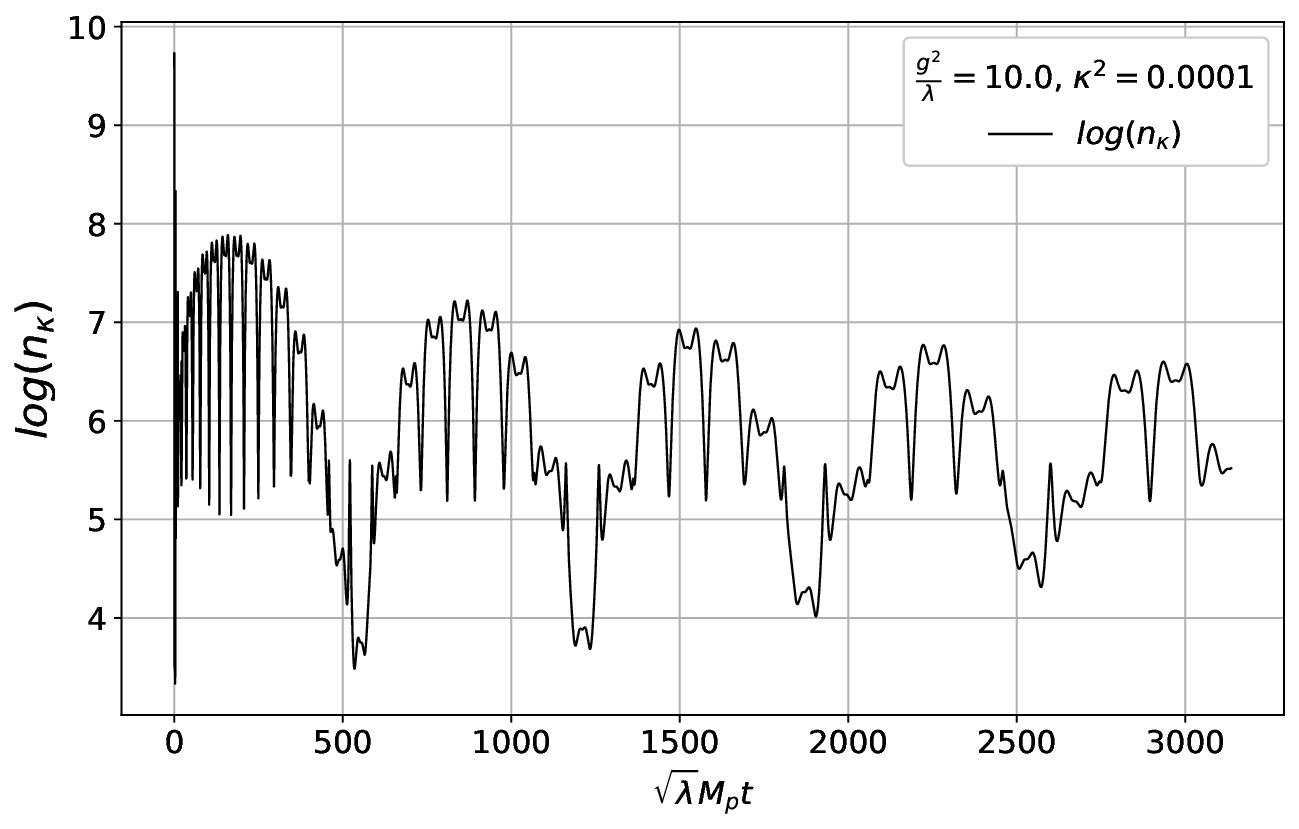}
    \includegraphics[width=.47\textwidth, height=5cm]{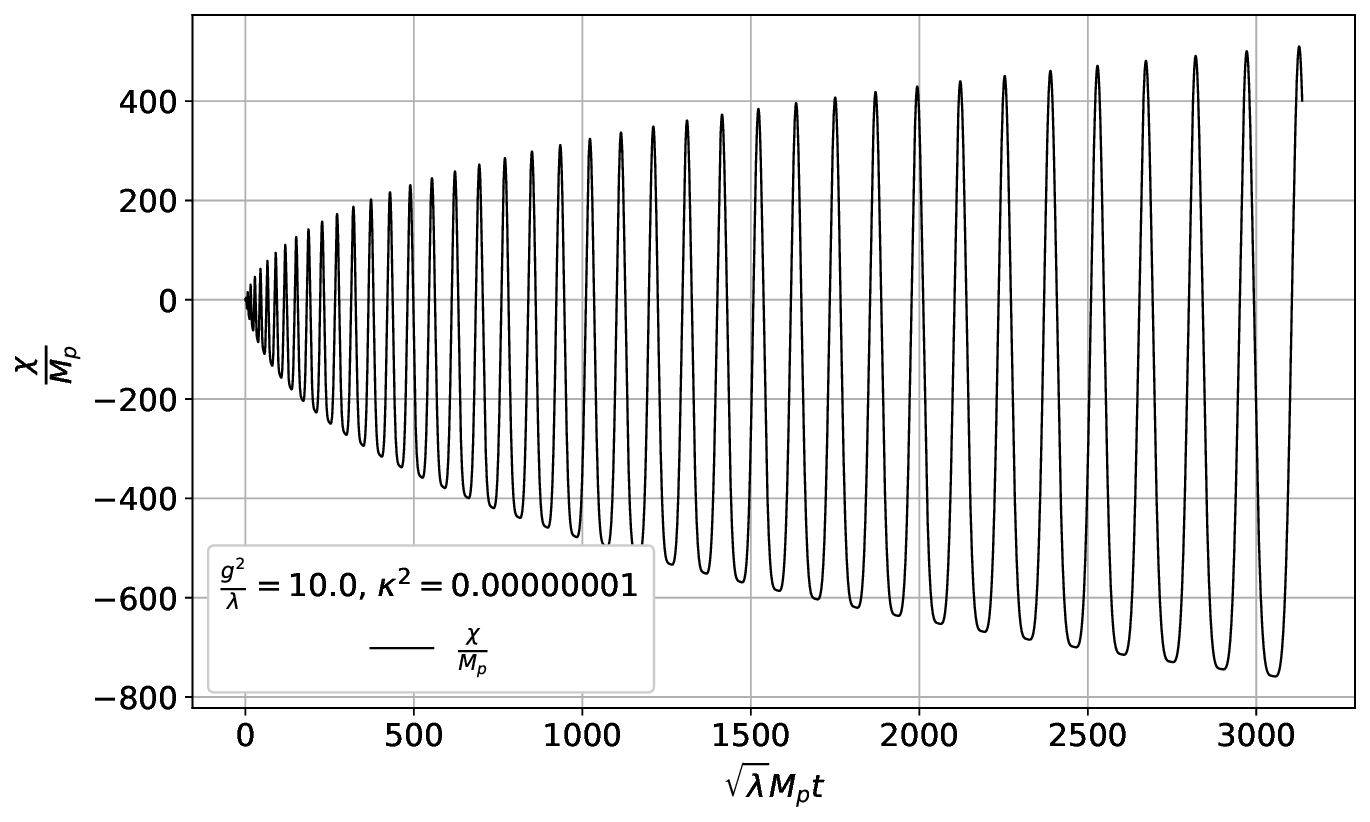}
    \includegraphics[width=.47\textwidth, height=5cm]{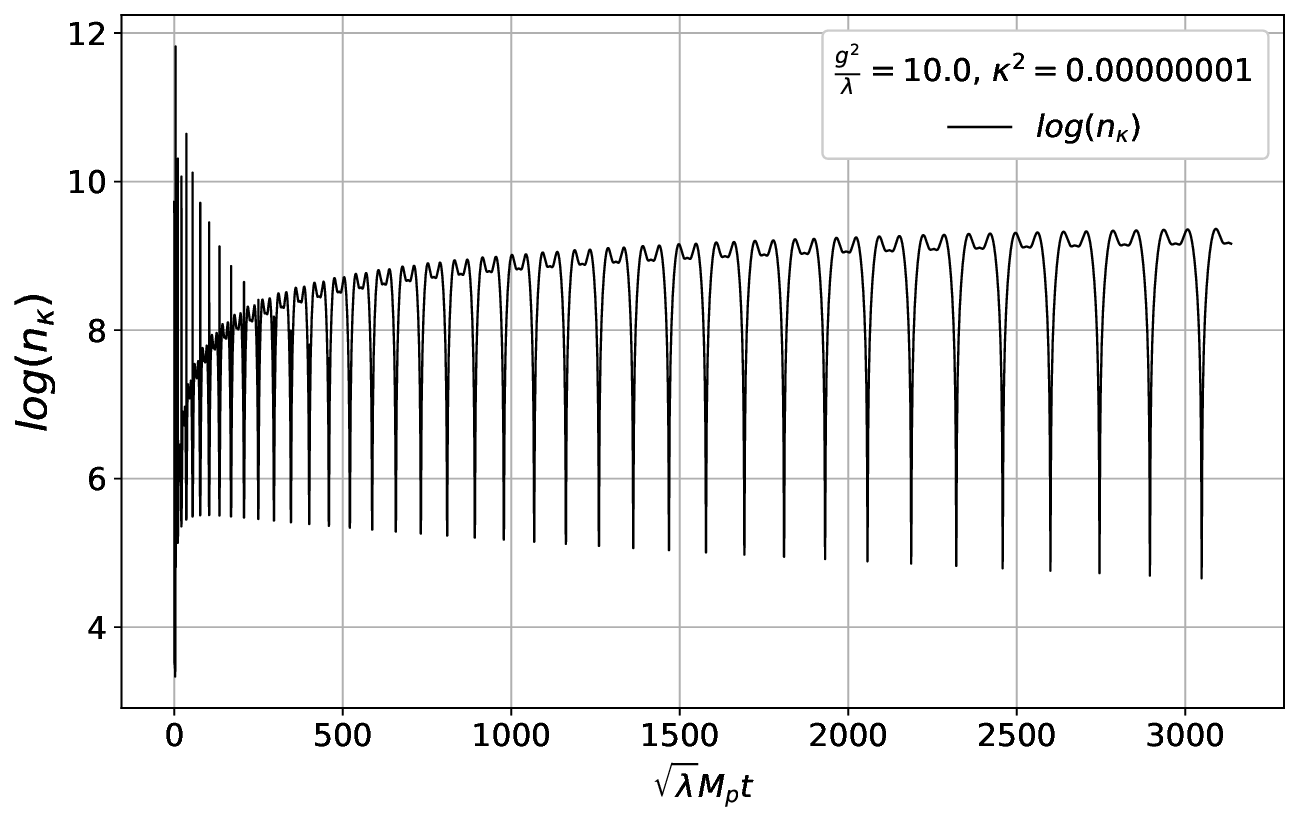}
    \caption{Left panels: Time evolution of the mode function $\chi_k(t)$ for different values of $\kappa^2$. Right panels: Time evolution of the log of occupation number $n_k(t)$ for different values of $\kappa^2$. In all panels, $\frac{g^2}{\lambda}= 10.0$.}
    \label{fig:g_10_k_0.1ZOOM}
\end{figure}
\begin{figure}[H]
\centering
    \includegraphics[width=0.7\textwidth]{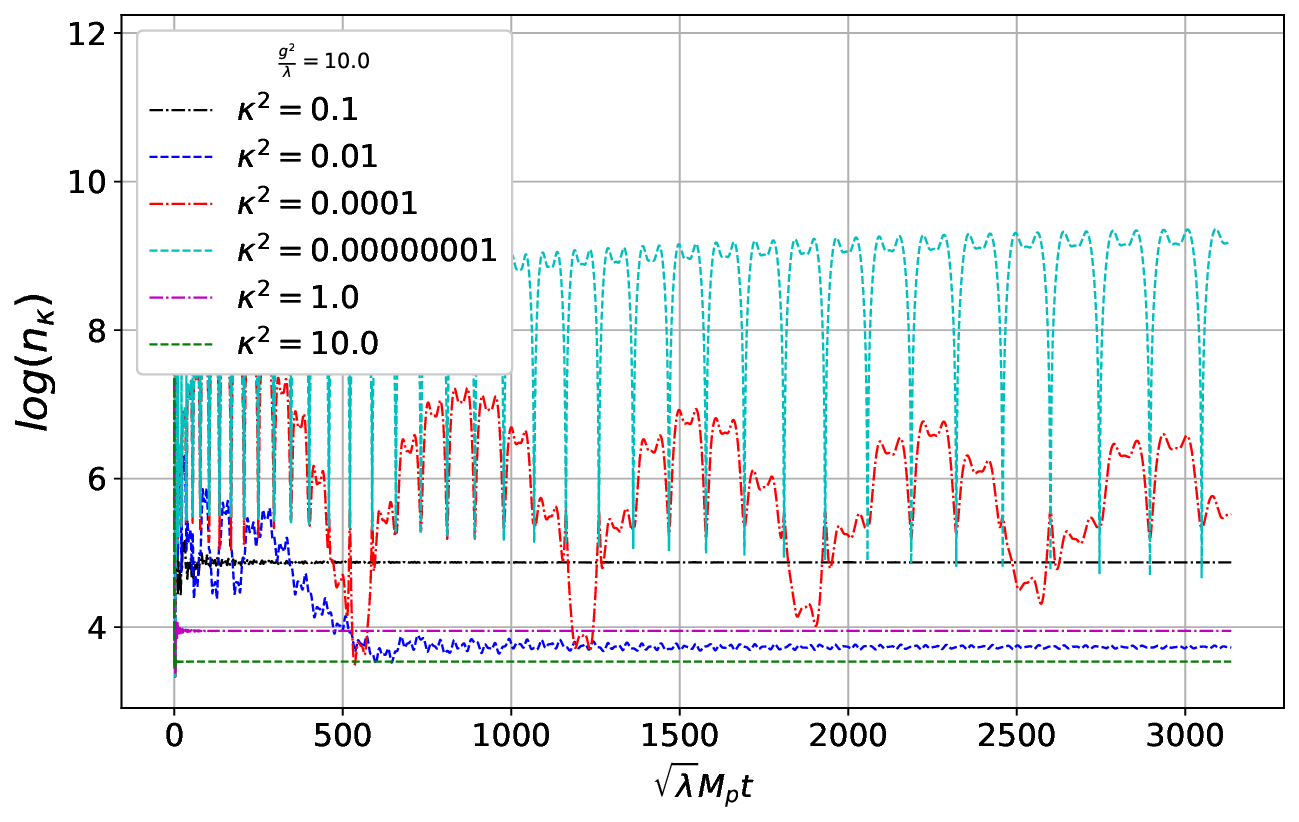}
    \caption{Comparison of the evolution of occupation number $n_k(t)$ for different values of $\kappa^2$, with $\frac{g^2}{\lambda}= 10.0$.}
    \label{fig:g_10_PARTICLEno}
\end{figure}
\begin{equation}
\label{eqn:Fried1b}
H^2 = \frac{1}{3M_{p}^2}\left[\frac{1}{2}\dot{\phi}^2 + \frac{1}{4}\lambda \phi^4\right],
\end{equation}
\begin{equation}
\label{eqn:Fried2b}
    \frac{\ddot{a}}{a} = -\frac{1}{3M_{p}^2}\left[\dot{\phi}^2 - \frac{1}{4}\lambda \phi^4\right],
\end{equation}
and
\begin{equation}
\label{eqn:KGb}
\ddot{\phi}+ 3H\dot{\phi} + \lambda \phi^3=0.
\end{equation}

\section{Dynamics of the created modes}
\label{sec_phi4}

Following Greene et al.~\cite{Greene1997}, we take the Lagrangian, where the inflaton field $\phi$ interacts with a massless bosonic field $\chi$, as
\begin{equation}
\label{eqn:L4}
    \mathcal{L} = \frac{M_P^2}{2} R - \frac{1}{2}g^{\mu\nu}\phi_{,\mu}\phi_{,\nu} - \frac{\lambda}{4}\phi^4
    - \frac{1}{2} g^{\mu\nu}\chi_{,\mu}\chi_{,\nu} - \frac{1}{2} g^2\phi^2\chi^2.
\end{equation}

In the context of quantum field theory, the scalar field $\chi$ may be treated as an operator, and in the Heisenberg representation, it can be expanded as
\begin{equation}
\label{eqn:n20}
    \hat{\chi}(t,\mathbf{x})=\frac{1}{(2\pi)^{3/2}}\int d^3k\left[\hat{a}_k\chi_k(t)e^{-i\mathbf{k\cdot x}}+\hat{a}_k^\dag \chi_k^*(t)e^{i\mathbf{k\cdot x}}\right],
\end{equation}
where $\hat{a}_k$ and $\hat{a}_k^\dag$ are the annihilation and creation operators, respectively. In a flat FLRW background, the equation of motion for the mode function $\chi_k$ is given by
\begin{equation}
\label{eqn:chi4}
    \ddot{\chi}_k + 3\frac{\dot{a}}{a} \dot{\chi}_k + \left( \frac{k^2}{a^2} + g^2\phi^2 \right) \chi_k = 0.
\end{equation}

\begin{figure}[H]

    \includegraphics[width=.47\textwidth, height=5cm]{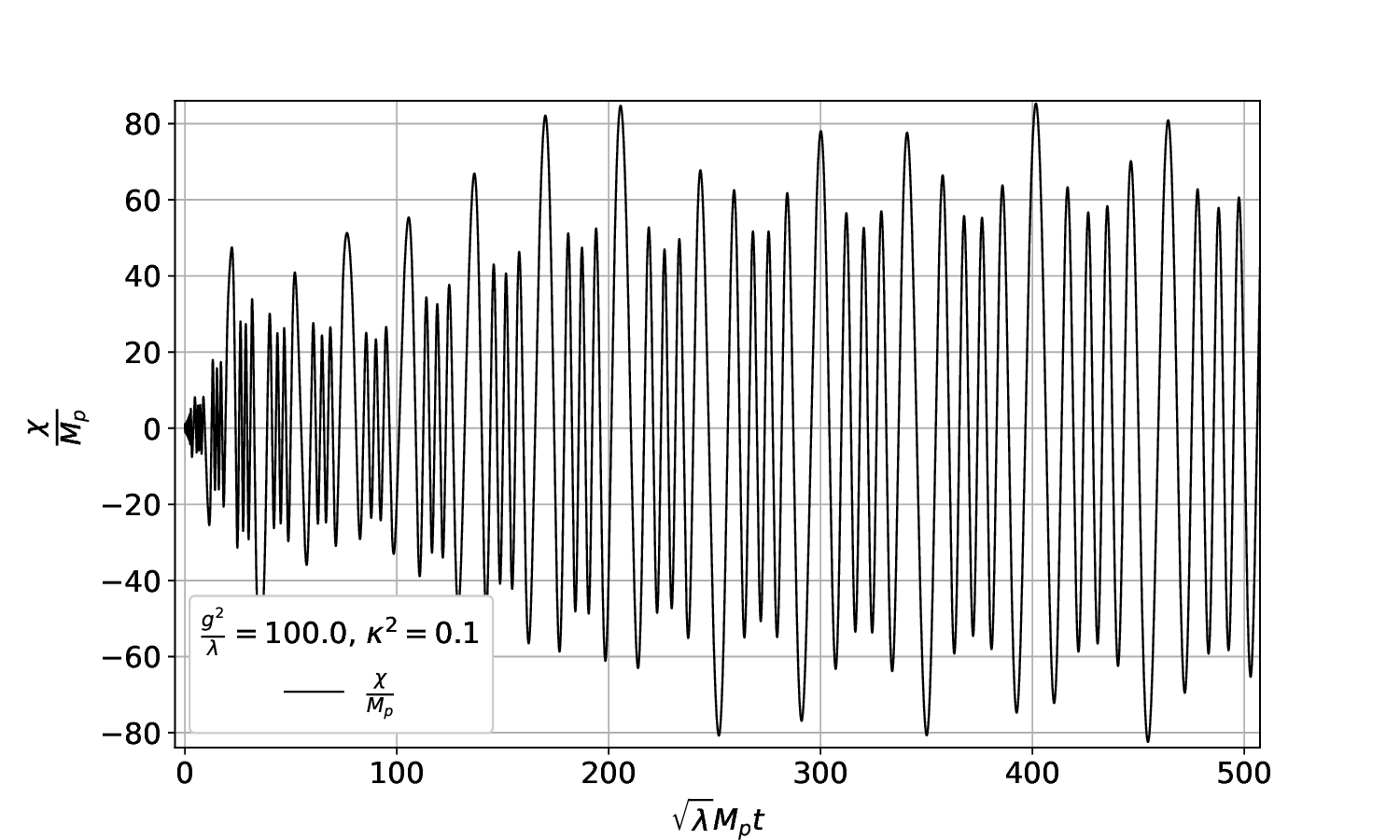}
    \includegraphics[width=.47\textwidth, height=5cm]{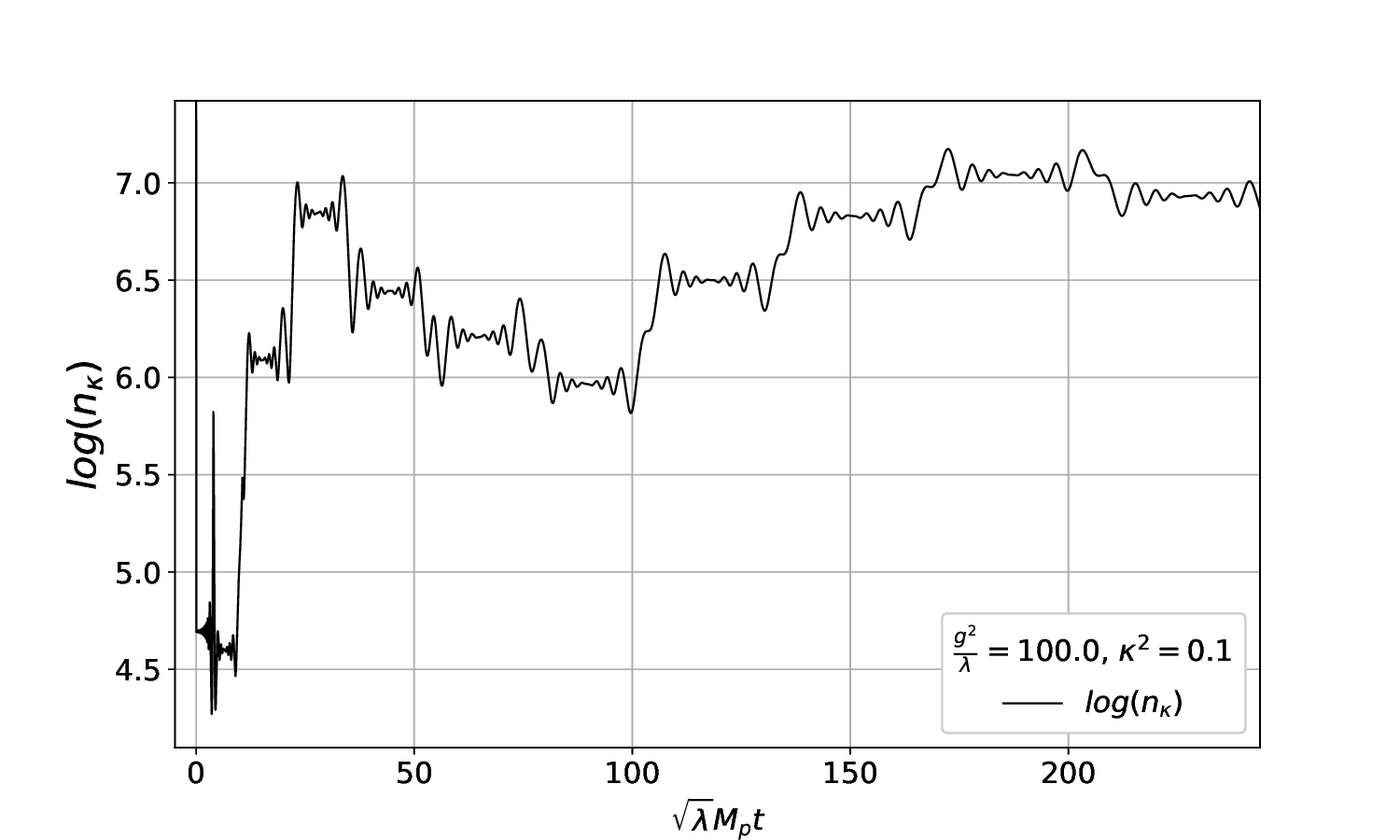}
    \includegraphics[width=.47\textwidth, height=5cm]{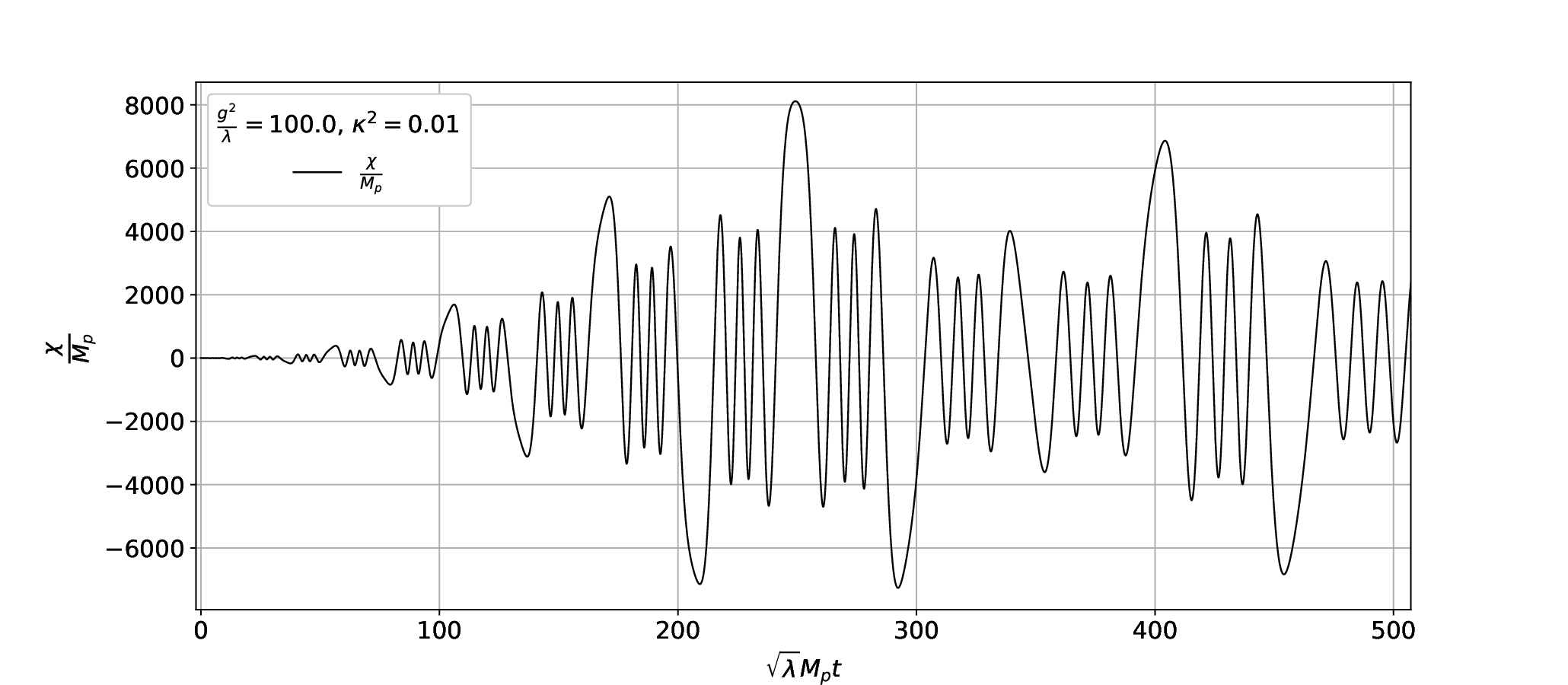}
    \includegraphics[width=.47\textwidth, height=5cm]{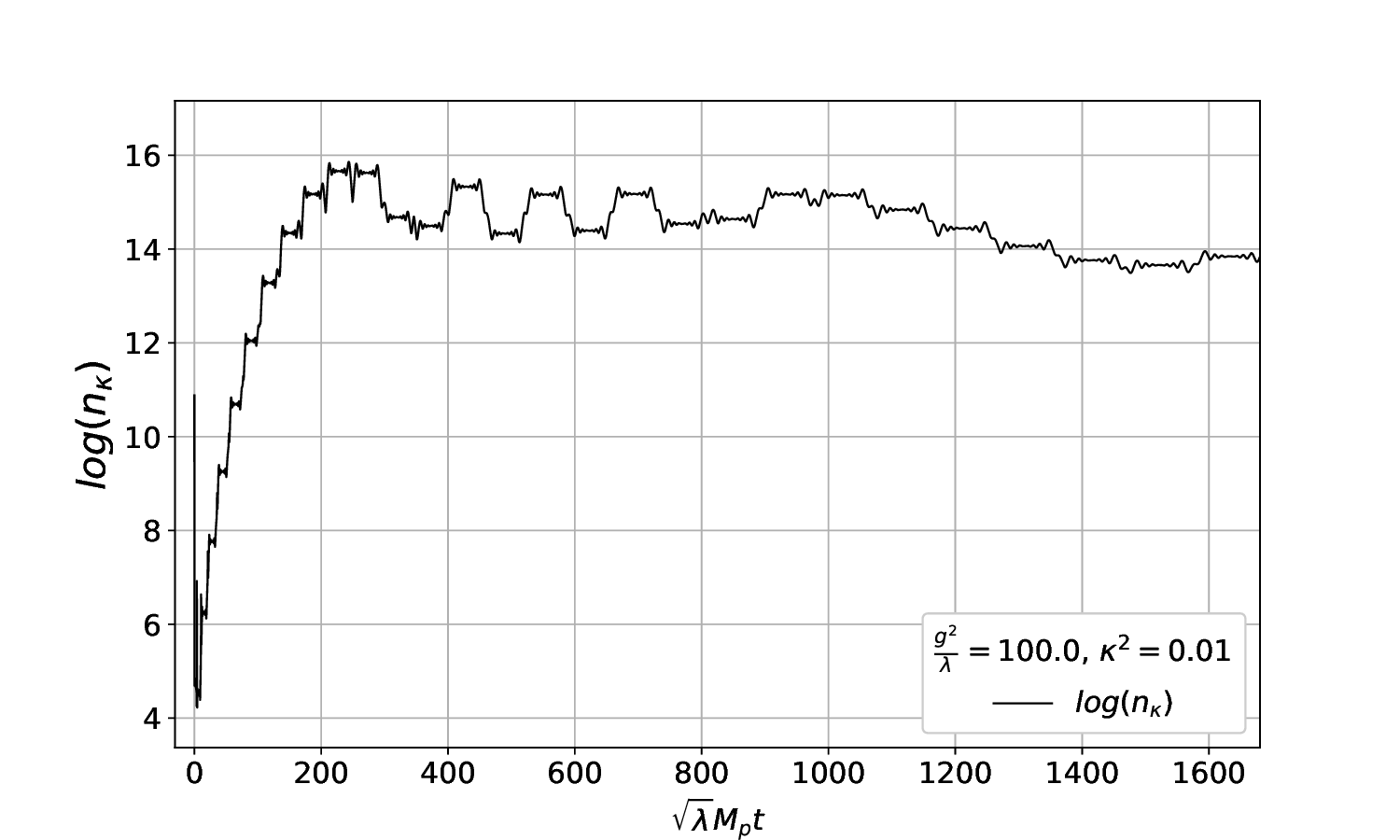}
    \includegraphics[width=.47\textwidth, height=5cm]{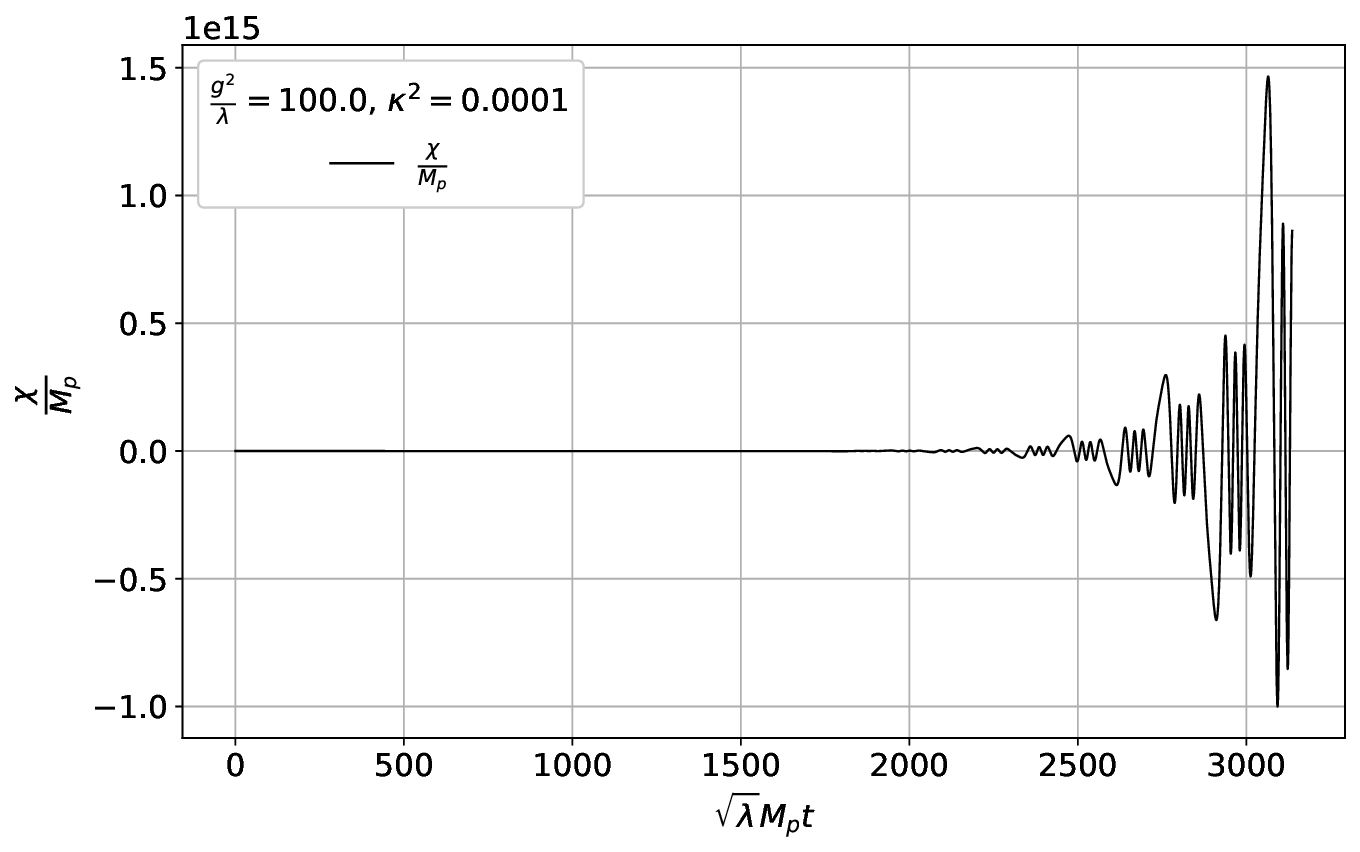}
    \includegraphics[width=.47\textwidth, height=5cm]{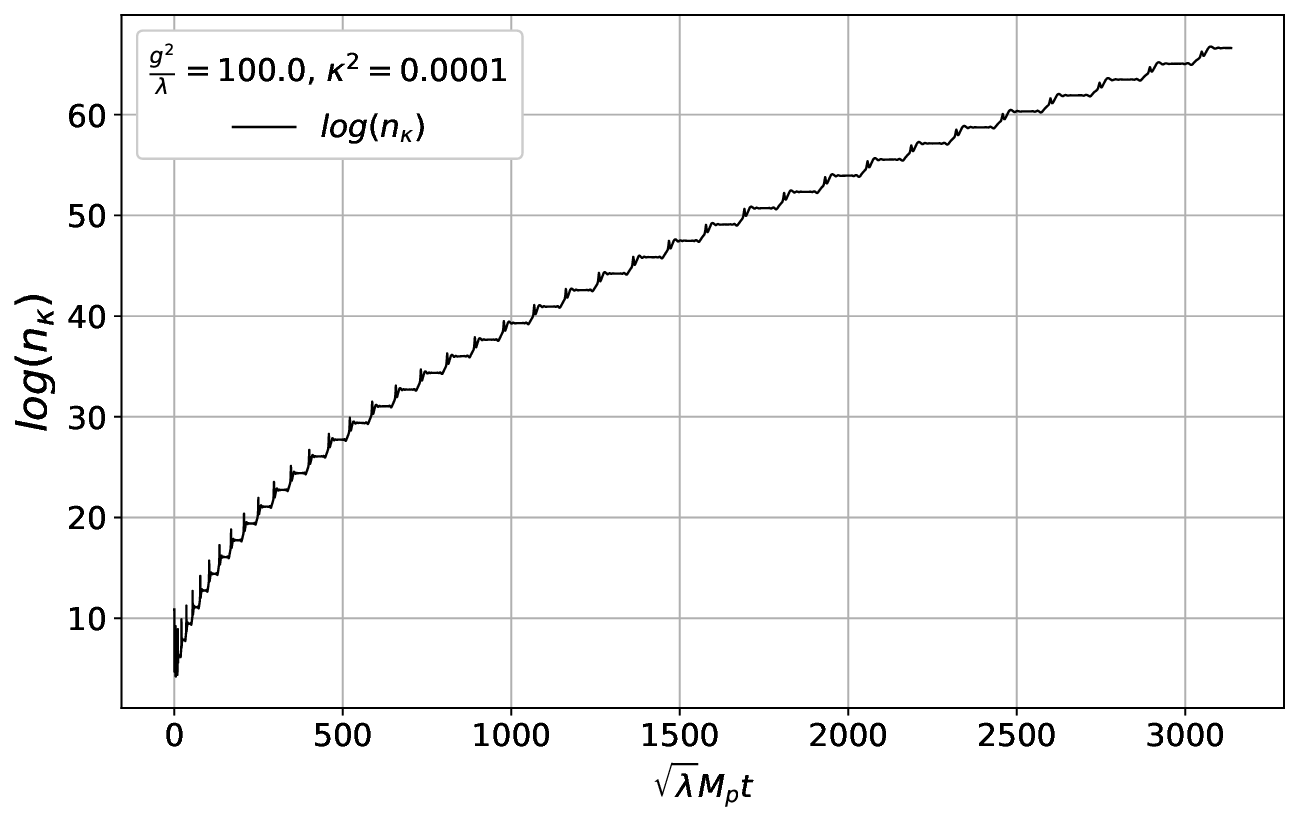}
    \includegraphics[width=.47\textwidth, height=5cm]{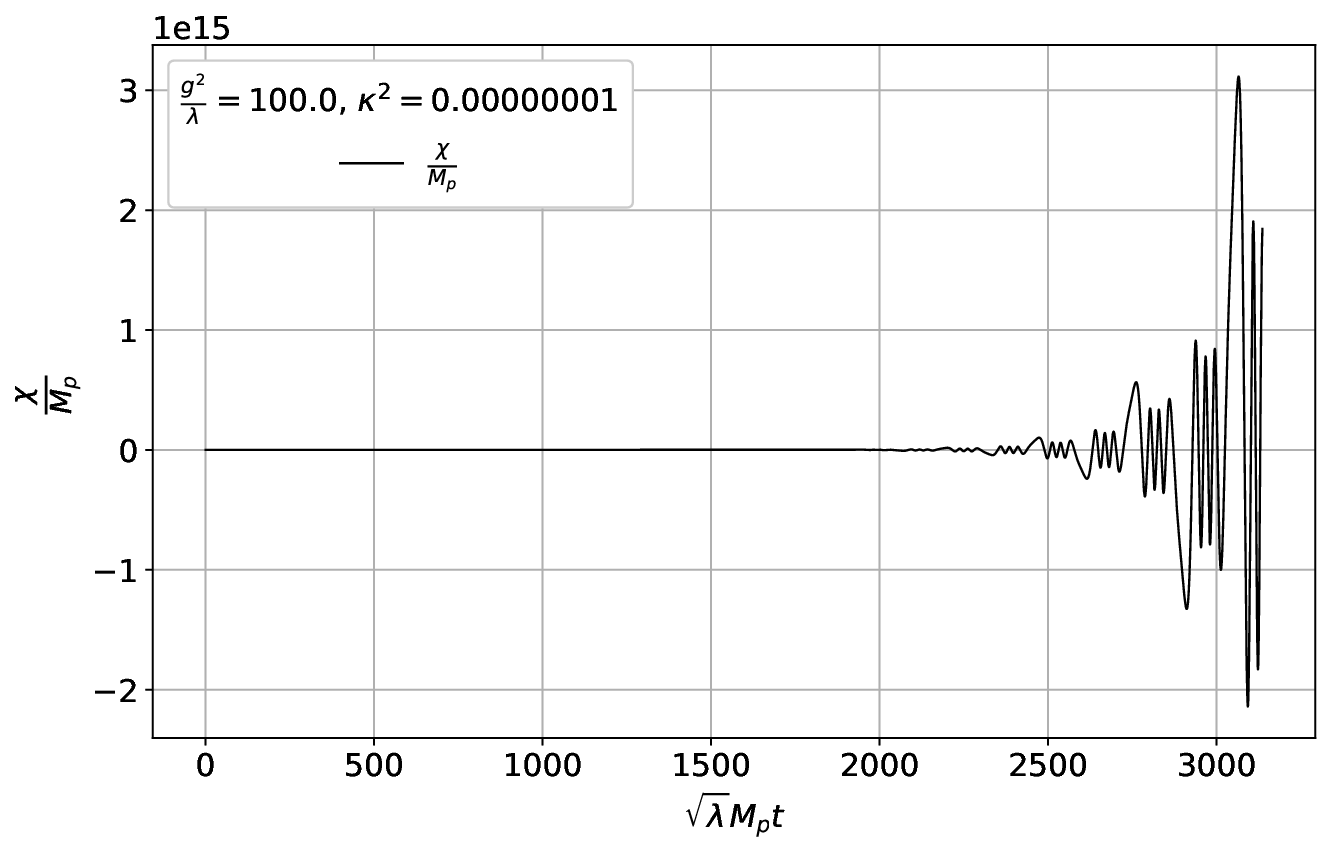}
    \includegraphics[width=.47\textwidth, height=5cm]{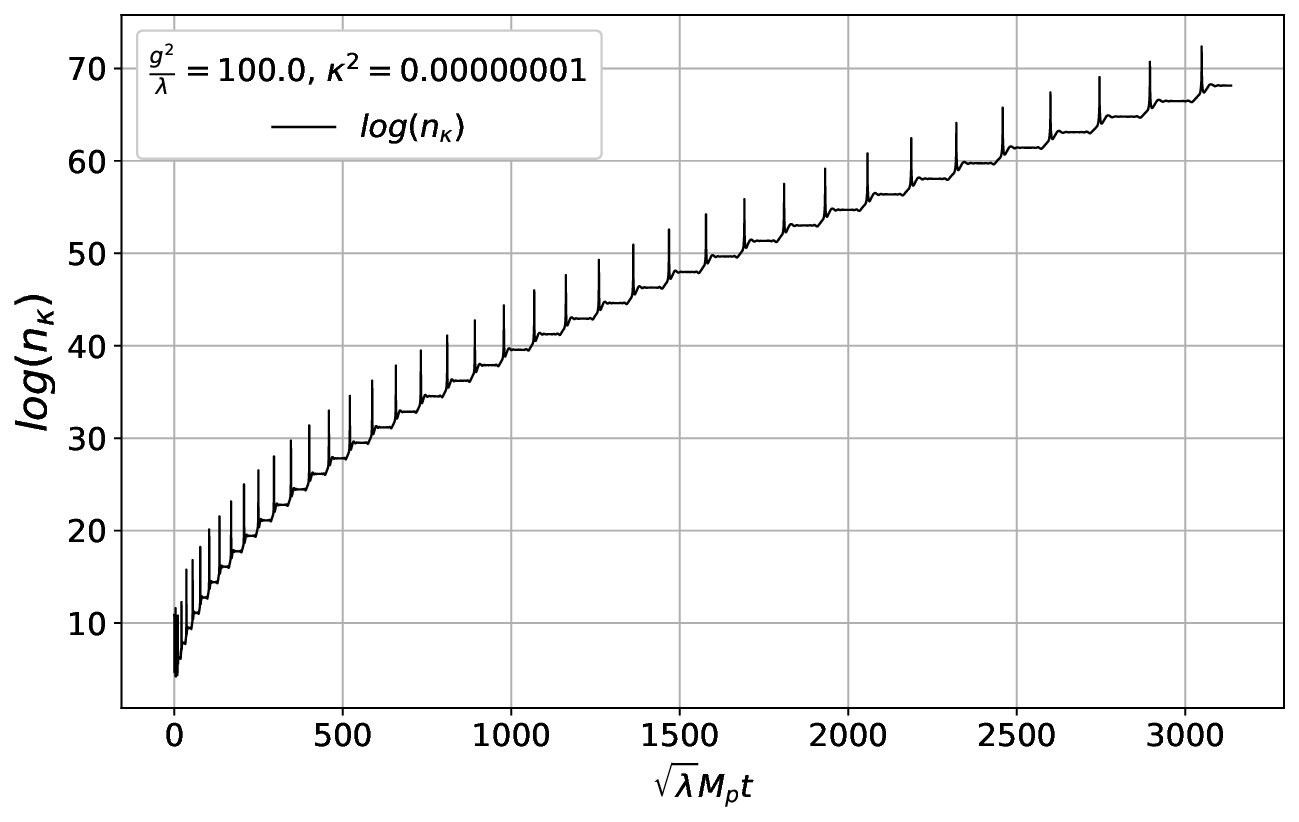}
    \caption{Left panels: Time evolution of the mode function $\chi_k(t)$ for different values of $\kappa^2$. Right panels: Time evolution of the log of occupation number $n_k(t)$ for different values of $\kappa^2$. In all panels, $\frac{g^2}{\lambda}= 100.0$.}
    \label{fig:g_100_k_0.1ZOOM}

\end{figure}

\begin{figure}[H]
\centering
    \includegraphics[width=0.7\textwidth]{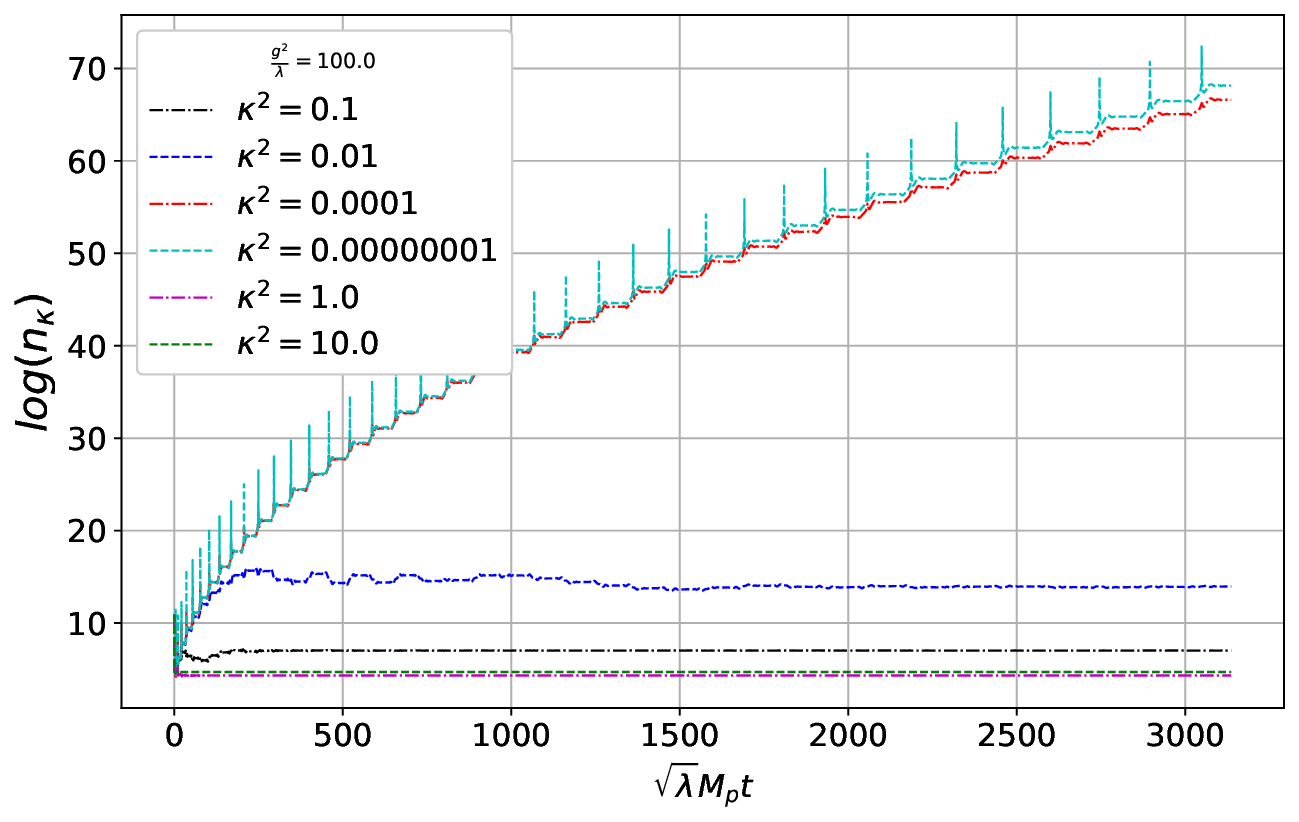}
    \caption{Comparison of the evolution of occupation number $n_k(t)$ for different values of $\kappa^2$, with $\frac{g^2}{\lambda}= 100.0$.}
    \label{fig:g_100_PARTICLEno}
\end{figure}

In order to compare the results with the quadratic case, we take $a=1$, and with the redefinitions $\frac{\chi}{M_P} =  X$, $(\sqrt{\lambda}M_P)t = \tau$ and $\kappa^2 = \frac{k^2}{\lambda M_P^2}$, so that equation (\ref{eqn:chi4}) becomes
\begin{equation}
    \label{eqn:chi4r}
    \ddot{X}_k + \left( \kappa^2 + \frac{g^2}{\lambda}\phi^2 \right) X_k = 0.
\end{equation}

Importantly, this equation \ref{eqn:chi4r} contains all features of the nonlinearity in the $\phi^4$ potential via the last term, $ \frac{g^2}{\lambda}\phi^2 X_k$. Solving this differential equation {\em exactly} will thus disclose the effect of the nonlinearity on the resonance behaviour of the model in a straightforward manner.

Noting that the energy of the mode in (\ref{eqn:chi4r}) can be written as
\begin{equation*}
    E_k = \frac{1}{2}|\dot{X}_k|^2+\frac{1}{2}\omega_k^2|X_k|^2 = \left(n_k + \frac{1}{2}\right) \omega_k,
\end{equation*}
where $\omega_k^2 (t) = \kappa^2 + \frac{g^2}{\lambda}\phi^2(t)$, the occupation number $n_k(t)$ of the created particles in the mode $k$ takes the form
\begin{equation}
\label{eqn:lognk}
    n_k = \frac{\omega_k}{2}\left(\frac{|\dot{X}_k|^2}{\omega_k^2}+|X_k|^2 \right)- \frac{1}{2},
\end{equation}
which can be evaluated from the solutions of equation \ref{eqn:chi4r}.

In order to obtain an exact solution of equation \ref{eqn:chi4r}, we need the exact behaviour of the driving field $\phi(t)$. This can be achieved only by exact numerical integration of the inflaton dynamics given by equation \ref{eqn:KGb} coupled with the Freidmann equation \ref{eqn:Fried1b}. This is because the non-linear nature of the differential equation \ref{eqn:chi4r} forbids any exact analytical treatment in a straightforward manner to obtain the resonance behaviour.

\section{Numerical Integration}\label{sec_NumericalIntegration}

We numerically integrate Eqs.~\ref{eqn:Fried1b}, \ref{eqn:KGb}, and \ref{eqn:chi4r} simultaneously in order to obtain the exact time evolution of the mode function $ X_k(t) $ for different values of $\kappa$ and $\frac{g^2}{\lambda}$. These solutions are then used to compute the corresponding occupation numbers $ n_k(t) $ directly from Eq.~\ref{eqn:lognk}, thereby enabling a fully non-perturbative analysis of particle production during preheating.\\

Figure \ref{fig:p4vt} illustrates the temporal evolution of the inflaton field $ \phi(t) $, clearly demonstrating that it undergoes damped oscillation in the preheating regime.\\

For the coupling ratio $ g^2/\lambda = 0.1 $, which characterizes the relative strength of the interaction terms, Figure \ref{fig:g_0.1_k_0.1ZOOM} presents the time evolution of the mode function $ \chi_k(t) $ [or equivalently $ X_k(t) $] along with the associated occupation numbers $ n_k(t) $, for several values of the dimensionless wavenumber $ \kappa^2 = \frac{k^2}{\lambda M_p^2} $.

From Figure \ref{fig:g_0.1_k_0.1ZOOM}, it is evident that for larger values of the wavenumber $ k $, the mode function $ \chi_k(t) $ settles into oscillations with nearly constant amplitude shortly after the onset of preheating. In contrast, for smaller values of $ k $, the amplitude of oscillation grows, indicating enhanced resonance effects in the long-wavelength regime.

The corresponding evolution of the occupation number $ n_k(t) $ exhibits distinct behavior across length-scales. For higher $ k $-modes, $ n_k(t) $ approaches an approximately stationary value, suggesting a saturation of particle production. Conversely, as $ k $ decreases, the evolution of $ n_k(t) $ transitions into a non-linear oscillating behaviour, reflecting sustained energy transfer and non-linear mode coupling.

For clarity, the evolution of $ n_k(t) $ across different wavenumbers is superposed in Figure \ref{fig:g_0.1_PARTICLEno}, facilitating a direct comparison of their dynamical behavior.\\

We next consider a larger coupling ratio, $ g^2/\lambda = 1.0 $. As illustrated in Figure \ref{fig:g_1_k_0.1ZOOM}, the behavior of the mode function $ \chi_k(t) $ for higher wavenumbers $ k $ remains qualitatively similar to the previously discussed case. In contrast, for smaller values of $ k $, the amplitude of the mode function $ \chi_k(t) $ exhibits a pronounced growth over time, indicating a stronger resonance effect in the long-wavelength sector.

The corresponding evolution of the occupation number $ n_k(t) $ reveals a complementary trend. For lower $ k $-modes, $ n_k(t) $ approaches an approximately stationary value, suggesting saturation of particle production. However, for higher wavenumbers, the envelope of $ n_k(t) $ displays a gradual increase accompanied by nonlinear oscillations, reflecting sustained dynamical evolution in these modes.

A comprehensive comparison of the time evolution of $ n_k(t) $ across different wavenumbers is presented in Figure \ref{fig:g_1_PARTICLEno}.\\

For a slightly higher coupling ratio $ g^2/\lambda = 3 $, Figure \ref{fig:g_3_k_0.1ZOOM} presents the time evolution of the mode function $ \chi_k(t) $ and the corresponding occupation number $ n_k(t) $ for various values of the wavenumber $ k $. For larger $ k $, the mode function evolves into an oscillatory regime with nearly constant amplitude, while the associated occupation number $ n_k(t) $ approaches an approximately constant average value, indicating saturation of particle production in these modes.

In contrast, as the wavenumber $ k $ decreases, the dynamics becomes significantly more complicated. The mode function $ \chi_k(t) $ exhibits a transition to a stochastic oscillatory regime, reflecting the onset of strong nonlinear effects. Correspondingly, the evolution of $ n_k(t) $ also undergoes a transition: after passing through a stochastic phase, it settles into a regime characterized by an approximately constant envelope with superimposed nonlinear oscillation.

The time evolution of the occupation number $ n_k(t) $ for different wavenumbers is summarized in Figure \ref{fig:g_3_PARTICLEno}, enabling a direct comparison of their behavior across the spectrum.\\

For the coupling ratio $ g^2/\lambda = 10 $, Figure \ref{fig:g_10_k_0.1ZOOM} illustrates the time evolution of the mode function $ \chi_k(t) $ and the corresponding occupation numbers $ n_k(t) $ for a range of wavenumbers $ k $. In this regime, the stochastic feature in the evolution of $ \chi_k(t) $ becomes significantly more pronounced compared to the previous case. As the wavenumber decreases, this stochastic behavior intensifies, indicating the increasing influence of nonlinear dynamics. However, for sufficiently small $ k $, the stochasticity diminishes, and the mode function transitions into an oscillatory regime characterized by a slowly increasing amplitude with a well-defined envelope.

The corresponding evolution of the occupation number $ n_k(t) $ reflects a similarly rich structure. While high-$ k $ modes approach an approximately constant average value, lower-$ k $ modes exhibit a sequence of dynamical transitions. As the wavenumber is continually decreased, it enters into a nonlinear oscillatory regime, followed by the appearance of a beat-like pattern, and eventually settling into a phase with a nearly constant envelope superimposed with nonlinear oscillations.

A consolidated view of the time evolution of $ n_k(t) $ across different wavenumbers is presented in Figure \ref{fig:g_10_PARTICLEno}, facilitating a comparative analysis of these behaviors.\\

For the higher ratio $ g^2/\lambda = 100 $, Figure \ref{fig:g_100_k_0.1ZOOM} presents the time evolution of the mode function $ \chi_k(t) $ and the corresponding occupation number $ n_k(t) $ for a range of wavenumbers $ k $. In this strongly coupled regime, the dynamics of the high-$ k $ modes exhibit pronounced stochastic behavior. This stochasticity is clearly reflected in the evolution of the corresponding occupation number $ n_k(t) $, which displays a characteristic step-like (staircase) pattern, indicative of intermittent bursts of particle production.

However, in the low-$k$ regime, the onset of stochasticity in $ \chi_k(t) $ occurs only after a relatively prolonged period of evolution while $n_k(t)$ undergoes a monotonic evolution on an average with small fluctuations superimposed.

A comprehensive comparison of the time evolution of $ n_k(t) $ across different wavenumbers is provided in Figure \ref{fig:g_100_PARTICLEno}, enabling a clear comparison of the dynamical features in this strongly nonlinear regime.

\section{Discussion and Conclusion}\label{sec_conclusion}

The phenomenon of preheating as an efficient mechanism for particle production after inflation has been well established through the seminal work of Kofman, Linde, and Starobinsky, who demonstrated the role of parametric resonance in driving explosive particle creation. In the context of quadratic inflaton potentials, the dynamics is governed by the Mathieu equation, leading to well-defined narrow and broad resonance bands, as well as stochastic resonance in an expanding background. For the quartic potential, earlier studies based on approximate treatments reduced the dynamics to a Lam\'e-type equation, revealing resonance structures qualitatively distinct from the quadratic case.

In this work, we have reconsidered the quartic inflaton model with the aim of providing an exact characterization of parametric resonance during the preheating phase, without resorting to approximations. By numerically solving the coupled system of equations governing the inflaton dynamics and the evolution of particle mode functions, we have avoided the approximations inherent in earlier analyses, such as Hartree-type or Lam\'e-type approximations, and thereby captured the full nonlinear behavior of the system.

Our results reveal a rich and coupling-dependent structure of resonance dynamics. In the weak coupling regime, short-wavelength modes rapidly approach a regime of nearly constant-amplitude oscillations, with the associated occupation numbers saturating to approximately steady values. In contrast, long-wavelength modes exhibit a gradual growth in amplitude, accompanied by nonlinear oscillations in the occupation number, indicating sustained energy transfer over extended timescales. As the coupling strength increases, nonlinear effects become increasingly significant, and the system transitions into a regime characterized by stochastic dynamics. In this strongly coupled regime, short-wavelength modes display intermittent bursts of particle production, reflected in a distinctive step-like (staircase) evolution of the occupation number. Meanwhile, long-wavelength occupation numbers follow a smoother, monotonic growth with superimposed fluctuations, highlighting a clear scale-dependent distinctive feature in the dynamics.

These findings differ substantially from earlier results obtained within approximate frameworks for the quartic model, underscoring the importance of an exact numerical treatment in accurately capturing the structure of parametric resonance. In particular, the emergence of stochastic behavior and the distinct evolution patterns across different wavelength regimes emphasize the limitations of approximate analytical descriptions in strongly nonlinear settings.

Although the quartic inflaton potential considered here is not favored by current observational constraints from Planck, BICEP, and Keck experiments, it provides a valuable theoretical setting for developing and testing methods to study nonlinear preheating dynamics. The relative simplicity of the $\phi^4$ model enables the construction of a robust numerical framework that fully incorporates the coupled evolution of the inflaton and the produced field-mode without resorting to approximations. Importantly, the methodology developed in this work is readily extendible to more realistic inflationary scenarios, including models with more complicated potentials such as the Starobinsky model and $\alpha$-attractor frameworks.

In this sense, the present study not only clarifies the detailed resonance structure in quartic preheating but also establishes a foundation for future investigations of particle production in observationally viable inflationary models, where exact treatments are expected to play a crucial role in understanding the nonlinear dynamics of the early Universe.

\section*{Acknowledgments}\label{sec_acknowledgments}

Hrisikesh Thakur is supported through a CSIR Research Fellowship (Award No. 09/731(0195)/2021-EMR-I), Government of India. The Authors are thankful to the Indian Institute of Technology Guwahati for providing access to computing and supercomputing facilities.


\begin{thebibliography}{10}

\bibitem{Guth1981}
Alan~H. Guth.
\newblock Inflationary universe: A possible solution to the horizon and
  flatness problems.
\newblock {\em Phys. Rev. D}, 23:347--356, Jan 1981.
\newblock URL: \url{https://link.aps.org/doi/10.1103/PhysRevD.23.347}, \href
  {https://doi.org/10.1103/PhysRevD.23.347}
  {\path{doi:10.1103/PhysRevD.23.347}}.

\bibitem{Starobinsky1980}
A.A. Starobinsky.
\newblock A new type of isotropic cosmological models without singularity.
\newblock {\em Physics Letters B}, 91(1):99--102, 1980.
\newblock URL:
  \url{https://www.sciencedirect.com/science/article/pii/037026938090670X},
  \href {https://doi.org/10.1016/0370-2693(80)90670-X}
  {\path{doi:10.1016/0370-2693(80)90670-X}}.

\bibitem{Kazanas1980}
D.~{Kazanas}.
\newblock {Dynamics of the universe and spontaneous symmetry breaking}.
\newblock {\em Astrophysical Journal}, 241:L59--L63, October 1980.
\newblock \href {https://doi.org/10.1086/183361} {\path{doi:10.1086/183361}}.

\bibitem{Sato1981}
Katsuhiko Sato.
\newblock Cosmological baryon-number domain structure and the first order phase
  transition of a vacuum.
\newblock {\em Physics Letters B}, 99(1):66--70, 1981.
\newblock URL:
  \url{https://www.sciencedirect.com/science/article/pii/0370269381908054},
  \href {https://doi.org/10.1016/0370-2693(81)90805-4}
  {\path{doi:10.1016/0370-2693(81)90805-4}}.

\bibitem{Linde1982}
A.~D. {Linde}.
\newblock {A new inflationary universe scenario: A possible solution of the
  horizon, flatness, homogeneity, isotropy and primordial monopole problems}.
\newblock {\em Physics Letters B}, 108(6):389--393, February 1982.
\newblock \href {https://doi.org/10.1016/0370-2693(82)91219-9}
  {\path{doi:10.1016/0370-2693(82)91219-9}}.

\bibitem{Albrecht1982}
Andreas Albrecht and Paul~J. Steinhardt.
\newblock Cosmology for grand unified theories with radiatively induced
  symmetry breaking.
\newblock {\em Phys. Rev. Lett.}, 48:1220--1223, Apr 1982.
\newblock URL: \url{https://link.aps.org/doi/10.1103/PhysRevLett.48.1220},
  \href {https://doi.org/10.1103/PhysRevLett.48.1220}
  {\path{doi:10.1103/PhysRevLett.48.1220}}.

\bibitem{Linde1983}
A.D. Linde.
\newblock Chaotic inflation.
\newblock {\em Physics Letters B}, 129(3):177--181, 1983.
\newblock URL:
  \url{https://www.sciencedirect.com/science/article/pii/0370269383908377},
  \href {https://doi.org/10.1016/0370-2693(83)90837-7}
  {\path{doi:10.1016/0370-2693(83)90837-7}}.

\bibitem{Hawking1982}
S.W. Hawking and I.L. Moss.
\newblock Supercooled phase transitions in the very early universe.
\newblock {\em Physics Letters B}, 110(1):35--38, 1982.
\newblock URL:
  \url{https://www.sciencedirect.com/science/article/pii/0370269382909467},
  \href {https://doi.org/10.1016/0370-2693(82)90946-7}
  {\path{doi:10.1016/0370-2693(82)90946-7}}.

\bibitem{Steinhardt1984}
Paul~J. Steinhardt and Michael~S. Turner.
\newblock Prescription for successful new inflation.
\newblock {\em Phys. Rev. D}, 29:2162--2171, May 1984.
\newblock URL: \url{https://link.aps.org/doi/10.1103/PhysRevD.29.2162}, \href
  {https://doi.org/10.1103/PhysRevD.29.2162}
  {\path{doi:10.1103/PhysRevD.29.2162}}.

\bibitem{Liddle1992}
Andrew~R. Liddle and David~H. Lyth.
\newblock Cobe, gravitational waves, inflation and extended inflation.
\newblock {\em Physics Letters B}, 291(4):391--398, 1992.
\newblock URL:
  \url{https://www.sciencedirect.com/science/article/pii/037026939291393N},
  \href {https://doi.org/10.1016/0370-2693(92)91393-N}
  {\path{doi:10.1016/0370-2693(92)91393-N}}.

\bibitem{Liddle1994}
Andrew~R. Liddle, Paul Parsons, and John~D. Barrow.
\newblock Formalizing the slow-roll approximation in inflation.
\newblock {\em Phys. Rev. D}, 50:7222--7232, Dec 1994.
\newblock URL: \url{https://link.aps.org/doi/10.1103/PhysRevD.50.7222}, \href
  {https://doi.org/10.1103/PhysRevD.50.7222}
  {\path{doi:10.1103/PhysRevD.50.7222}}.

\bibitem{Coughlan1985}
G.D. Coughlan and G.G. Ross.
\newblock Initial conditions for inflation.
\newblock {\em Physics Letters B}, 157(2):151--156, 1985.
\newblock URL:
  \url{https://www.sciencedirect.com/science/article/pii/0370269385915369},
  \href {https://doi.org/10.1016/0370-2693(85)91536-9}
  {\path{doi:10.1016/0370-2693(85)91536-9}}.

\bibitem{Linde1985}
A.D. Linde.
\newblock Initial conditions for inflation.
\newblock {\em Physics Letters B}, 162(4):281--286, 1985.
\newblock URL:
  \url{https://www.sciencedirect.com/science/article/pii/0370269385909232},
  \href {https://doi.org/10.1016/0370-2693(85)90923-2}
  {\path{doi:10.1016/0370-2693(85)90923-2}}.

\bibitem{Albrecht1985}
Andreas Albrecht and Robert~H. Brandenberger.
\newblock Realization of new inflation.
\newblock {\em Phys. Rev. D}, 31:1225--1231, Mar 1985.
\newblock URL: \url{https://link.aps.org/doi/10.1103/PhysRevD.31.1225}, \href
  {https://doi.org/10.1103/PhysRevD.31.1225}
  {\path{doi:10.1103/PhysRevD.31.1225}}.

\bibitem{Albrecht1985b}
Andreas Albrecht, Robert~H. Brandenberger, and Richard~A. Matzner.
\newblock Numerical analysis of inflation.
\newblock {\em Phys. Rev. D}, 32:1280--1289, Sep 1985.
\newblock URL: \url{https://link.aps.org/doi/10.1103/PhysRevD.32.1280}, \href
  {https://doi.org/10.1103/PhysRevD.32.1280}
  {\path{doi:10.1103/PhysRevD.32.1280}}.

\bibitem{Albrecht1987}
Andreas Albrecht, Robert Brandenberger, and Richard Matzner.
\newblock Inflation with generalized initial conditions.
\newblock {\em Phys. Rev. D}, 35:429--434, Jan 1987.
\newblock URL: \url{https://link.aps.org/doi/10.1103/PhysRevD.35.429}, \href
  {https://doi.org/10.1103/PhysRevD.35.429}
  {\path{doi:10.1103/PhysRevD.35.429}}.

\bibitem{Linde2005}
Andrei Linde.
\newblock Particle physics and inflationary cosmology, 2005.
\newblock URL: \url{https://arxiv.org/abs/hep-th/0503203}, \href
  {https://arxiv.org/abs/hep-th/0503203} {\path{arXiv:hep-th/0503203}}.

\bibitem{Linde2007}
Andrei Linde.
\newblock {\em Inflationary Cosmology}, pages 1--54.
\newblock Springer Berlin Heidelberg, Berlin, Heidelberg, 2007.
\newblock \href {https://doi.org/10.1007/978-3-540-74353-8_1}
  {\path{doi:10.1007/978-3-540-74353-8_1}}.

\bibitem{Olive1990}
Keith~A. Olive.
\newblock Inflation.
\newblock {\em Physics Reports}, 190(6):307--403, 1990.
\newblock URL:
  \url{https://www.sciencedirect.com/science/article/pii/037015739090144Q},
  \href {https://doi.org/10.1016/0370-1573(90)90144-Q}
  {\path{doi:10.1016/0370-1573(90)90144-Q}}.

\bibitem{Baumann2011}
DANIEL BAUMANN.
\newblock {\em INFLATION}, pages 523--686.
\newblock URL:
  \url{https://www.worldscientific.com/doi/abs/10.1142/9789814327183_0010},
  \href
  {https://arxiv.org/abs/https://www.worldscientific.com/doi/pdf/10.1142/9789814327183_0010}
  {\path{arXiv:https://www.worldscientific.com/doi/pdf/10.1142/9789814327183_0010}},
  \href {https://doi.org/10.1142/9789814327183_0010}
  {\path{doi:10.1142/9789814327183_0010}}.

\bibitem{Martin2014}
Jérôme Martin, Christophe Ringeval, and Vincent Vennin.
\newblock Encyclopædia inflationaris.
\newblock {\em Physics of the Dark Universe}, 5-6:75--235, 2014.
\newblock Hunt for Dark Matter.
\newblock URL:
  \url{https://www.sciencedirect.com/science/article/pii/S2212686414000053},
  \href {https://doi.org/10.1016/j.dark.2014.01.003}
  {\path{doi:10.1016/j.dark.2014.01.003}}.

\bibitem{Kurki-Suonio1987}
Hannu Kurki-Suonio, Joan Centrella, Richard~A. Matzner, and James~R. Wilson.
\newblock Inflation from inhomogeneous initial data in a one-dimensional
  back-reacting cosmology.
\newblock {\em Phys. Rev. D}, 35:435--448, Jan 1987.
\newblock URL: \url{https://link.aps.org/doi/10.1103/PhysRevD.35.435}, \href
  {https://doi.org/10.1103/PhysRevD.35.435}
  {\path{doi:10.1103/PhysRevD.35.435}}.

\bibitem{Goldwirth1990}
Dalia~S. Goldwirth and Tsvi Piran.
\newblock Inhomogeneity and the onset of inflation.
\newblock {\em Phys. Rev. Lett.}, 64:2852--2855, Jun 1990.
\newblock URL: \url{https://link.aps.org/doi/10.1103/PhysRevLett.64.2852},
  \href {https://doi.org/10.1103/PhysRevLett.64.2852}
  {\path{doi:10.1103/PhysRevLett.64.2852}}.

\bibitem{Goldwirth1992}
Dalia~S Goldwirth and Tsvi Piran.
\newblock Initial conditions for inflation.
\newblock {\em Physics Reports}, 214(4):223--292, 1992.
\newblock URL:
  \url{https://www.sciencedirect.com/science/article/pii/0370157392900739},
  \href {https://doi.org/10.1016/0370-1573(92)90073-9}
  {\path{doi:10.1016/0370-1573(92)90073-9}}.

\bibitem{Kurki-Suonio-1993}
Hannu Kurki-Suonio, Pablo Laguna, and Richard~A. Matzner.
\newblock Inhomogeneous inflation: Numerical evolution.
\newblock {\em Phys. Rev. D}, 48:3611--3624, Oct 1993.
\newblock URL: \url{https://link.aps.org/doi/10.1103/PhysRevD.48.3611}, \href
  {https://doi.org/10.1103/PhysRevD.48.3611}
  {\path{doi:10.1103/PhysRevD.48.3611}}.

\bibitem{Iguchi1997}
Osamu Iguchi and Hideki Ishihara.
\newblock Onset of inflation in inhomogeneous cosmology.
\newblock {\em Phys. Rev. D}, 56:3216--3224, Sep 1997.
\newblock URL: \url{https://link.aps.org/doi/10.1103/PhysRevD.56.3216}, \href
  {https://doi.org/10.1103/PhysRevD.56.3216}
  {\path{doi:10.1103/PhysRevD.56.3216}}.

\bibitem{Vachaspati1999}
Tanmay Vachaspati and Mark Trodden.
\newblock Causality and cosmic inflation.
\newblock {\em Phys. Rev. D}, 61:023502, Dec 1999.
\newblock URL: \url{https://link.aps.org/doi/10.1103/PhysRevD.61.023502}, \href
  {https://doi.org/10.1103/PhysRevD.61.023502}
  {\path{doi:10.1103/PhysRevD.61.023502}}.

\bibitem{Easther2014}
Richard Easther, Layne~C. Price, and Javier Rasero.
\newblock Inflating an inhomogeneous universe.
\newblock {\em Journal of Cosmology and Astroparticle Physics}, 2014(08):041,
  aug 2014.
\newblock URL: \url{https://dx.doi.org/10.1088/1475-7516/2014/08/041}, \href
  {https://doi.org/10.1088/1475-7516/2014/08/041}
  {\path{doi:10.1088/1475-7516/2014/08/041}}.

\bibitem{East2016}
William~E. East, Matthew Kleban, Andrei Linde, and Leonardo Senatore.
\newblock Beginning inflation in an inhomogeneous universe.
\newblock {\em Journal of Cosmology and Astroparticle Physics}, 2016(09):010,
  sep 2016.
\newblock URL: \url{https://dx.doi.org/10.1088/1475-7516/2016/09/010}, \href
  {https://doi.org/10.1088/1475-7516/2016/09/010}
  {\path{doi:10.1088/1475-7516/2016/09/010}}.

\bibitem{Kofman1997}
Lev Kofman, Andrei Linde, and Alexei~A. Starobinsky.
\newblock Towards the theory of reheating after inflation.
\newblock {\em Phys. Rev. D}, 56:3258--3295, Sep 1997.
\newblock URL: \url{https://link.aps.org/doi/10.1103/PhysRevD.56.3258}, \href
  {https://doi.org/10.1103/PhysRevD.56.3258}
  {\path{doi:10.1103/PhysRevD.56.3258}}.

\bibitem{Greene1997}
Patrick~B. Greene, Lev Kofman, Andrei Linde, and Alexei~A. Starobinsky.
\newblock Structure of resonance in preheating after inflation.
\newblock {\em Phys. Rev. D}, 56:6175--6192, Nov 1997.
\newblock URL: \url{https://link.aps.org/doi/10.1103/PhysRevD.56.6175}, \href
  {https://doi.org/10.1103/PhysRevD.56.6175}
  {\path{doi:10.1103/PhysRevD.56.6175}}.

\bibitem{kaiser1998}
David~I. Kaiser.
\newblock Resonance structure for preheating with massless fields.
\newblock {\em Phys. Rev. D}, 57:702--711, Jan 1998.
\newblock URL: \url{https://link.aps.org/doi/10.1103/PhysRevD.57.702}, \href
  {https://doi.org/10.1103/PhysRevD.57.702}
  {\path{doi:10.1103/PhysRevD.57.702}}.

\bibitem{kaiser1997}
David~I. Kaiser.
\newblock Preheating in an expanding universe: Analytic results for the
  massless case.
\newblock {\em Phys. Rev. D}, 56:706--716, Jul 1997.
\newblock URL: \url{https://link.aps.org/doi/10.1103/PhysRevD.56.706}, \href
  {https://doi.org/10.1103/PhysRevD.56.706}
  {\path{doi:10.1103/PhysRevD.56.706}}.

\bibitem{Calzetta1995}
E.~Calzetta and B.~L. Hu.
\newblock Quantum fluctuations, decoherence of the mean field, and structure
  formation in the early universe.
\newblock {\em Phys. Rev. D}, 52:6770--6788, Dec 1995.
\newblock URL: \url{https://link.aps.org/doi/10.1103/PhysRevD.52.6770}, \href
  {https://doi.org/10.1103/PhysRevD.52.6770}
  {\path{doi:10.1103/PhysRevD.52.6770}}.

\bibitem{Bassett1999}
Bruce~A. Bassett, Fabrizio Tamburini, David~I. Kaiser, and Roy Maartens.
\newblock Metric preheating and limitations of linearized gravity.
\newblock {\em Nuclear Physics B}, 561(1):188--240, 1999.
\newblock URL:
  \url{https://www.sciencedirect.com/science/article/pii/S0550321399004952},
  \href {https://doi.org/10.1016/S0550-3213(99)00495-2}
  {\path{doi:10.1016/S0550-3213(99)00495-2}}.

\bibitem{Easther2000}
Richard Easther and Matthew Parry.
\newblock Gravity, parametric resonance, and chaotic inflation.
\newblock {\em Phys. Rev. D}, 62:103503, Oct 2000.
\newblock URL: \url{https://link.aps.org/doi/10.1103/PhysRevD.62.103503}, \href
  {https://doi.org/10.1103/PhysRevD.62.103503}
  {\path{doi:10.1103/PhysRevD.62.103503}}.

\bibitem{Jin_2006}
Yoshida Jin and Shinji Tsujikawa.
\newblock Chaotic dynamics in preheating after inflation.
\newblock {\em Classical and Quantum Gravity}, 23(2):353, dec 2005.
\newblock URL: \url{https://dx.doi.org/10.1088/0264-9381/23/2/006}, \href
  {https://doi.org/10.1088/0264-9381/23/2/006}
  {\path{doi:10.1088/0264-9381/23/2/006}}.

\bibitem{Podolsky2002}
D.~I. Podolsky and A.~A. Starobinsky.
\newblock Chaotic reheating, 2002.
\newblock URL: \url{https://arxiv.org/abs/astro-ph/0204327}, \href
  {https://arxiv.org/abs/astro-ph/0204327} {\path{arXiv:astro-ph/0204327}}.

\bibitem{Nambu_2006}
Yasusada Nambu and Yohei Araki.
\newblock Evolution of nonlinear fluctuations in preheating after inflation.
\newblock {\em Classical and Quantum Gravity}, 23(2):511, dec 2005.
\newblock URL: \url{https://dx.doi.org/10.1088/0264-9381/23/2/015}, \href
  {https://doi.org/10.1088/0264-9381/23/2/015}
  {\path{doi:10.1088/0264-9381/23/2/015}}.

\bibitem{Suyama_2007}
Teruaki Suyama and Shuichiro Yokoyama.
\newblock Analysis of the evolution of curvature perturbations in preheating by
  using $\delta n$ formalism.
\newblock {\em Classical and Quantum Gravity}, 24(6):1615, mar 2007.
\newblock URL: \url{https://dx.doi.org/10.1088/0264-9381/24/6/015}, \href
  {https://doi.org/10.1088/0264-9381/24/6/015}
  {\path{doi:10.1088/0264-9381/24/6/015}}.

\bibitem{Jedamzik_2010}
Karsten Jedamzik, Martin Lemoine, and Jérôme Martin.
\newblock Collapse of small-scale density perturbations during preheating in
  single field inflation.
\newblock {\em Journal of Cosmology and Astroparticle Physics}, 2010(09):034,
  sep 2010.
\newblock URL: \url{https://dx.doi.org/10.1088/1475-7516/2010/09/034}, \href
  {https://doi.org/10.1088/1475-7516/2010/09/034}
  {\path{doi:10.1088/1475-7516/2010/09/034}}.

\bibitem{Planck2018}
Y. Akrami~et~al.
\newblock Planck 2018 results - X. constraints on inflation.
\newblock {\em A\&A}, 641:A10, 2020.
\newblock \href {https://doi.org/10.1051/0004-6361/201833887}
  {\path{doi:10.1051/0004-6361/201833887}}.

\bibitem{Bicep2015}
P.~A. R. Ade et~al.
\newblock Joint analysis of BICEP2/Keck array and Planck data.
\newblock {\em Phys. Rev. Lett.}, 114:101301, Mar 2015.
\newblock URL: \url{https://link.aps.org/doi/10.1103/PhysRevLett.114.101301},
  \href {https://doi.org/10.1103/PhysRevLett.114.101301}
  {\path{doi:10.1103/PhysRevLett.114.101301}}.

\end{thebibliography}

\end{document}